\pdfoutput=1

\documentclass[11pt,twoside,a4paper,cmspaper,final,collab]{cms-tdr}
\def\svnVersion{463271P}\def\cmsCernNoTag{CERN-EP-2017-292}\def\cmsCernDate{\today}\def\cmsMessage{Published in the Journal of High Energy Physics as \href{http://dx.doi.org/10.007/JHEP06(2018)001}{\doi{10.007/JHEP06(2018)001.}}}

\begin{document}\cmsNoteHeader{HIG-17-001}

\hyphenation{had-ron-i-za-tion}
\hyphenation{cal-or-i-me-ter}
\hyphenation{de-vices}
\RCS$Revision: 462162 $
\RCS$HeadURL: svn+ssh://svn.cern.ch/reps/tdr2/papers/HIG-17-001/trunk/HIG-17-001.tex $
\RCS$Id: HIG-17-001.tex 462162 2018-05-29 14:29:12Z taroni $
\newlength\cmsFigWidth
\ifthenelse{\boolean{cms@external}}{\setlength\cmsFigWidth{0.85\columnwidth}}{\setlength\cmsFigWidth{0.4\textwidth}}
\ifthenelse{\boolean{cms@external}}{\providecommand{\cmsLeft}{top\xspace}}{\providecommand{\cmsLeft}{left\xspace}}
\ifthenelse{\boolean{cms@external}}{\providecommand{\cmsRight}{bottom\xspace}}{\providecommand{\cmsRight}{right\xspace}}
\providecommand{\NA}{\ensuremath{\text{---}}}
\newlength\cmsTabSkip
\setlength\cmsTabSkip{2ex}
\cmsNoteHeader{HIG-17-001}
\title{Search for lepton flavour violating  decays of the Higgs boson to $\mu\tau$  and  $\Pe\tau$ in proton-proton collisions at $\sqrt{s}=13\TeV$}

\date{\today}

\abstract{
A search for lepton flavour violating decays of the Higgs boson in the $\mu\tau$ and $\Pe\tau$ decay modes is presented. The search is based on a data set corresponding to an integrated luminosity of 35.9\fbinv of  proton-proton collisions collected with the CMS detector in 2016, at a centre-of-mass energy of 13\TeV. No significant excess over the standard model expectation is observed. The observed (expected) upper limits on the lepton flavour violating branching fractions of the Higgs boson are $\mathcal{B}(\PH\to\mu\tau) < 0.25\%\,(0.25\%)$ and $\mathcal{B}(\PH\to\Pe\tau) < 0.61\%\,(0.37\%)$, at 95\% confidence level. These results are used to derive upper limits on the off-diagonal $\mu\tau$ and $\Pe\tau$ Yukawa couplings $\sqrt{\smash[b]{|{Y_{\mu\tau}}|^{2}+|{Y_{\tau\mu}}|^{2}}}<1.43\times 10^{-3}$ and $\sqrt{|{Y_{\Pe\tau}}|^{2}+|{Y_{\tau\Pe}}|^{2}}<2.26\times 10^{-3}$  at 95\% confidence level. The limits on the lepton flavour violating branching fractions of the Higgs boson and on the associated Yukawa couplings are the
most stringent to date.
}

\hypersetup{%
pdfauthor={CMS Collaboration},%
pdftitle={Search for lepton flavour violating  decays of the Higgs boson to mu tau  and  e tau in proton-proton collisions at sqrt(s)=13 TeV},%
pdfsubject={CMS},%
pdfkeywords={Higgs, electrons, muons, taus, lepton-flavour-violation}}

\providecommand{\mt}{\ensuremath{M_\mathrm{T}}\xspace}
\newcommand{\Htt}{\ensuremath{\PH \to \Pgt \Pgt}\xspace}
\newcommand{\Het}{\ensuremath{\PH \to \Pe \Pgt}\xspace}
\newcommand{\Hmue}{\ensuremath{\PH \to \Pgm \Pgt_{\Pe}}\xspace}
\newcommand{\Hmuhad}{\ensuremath{\PH \to \Pgm \Pgt_{\text{h}}}\xspace}
\newcommand{\Hehad}{\ensuremath{\PH \to \Pe \Pgt_{\text{h}}}\xspace}
\newcommand{\Hemu}{\ensuremath{\PH \to \Pe \Pgt_{\Pgm}}\xspace}
\newcommand{\mue}{\ensuremath{\Pgm \Pgt_{\Pe}}\xspace}
\newcommand{\emu}{\ensuremath{\Pe \Pgt_{\Pgm}}\xspace}
\newcommand{\muhad}{\ensuremath{\Pgm \Pgt_{\text{h}}}\xspace}
\newcommand{\ehad}{\ensuremath{\Pe \Pgt_{\text{h}}}\xspace}
\newcommand{\mcol}{\ensuremath{M_{\text{col}}}\xspace}
\newcommand{\mvis}{\ensuremath{M_{\text{vis}}}\xspace}
\newcommand{\wjets}{\ensuremath{\PW+\text{jets}}\xspace}
\newcommand{\zjets}{\ensuremath{\cPZ+\text{jets}}\xspace}
\newcommand{\aMCATNLO} {\textsc{MG5}\_a\MCATNLO\xspace}

\maketitle

\section{Introduction}

The discovery of the Higgs boson (H) at the CERN LHC~\cite{Aad:2012tfa, Chatrchyan:2012ufa, Chatrchyan:2013lba} has stimulated further precision measurements of the properties of the new particle. A combined study of the 7 and 8\TeV data sets collected by the CMS and ATLAS collaborations shows consistency between the measured couplings of the Higgs boson and the standard model (SM) predictions~\cite{Khachatryan:2016vau}. However, the constraint on the branching fraction to non-SM decay modes derived from these measurements, $\mathcal{B}(\text{non-SM})<34$\% at 95\% confidence level (CL), still  allows for a significant contribution from exotic decays~\cite{Khachatryan:2016vau}.

In this paper a search for lepton flavour violating (LFV) decays of the Higgs boson in the  $\Pgm\Pgt$ and $\Pe\Pgt$ channels is presented. These decays are  forbidden in the SM but occur in many new physics scenarios.
These include supersymmetric~\cite{DiazCruz:1999xe,Han:2000jz,Arganda:2004bz,Arhrib:2012ax,Arana-Catania:2013xma,Arganda:2015uca,Arganda:2015naa,Gomez:2017dhl,Zhang:2015csm}, composite Higgs~\cite{Agashe:2009di,Azatov:2009na}, or Randall--Sundrum models~\cite{Perez:2008ee,Casagrande:2008hr,Buras:2009ka}, SM extensions with more than one Higgs boson doublet~\cite{PhysRevLett.38.622,Lee:2016dcb} or with flavour
symmetries~\cite{Ishimori:2010au}, and many other scenarios~\cite{Blanke:2008zb,Giudice:2008uua,AguilarSaavedra:2009mx,Albrecht:2009xr,Goudelis:2011un,McKeen:2012av, Pilaftsis199268,PhysRevD.47.1080,Arganda:2014dta,Herrero-Garcia:2016uab,delAguila:2017ugt,Thao:2017qtn,Galon:2017qes,Arganda:2017vdb,Choudhury:2016ulr}.
The presence of LFV Higgs boson couplings would allow $\Pgt \to \Pgm$ and $\Pgt \to \Pe$  to proceed via a virtual Higgs
boson~\cite{McWilliams:1980kj,Shanker:1981mj}. Consequently the  experimental limits on rare \Pgt\  lepton decays, such as $\Pgt \to \Pe \gamma$ and $\Pgt \to \Pgm \gamma$~\cite{Celis:2013xja}, provide upper limits on  $\mathcal{B}(\PH \to \Pgm \Pgt)$ and $\mathcal{B}(\PH \to \Pe \Pgt)$~\cite{Blankenburg:2012ex,Harnik:2012pb} of
$\mathcal{O}(10\%)$. Measurements of the electron and muon magnetic moments, and exclusion limits on the electric dipole moment of the electron also provide complementary constraints~\cite{Barr:1990vd}. The LFV Higgs boson decay to $\Pgm\Pe$ is strongly constrained by the $\Pgm \to \Pe \gamma$ limit,  $\mathcal{B}(\PH \to \Pe \Pgm) < \mathcal{O}(10^{-9})$~\cite{TheMEG:2016wtm}.

The CMS experiment published the first direct search for  $\PH \to \Pgm \Pgt$
~\cite{Khachatryan:2015kon}, followed by searches for $\PH \to \Pe \Pgt$ and $\PH \to \Pe \Pgm$ decays~\cite{HIG-14-040}, using  proton-proton ($\Pp\Pp$) collision data corresponding to an integrated luminosity of 19.7\fbinv at a centre-of-mass energy of 8\TeV.  A small excess of data with respect to the SM background-only hypothesis at
$m_{\PH} =125$\GeV was observed in the $\PH \to \Pgm \Pgt$ channel, with a significance of $2.4$ standard deviations ($\sigma$), and the best fit for the branching fraction was found to be  $\mathcal{B}(\PH \to \Pgm \Pgt)=(0.84^{+0.39}_{-0.37})\%$. A constraint was set on the observed (expected) branching fraction  $\mathcal{B}(\PH \to \Pgm \Pgt)<1.51\%$ (0.75\%)
at 95\% CL. No excess of events over the estimated background was observed in the $\PH \to \Pe \Pgt$ or $\PH \to \Pe \Pgm$ channels, and observed (expected) upper limits on the branching fractions  $\mathcal{B}(\PH \to \Pe \Pgt )<0.69\%$ (0.75\%) and $\mathcal{B}(\PH\to\Pe\Pgm )<0.035\%$ (0.048\%) at 95\% CL were set.
The ATLAS Collaboration reported  searches for $\PH \to \Pe \Pgt$ and $\PH \to \Pgm \Pgt$ using  $\Pp\Pp$ collision data at a centre-of-mass energy of 8\TeV, finding no significant excess of events over the background expectation, and set observed (expected) limits of $\mathcal{B}(\PH \to \Pgm \Pgt)<1.43\%$ (1.01\%) and $\mathcal{B}(\PH \to \Pe \Pgt)<1.04\%$ (1.21\%) at 95\% CL~\cite{Aad:2016blu,Aad:2015gha}.

The search described in this paper is performed in  four  decay channels, $\PH \to \Pgm \tauh$,   $\PH \to \Pgm \Pgt_{\Pe}$,
$\PH \to \Pe \tauh$,  $\PH \to \Pe \Pgt_{\Pgm}$, where $\tauh$, $\Pgt_{\Pe}$, and  $\Pgt_{\Pgm}$ correspond to the hadronic, electronic, and muonic  decay channels of $\Pgt$ leptons, respectively.  The decay channels $\PH \to \Pe \Pgt_{\Pe}$ and $\PH \to \Pgm \Pgt_{\Pgm}$, are
not considered because of the large background contribution from $\cPZ$ boson decays. The expected final state signatures are  very similar to those for the SM $\PH \to \Pgt\Pgt$ decays, studied by CMS~\cite{Chatrchyan:2014vua,CMS-PAPERS-HIG-13-004,CMS-PAS-HIG-16-043} and ATLAS~\cite{Aad:2015vsa},
but with some significant kinematic differences. The electron (muon) in the LFV $\PH \to \Pe(\Pgm) \Pgt$ decay is produced promptly, and tends to have a larger momentum than in the SM $\PH \to \Pgt_{\Pe(\Pgm)}\tauh$ decay. The search reported  in this paper improves upon the sensitivity of the
earlier CMS searches~\cite{Khachatryan:2015kon,HIG-14-040} by using a boosted decision trees (BDT) discriminator to distinguish signal from background events. A separate analysis, similar in strategy to the previous CMS publications, is performed as cross check. The results of both strategies are reported in this paper.

This paper is organized as follows. After a description of the CMS detector (Section~\ref{cmsdet}) and of the collision data and simulated samples used in the analyses (Section~\ref{samples}), the event reconstruction is described in Section~\ref{reconst}. The event selection is  described separately for the two Higgs boson decay modes $\PH \to \Pe \Pgt$ and $\PH \to \Pgm \Pgt$ in Section~\ref{eventsel}. The backgrounds, which are common to all  channels but with different rates in each, are described in Section~\ref{backgrounds}. The systematic uncertainties are described in Section~\ref{sec:systematics} and the results are then presented in Section~\ref{results}.

\section{The CMS detector\label{cmsdet}}
The central feature of the CMS apparatus is a superconducting solenoid of 6\unit{m} internal diameter, providing a magnetic field of 3.8\unit{T}. Within the solenoid volume are a silicon pixel and strip tracker, a lead tungstate crystal electromagnetic calorimeter (ECAL), and a brass and scintillator hadron calorimeter (HCAL), each composed of a barrel and two endcap sections. Forward calorimeters extend the pseudorapidity ($\eta$) coverage provided by the barrel and endcap detectors. Muons are detected in gas-ionization chambers embedded in the steel flux-return yoke outside the solenoid.
 The two-level CMS trigger system selects events of interest for
permanent storage~\cite{Khachatryan:2016bia}. The first trigger level,
composed of custom hardware processors, uses information from the
calorimeters and muon detectors to select events at a rate of around 100\unit{kHz} within a time interval of less than 4\mus.
The software algorithms of the high-level trigger, executed on a farm of
commercial processors, reduce the event rate to about 1\unit{kHz} using information from all detector subsystems.
A detailed description of the CMS detector, together with a definition of the coordinate system used and the relevant kinematic variables, can be found in Ref.~\cite{CMS-JINST}.

\section{Collision data and simulated events \label{samples}}
The analyses presented here  use samples of $\Pp\Pp$ collisions collected in 2016 by the CMS experiment
at the LHC at a centre-of-mass energy of $\sqrt{s}=13$\TeV, corresponding
to an integrated luminosity of 35.9\fbinv. Isolated single muon triggers
are used to collect the data samples in the $\PH \to \Pgm\Pgt$ search. Triggers requiring a single isolated electron, or a combination of an electron and a muon, are used in the $\PH \to \Pe\tauh$ and $\PH \to \Pe\Pgt_\Pgm$ channels, respectively.
Simulated samples of signal and background events are produced with several event generators.
The Higgs bosons are produced in $\Pp\Pp$ collisions predominantly by gluon fusion ({\cPg\cPg}\PH)~\cite{Georgi:1977gs},
but also by vector boson fusion (VBF)~\cite{Cahn:1986zv}, and in
association with a $\PW$ or $\cPZ$ boson~\cite{Glashow:1978ab}.
The {\cPg\cPg\PH} and VBF Higgs boson samples
are generated with \POWHEG 2.0~\cite{Nason:2004rx,Frixione:2007vw, Alioli:2010xd, Alioli:2010xa, Alioli:2008tz, Bagnaschi:2011tu} while the
 \textsc{minlo hvJ}~\cite{Luisoni:2013kna} extension of \POWHEG~2.0 is used for the $\PW\PH$ and $\PZ\PH$ simulated samples.
The \aMCATNLO~\cite{Alwall:2014} generator is used for $\PZ+\text{jets}$ and $\PW+\text{jets}$ processes. They are simulated at leading order (LO) with the MLM jet matching and merging~\cite{Alwall:2007fs}. Diboson production is simulated at  next-to-LO (NLO) using  \aMCATNLO generator with the FxFx jet matching and merging~\cite{Frederix:2012ps}, whereas $\POWHEG$ 2.0 and 1.0 are used for $\ttbar$ and single top quark production, respectively.
The \POWHEG and \MADGRAPH generators are interfaced with \PYTHIA~8.212 ~\cite{Sjostrand:2014zea}  for parton showering, fragmentation, and decays.
The \PYTHIA parameters for the underlying event description are set
to the {CUETP8M1} tune~\cite{Khachatryan:2015pea}.
Due to the high instantaneous luminosities attained during data taking, many events have multiple
$\Pp\Pp$ interactions per bunch crossing (pileup). The effect is taken into account in simulated samples, by generating concurrent minimum bias events. All simulated samples are weighted
to match the pileup distribution observed in
data, that has an average of approximately 27 interactions per bunch crossing.
The CMS detector response is modelled using \GEANTfour~\cite{GEANT4}.

\section{Event reconstruction \label{reconst}}

The global event reconstruction is performed using a particle-flow (PF) algorithm, which reconstructs and identifies each individual particle with an optimized combination of all subdetector information~\cite{Sirunyan:2017ulk}.
In this process, the identification of the particle type (photon, electron, muon, charged or  neutral hadron) plays an important role in the determination of the particle direction and energy. The primary $\Pp\Pp$  vertex of the event is identified as the reconstructed vertex with the largest value of summed physics-object $\pt^2$, where \pt is the transverse momentum. The physics objects are returned by a jet finding algorithm~\cite{Cacciari:2008gp,Cacciari:2011ma} applied to all charged tracks associated with the vertex, plus the corresponding associated missing transverse momentum.

A muon is identified as a track in the silicon detectors, consistent with the primary $\Pp\Pp$  vertex and with either a track or several hits in the muon system,  associated with an energy deposit in the calorimeters compatible with the expectations for a muon~\cite{Chatrchyan:2012xi,Sirunyan:2017ulk}. Identification is based on the number of spacial points measured in the tracker and in the muon system, the track quality and its consistency with the event vertex location. The energy is obtained from the corresponding track momentum.

An electron is identified as a  charged particle track from the primary $\Pp\Pp$  vertex in combination with one or more ECAL energy clusters. These clusters  correspond to the track extrapolation to the ECAL and to possible bremsstrahlung photons emitted when interacting with the material of the tracker~\cite{Khachatryan:2015hwa}. Electron candidates  are accepted in the range $\abs{\eta}<2.5$, with the exception
of the region $1.44 < \abs{\eta} < 1.57$ where service infrastructure for the detector
is located. They are identified using a multivariate (MVA) discriminator  that combines observables sensitive to the
amount of bremsstrahlung along the electron trajectory, the
geometrical and momentum matching between the electron trajectory and
associated clusters as well as various shower shape observables in the calorimeters. Electrons from photon conversions are removed.
The energy of electrons is determined from a combination of the track momentum at the primary vertex, the corresponding ECAL cluster energy, and the energy sum of all bremsstrahlung photons attached to the track.

Hadronically decaying $\Pgt$ leptons are reconstructed and identified using the hadrons-plus-strips (HPS) algorithm~\cite{Khachatryan:2015dfa,CMS:2016gvn}.
The reconstruction starts from a jet and searches for the products of the main $\Pgt$ lepton decay modes: one charged hadron and up to two neutral pions, or three charged hadrons.
To improve the reconstruction efficiency in the case of conversion of the photons from neutral-pion decay, the algorithm considers the PF photons and electrons from a  strip along the azimuthal direction  $\phi$.
The charges of all the PF objects from tau lepton decay, except for the electrons from neutral pions, are summed to reconstruct the tau lepton charge.
An MVA discriminator, based on the information of the reconstructed tau lepton and of the charged particles in a cone around it,
is used to reduce the rate for quark- and gluon-initiated jets identified as \Pgt\ candidates.
The working point used in the analysis  has an efficiency of about 60\% for a genuine \tauh, with approximately a  0.5\% misidentification rate for quark and gluon jets~\cite{CMS:2016gvn}. Additionally, muons and electrons misidentified as tau leptons are rejected  using a dedicated set of selection criteria based on the consistency between the measurements in the tracker, calorimeters, and muon detectors.
The specific identification criteria depend on the final state studied and
on the background composition. The tau leptons that decay to muons and electrons are reconstructed
as prompt muons and electrons as described above.

Charged hadrons are identified as charged particle tracks from the primary $\Pp\Pp$  vertex  neither reconstructed as electrons nor as muons nor as $\tau$ leptons. Neutral hadrons are identified as HCAL energy clusters not assigned to any charged hadron, or as ECAL and HCAL energy excesses with respect to the expected charged-hadron energy deposit. All the PF candidates are clustered into hadronic jets
using the infrared and collinear safe anti-\kt algorithm~\cite{Cacciari:2008gp}, implemented in the \FASTJET package~\cite{Cacciari:fastjet},  with a distance parameter of 0.4. The jet momentum is determined as the vector sum of all particle momenta in this jet, and is found in the simulation to be on average within 10\% of the true momentum over the whole \pt spectrum and detector acceptance. An offset correction is applied to jet energies to take into account the contribution from pileup~\cite{CMS-JME-10-011}. Jet energy corrections are derived from the simulation, and are confirmed with in situ measurements of the energy balance of dijet, multijet,  $\text{photon}+\text{jet}$, and $\PZ+\text{jet}$ events~\cite{Khachatryan:2016kdb}.  The variable $\Delta R = \sqrt {\smash[b]{(\Delta\eta)^2 +(\Delta\phi)^2}}$
is used to measure the separation between reconstructed objects in the detector. Any jet within  $\Delta R = 0.4$ of the identified leptons is removed.

Jets misidentified as electrons, muons, or tau leptons are suppressed by imposing  isolation requirements.
The muon (electron) isolation is measured relative to its $\pt^{\ell}$ ($\ell = \Pe, \Pgm$), by summing over the $\pt$ of PF particles
in a cone with $\Delta R = 0.4$ (0.3) around the lepton:
\ifthenelse{\boolean{cms@external}}{
\begin{multline*}
I_\text{rel}^{\ell} = \Bigg( \sum  \pt^\text{charged}\\
+ \text{max}\Big[ 0, \sum \pt^\text{neutral}
+  \sum \pt^{\gamma} - \pt^\text{PU}\left(\ell\right)  \Big] \Bigg) \Big/  \pt^{\ell}.
\end{multline*}
}{
\begin{equation*}
I_\text{rel}^{\ell} = \left( \sum  \pt^\text{charged} + \text{max}\left[ 0, \sum \pt^\text{neutral}
                                 +  \sum \pt^{\gamma} - \pt^\text{PU}\left(\ell\right)  \right] \right) /  \pt^{\ell},
\end{equation*}
}
where $\pt^\text{charged}$, $\pt^\text{neutral}$, and $\pt^{\gamma}$  indicate the \pt of a charged particle, a neutral particle, and a photon within the cone, respectively. The neutral contribution to isolation from pileup, $\pt^\text{PU}\left(\ell\right)$, is estimated from  the area of the jet and the average energy density of the event~\cite{1126-6708-2008-04-005, CACCIARI2008119} for the electron or from the  sum of transverse momenta of charged hadrons not originating
from the primary vertex scaled by a factor of 0.5 for the muons.
The charged contribution to isolation from pileup is rejected requiring the tracks to originate from the primary vertex.

All the reconstructed particles in the event are used to estimate the missing transverse momentum, $\ptvecmiss$, which is defined as the negative of the vector \ptvec sum of all identified PF objects in the event~\cite{Khachatryan:2014gga}.
Its magnitude is referred to as \ptmiss.

The transverse mass $\mt(\ell)$ is a variable formed from the lepton momentum and the missing transverse momentum vectors:
$ {\mt}(\ell)=\sqrt{\smash[b]{2|\ptvec^{\ell}||\ptvecmiss|(1-\cos{\Delta \phi_{\ell - \ptmiss}})}} $, where $\Delta \phi_{\ell - \ptmiss}$ is the angle in the transverse plane between the lepton and the missing transverse momentum.
It is used to discriminate the Higgs boson signal candidates from the $\PW+\text{jets}$ background.
The collinear mass, $M_{\text{col}}$, provides an estimate of $m_{\PH}$ using the observed
decay products of the Higgs boson candidate. It is reconstructed using the collinear approximation based on
the observation that, since $m_{\PH}\gg m_{\Pgt}$, the $\Pgt$ lepton decay products are
highly Lorentz boosted in the direction of the  $\Pgt$ candidate~\cite{Ellis:1987xu}.
The neutrino momenta can be approximated to have the same
direction as the other visible decay products of the $\Pgt$ ($\vec{\Pgt}^\text{vis}$)
and the component of the $\ptvecmiss$ in the direction of the visible $\Pgt$ lepton  decay products is used to
estimate the transverse component of the neutrino momentum ($\pt^{\nu,\,\text{est}}$). The collinear
mass can then be derived from the visible mass of the $\Pgt$-$\Pgm$ or $\Pgt$-$\Pe$ system ($M_{\text{vis}}$) as
$M_{\text{col}}= M_{\text{vis}} / \sqrt{\smash[b]{x_{\Pgt}^\text{vis}}}$, where $x_{\Pgt}^\text{vis}$ is the
fraction of energy carried by the visible decay products of the $\Pgt$
($x_{\Pgt}^\text{vis}={\pt^{\vec{\Pgt}^{\text{vis}}}}/{(\pt^{\vec{\Pgt}^\text{vis}}+\pt^{\nu,\,\text{est}})}$), and  $M_\text{vis}$ is the invariant mass of the visible decay products.

\section{Event selection \label{eventsel}}
The signal contains a prompt isolated lepton, \Pgm\ or \Pe, along with an oppositely charged isolated lepton of different flavour ($\Pgt_{\Pgm}$, $\Pgt_{\Pe}$ or $\tauh$). In each decay mode
a loose selection of this signature is defined first. The events are then divided into categories within each sample according to the number of jets in the event. This is designed to enhance the contribution of different Higgs boson production mechanisms. The jets are required to have $\pt>30$\GeV and $\abs{\eta}<4.7$. The 0-jet category enhances the {\cPg\cPg\PH} contribution, while the 1-jet category enhances {\cPg\cPg}H production with initial-state radiation. The 2-jet {\cPg\cPg}H category has a further requirement that the invariant mass of the two jets $M_{jj}<550$\GeV while the 2-jet VBF category with the requirement $M_{jj}\geq550$\GeV enhances the VBF contribution. The threshold on $M_{jj}$ has been optimized to give the best expected exclusion limits. The definition of the categories is the same in all the channels except in the \Het\ channels where the $M_{jj}$ threshold is 500\GeV, which optimizes the expected limits for this channel.

After the loose selection,
a binned likelihood is used to fit the distribution of a BDT discriminator for
the signal and the background contributions. This is referred to as the BDT fit analysis.
As a cross-check an analysis using a tighter set of selection criteria is also presented. In this case, selection requirements are placed on the kinematic variables and a fit is performed to the  \mcol distribution. This is referred to as the \mcol fit analysis. Requirements on additional kinematic variables such as $\mt(\ell)$ are chosen to obtain the most stringent expected limits. The lepton \pt has been excluded from this optimization to avoid biasing the selection toward  energetic leptons that sculpt  the background \mcol distribution to mimic the signal peak. This effect would reduce  the shape discrimination power of the signal extraction procedure.

\subsection{\texorpdfstring{\Hmuhad}{H to mu tau[h]}}
The loose selection begins by requiring  an isolated \Pgm\ and an isolated \tauh of opposite charge and  separated by $\Delta R > 0.3$.  The muon candidate is  required to have $\pt^{\Pgm} > 26$\GeV,  $|\eta^{\Pgm}|<2.4$ and
$I_\text{rel}^{\Pgm} < 0.15$. The hadronic tau candidate is required to have $\pt^{\tauh}>30$\GeV and $|\eta^{\tauh}| < 2.3$. The isolation requirement for the \tauh candidates is included in the MVA used for the HPS identification algorithm described in Section~\ref{reconst}. Events with additional \Pe, \Pgm\ or \tauh candidates are vetoed. Events with
at least one jet identified by the combined secondary vertex b-tagging algorithm~\cite{CMS-PAS-BTV-15-001} as arising from a b quark, are also vetoed in order to suppress the \ttbar\ background.  The tighter selection used for the \mcol fit analysis further requires $\mt(\tauh) <  105$\GeV in the 0-, 1- and 2-jet {\cPg\cPg}H categories, and $\mt(\tauh) <  85$\GeV in the 2-jet VBF category. The selections are summarized in Table~\ref{mutauSel}.

\begin{figure*}[!htpb]\centering
 \includegraphics[width=0.315\textwidth]{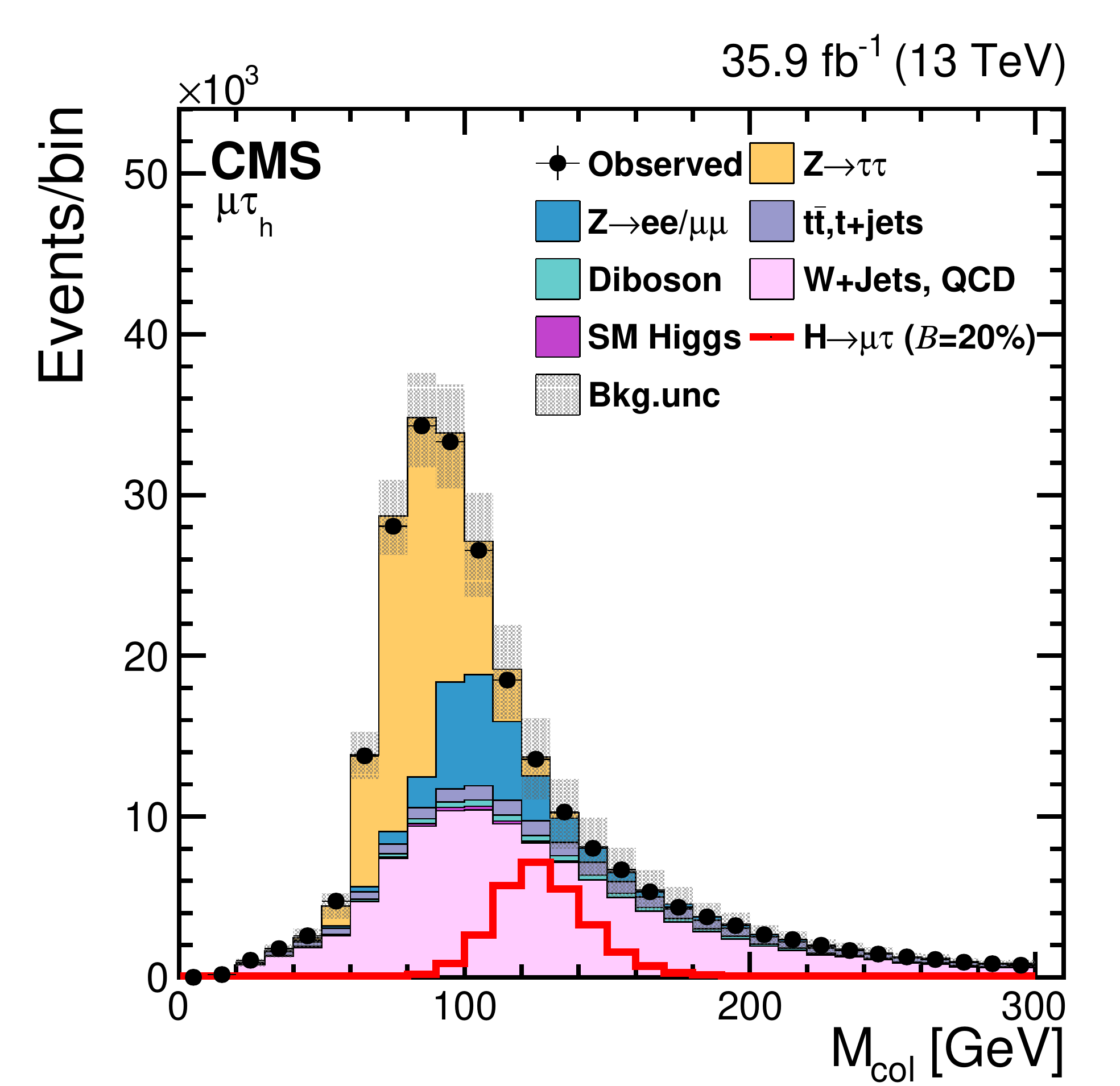}
 \includegraphics[width=0.315\textwidth]{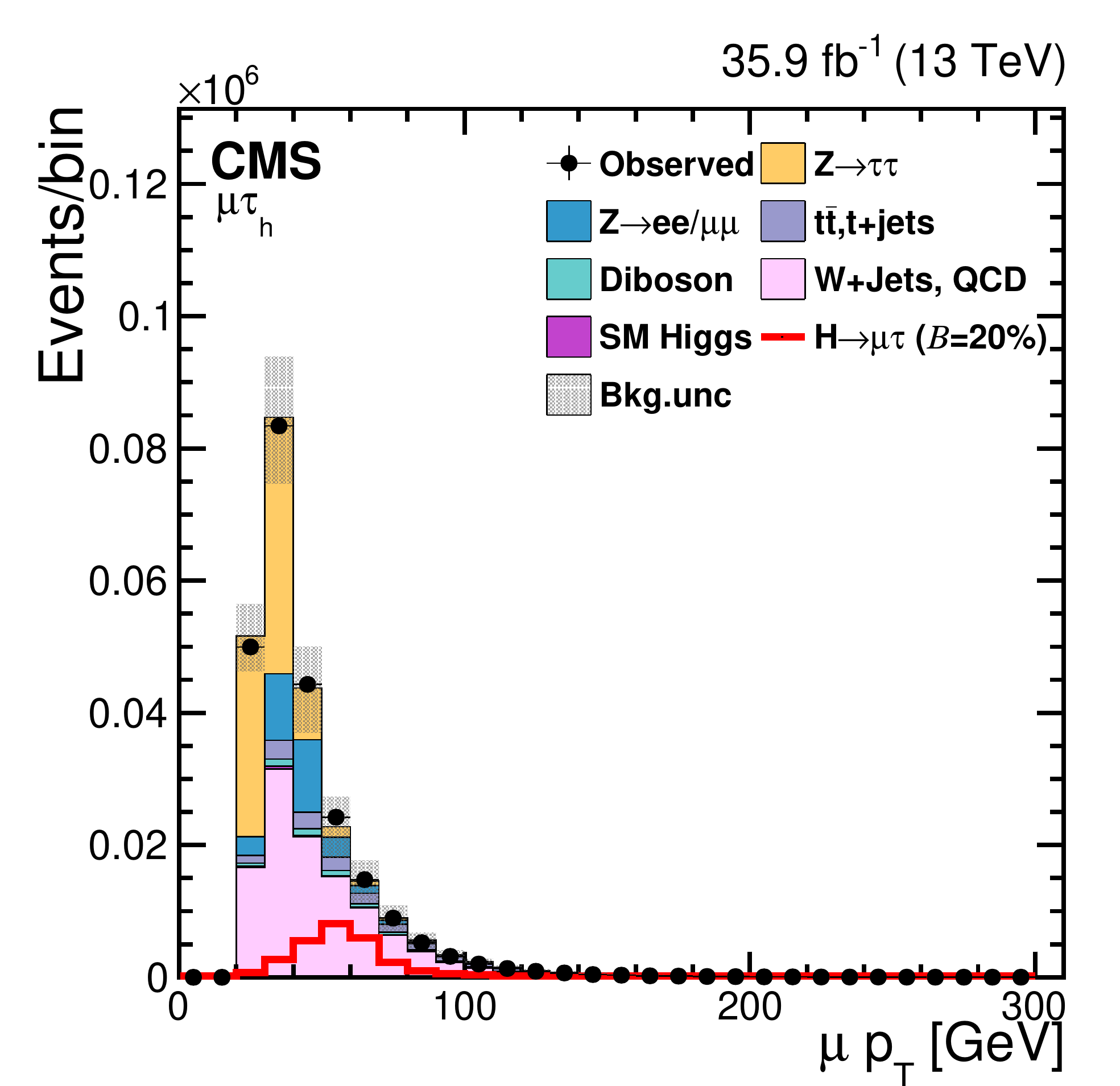} \\
 \includegraphics[width=0.315\textwidth]{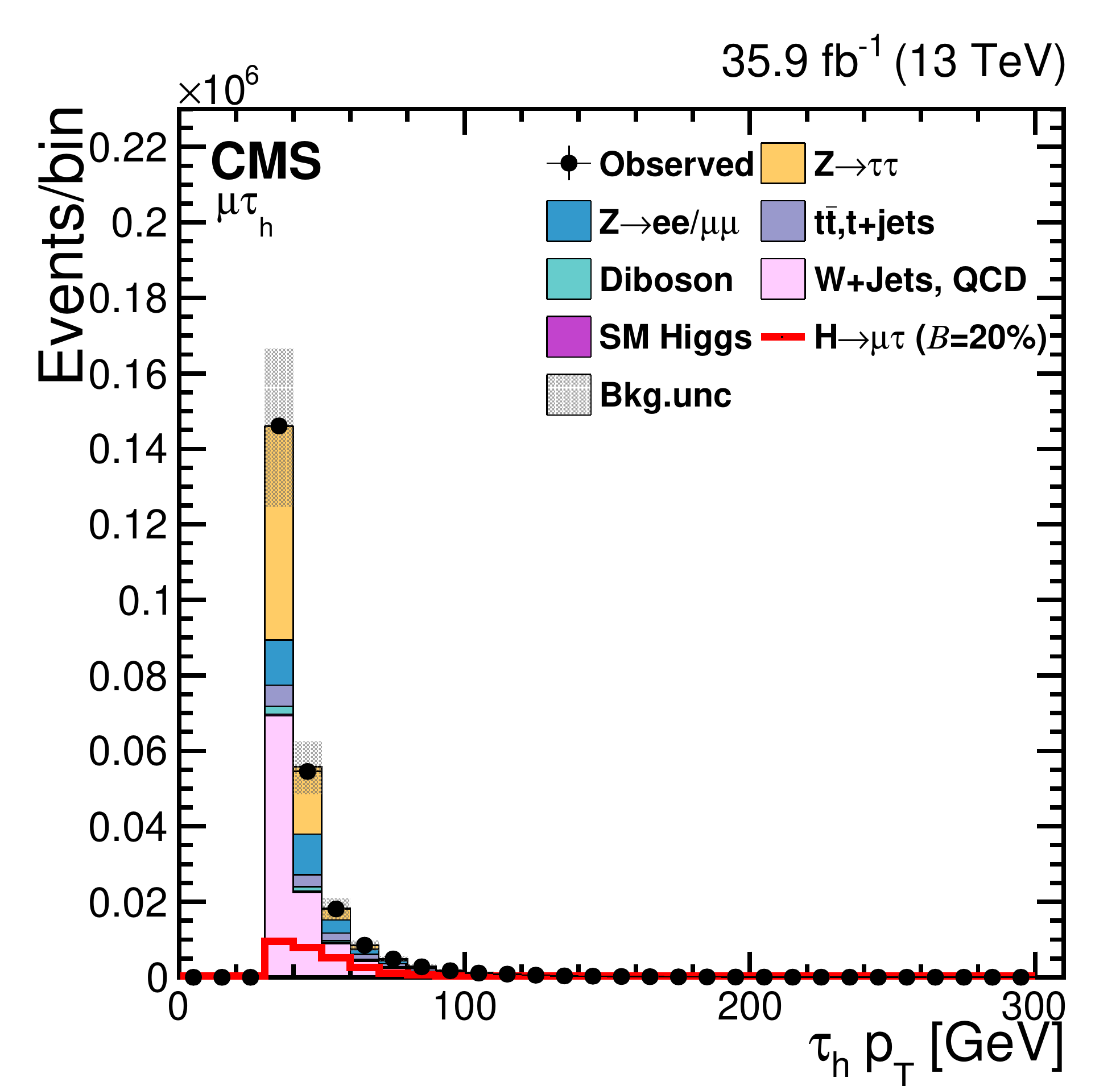}
 \includegraphics[width=0.315\textwidth]{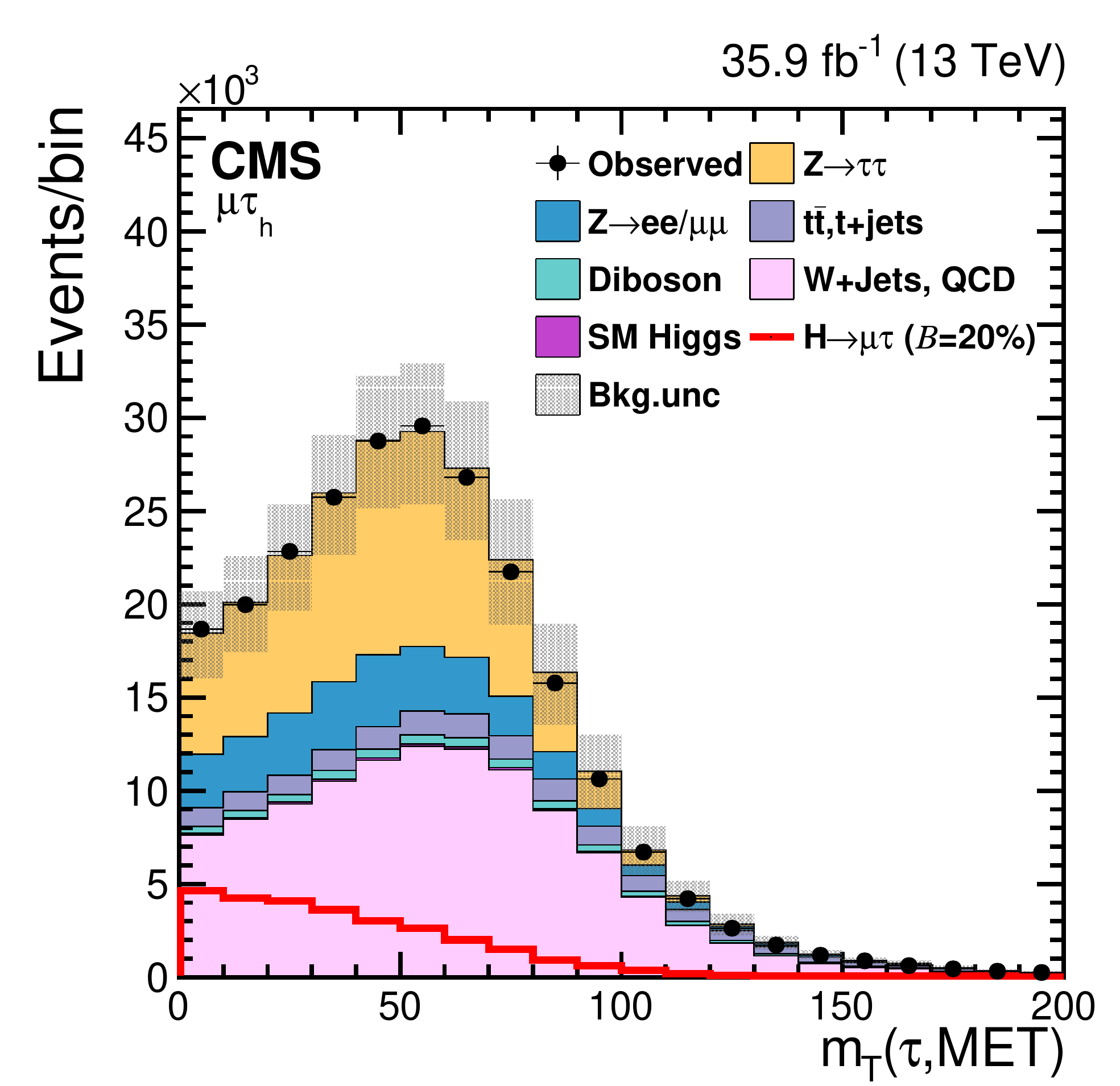}  \\
 \includegraphics[width=0.315\textwidth]{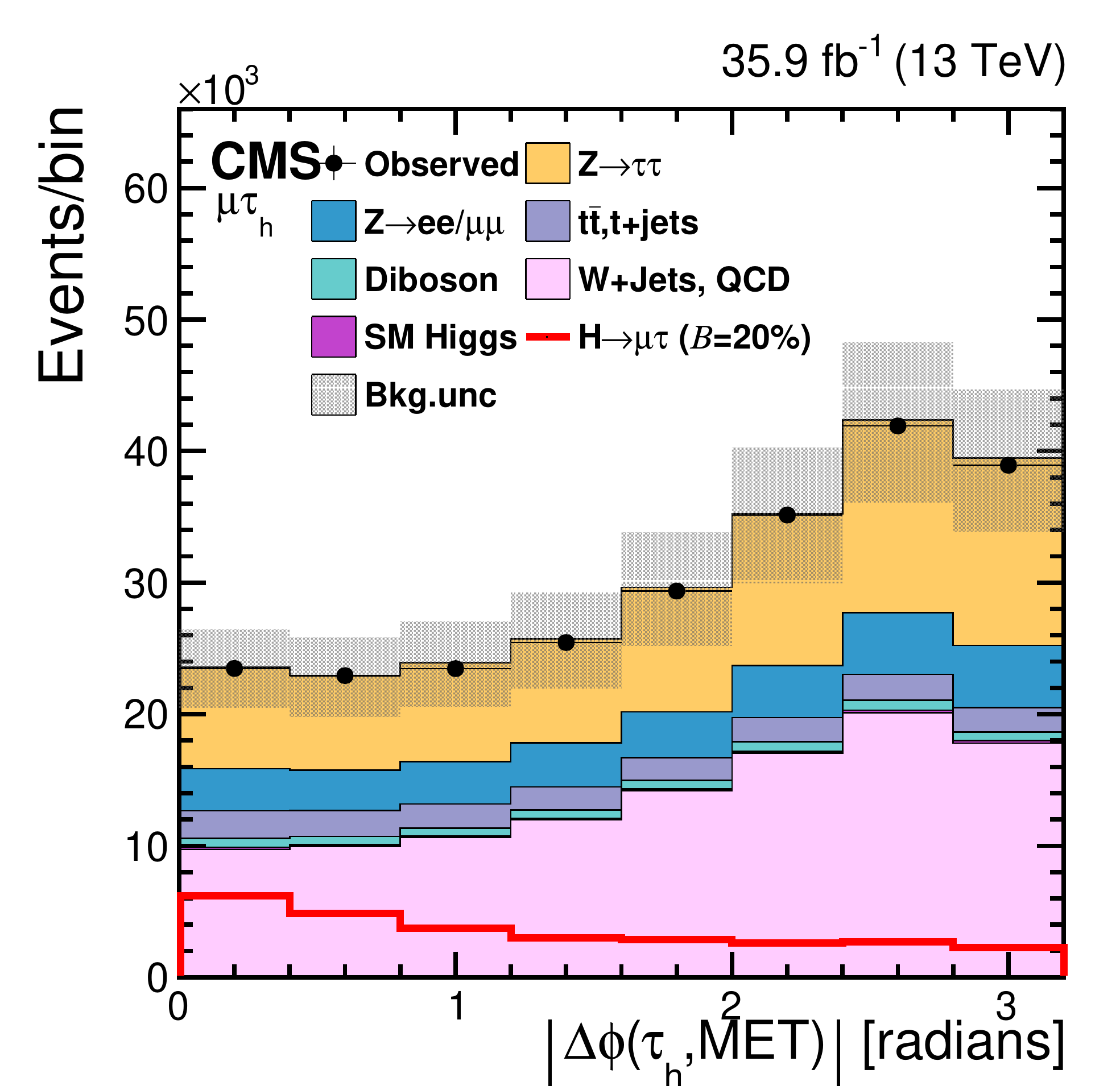}
 \includegraphics[width=0.315\textwidth]{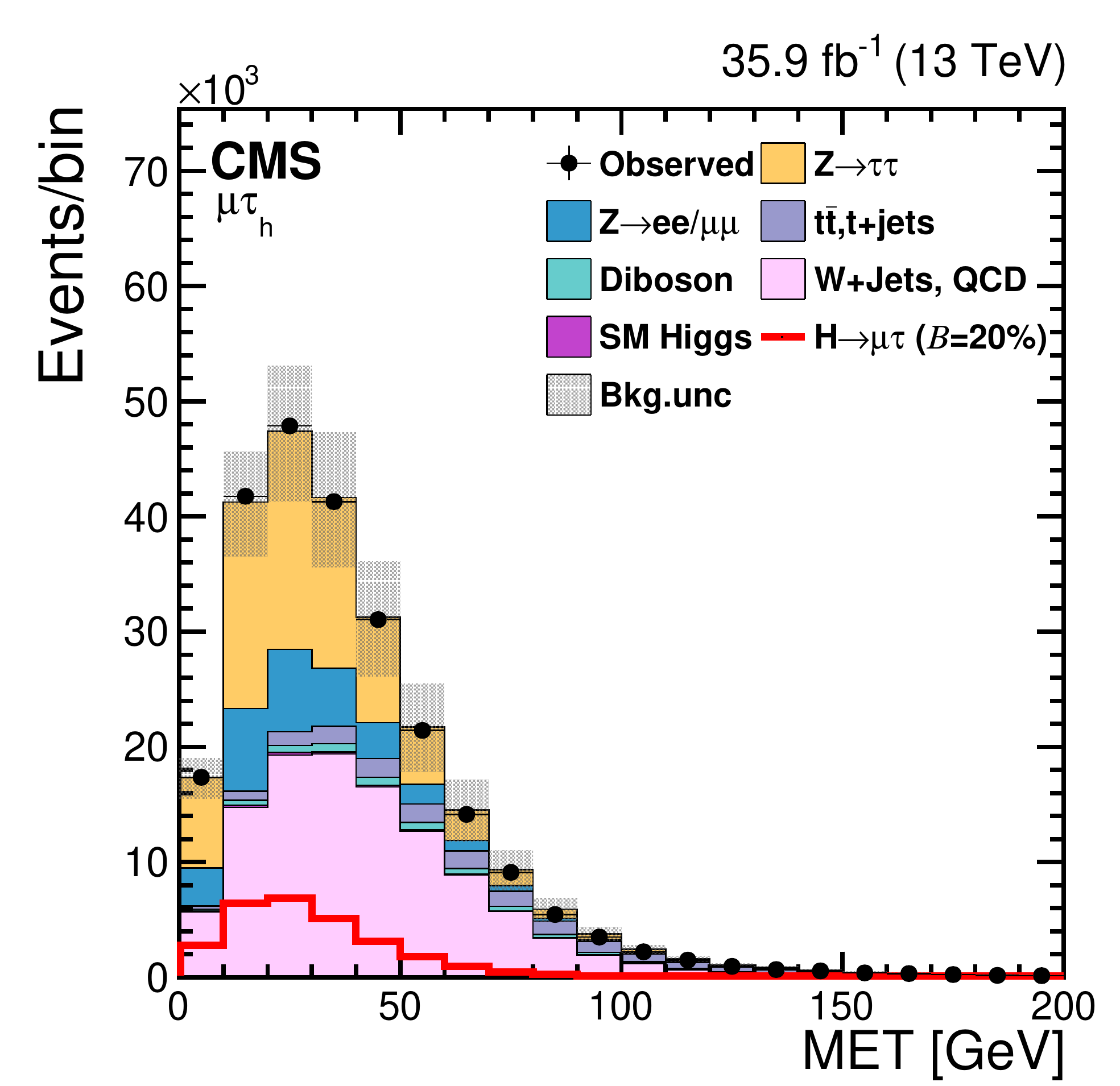} \\
 \includegraphics[width=0.315\textwidth]{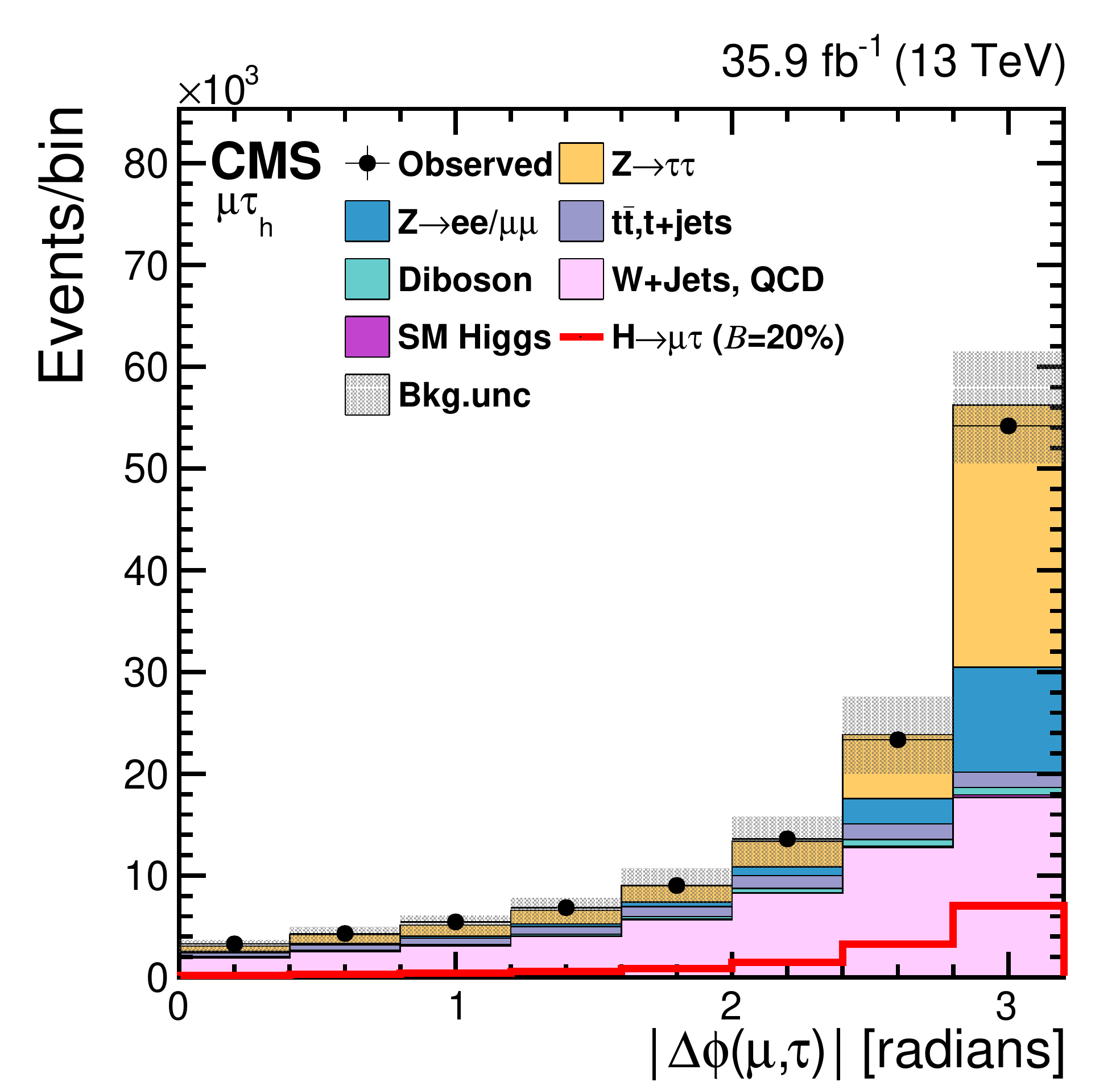}
 \includegraphics[width=0.315\textwidth]{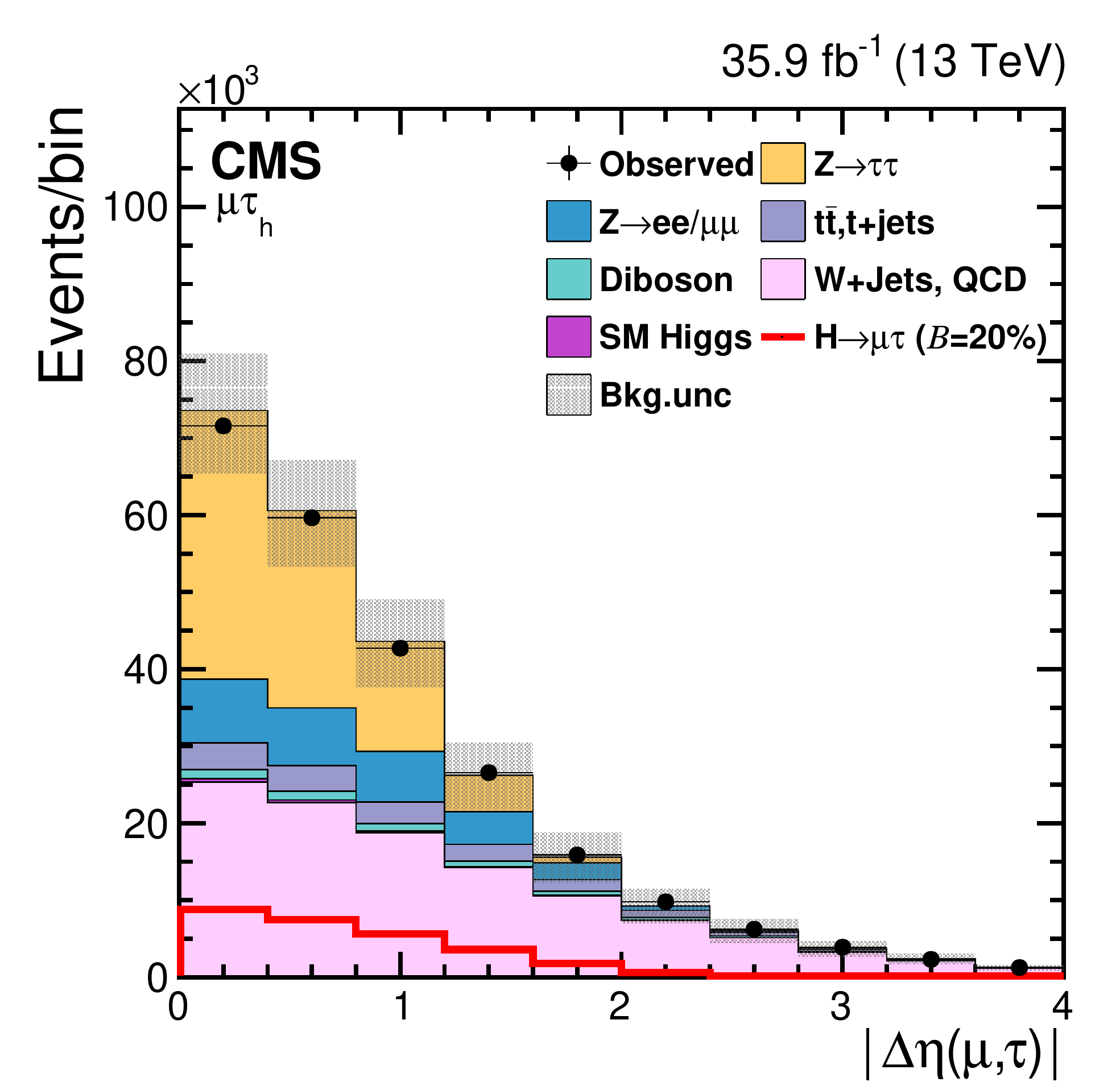}
\caption{Distributions of the  input variables to the BDT for the \Hmuhad channel. The background from SM Higgs boson production is  small and not visible in the plots.}
 \label{fig:BDT_input_var_mutauhad}
\end{figure*}

A BDT is trained after the loose selection combining all categories. The signal training sample  used is a mixture of simulated {\cPg\cPg}H and VBF events, weighted according to their respective SM production cross sections.
The background training sample is a set of collision events with  misidentified leptons, as this is the dominant background in this channel.
The leptons are required  to satisfy the same kinematic selection of the signal sample, be like-sign and  not isolated in order to select an orthogonal  data set to the signal sample, and  yet have the same kinematic properties.
The input variables to the BDT are: $\pt^{\Pgm}$, $\pt^{\tauh}$, $\mcol$, \ptmiss, $\mt(\tauh)$, $\Delta\eta(\Pgm, \tauh)$, $\Delta\phi(\Pgm, \tauh)$, and $\Delta\phi(\tauh, \ptvecmiss)$. The neutrino in the $\Pgt$ lepton decay leads to the presence of significant missing momentum motivating the
inclusion of the $\ptmiss$ variables. The neutrino is also approximately collinear with the visible \Pgt\ decay products while the two leptons tend to be azimuthally opposite leading to the inclusion of the $\Delta\phi$ variables.  The BDT input variables  are shown for signal and background in Fig.~\ref{fig:BDT_input_var_mutauhad}.

\subsection{\texorpdfstring{\Hmue}{H to mu tau[e]}}
The loose selection begins by requiring  an isolated  $\Pgm$ and an isolated $\Pe$ of opposite charge and separated by $\Delta R > 0.3$. The muon candidate is  required to have $\pt^{\Pgm} > 26$\GeV,  $|\eta^{\Pgm}|<2.4$, and $I_\text{rel}^{\Pgm} < 0.15$. The electron candidate is required to have $\pt^{\Pe}>10$ \GeV,  $|\eta^{\Pe}| < 2.4$, and $I_\text{rel}^{\Pe} < 0.1$.
Events with additional \Pe, \Pgm\ or \tauh  candidates, or with at least one  b-tagged jet are vetoed.

The tighter selection used in the \mcol fit analysis  requires $\pt^{\Pgm}>30$\GeV for the 0-jet category
and $\pt^{\Pgm}>26$\GeV in the other categories. In the 0-, 1- , 2-jet {\cPg\cPg}H and 2-jet VBF categories, $\mt(\Pgm)$ is required to be greater than 60, 40, 15, and 15\GeV respectively. A requirement is made on the azimuthal angle
between the electron and the \ptvecmiss:  $\Delta\phi(\Pe, \ptvecmiss) <$ 0.7, 0.7, 0.5, 0.3
for the 0-, 1-, 2-jet {\cPg\cPg}H,  and 2-jet VBF categories, respectively.
In the 0- and 1-jet categories it is further required that  $\Delta\phi(\Pe, \Pgm) >2.5$ and  1.0, respectively. The selections are summarized in Table~\ref{mutauSel}.

\begin{figure*}[!htpb]\centering
 \includegraphics[width=0.315\textwidth]{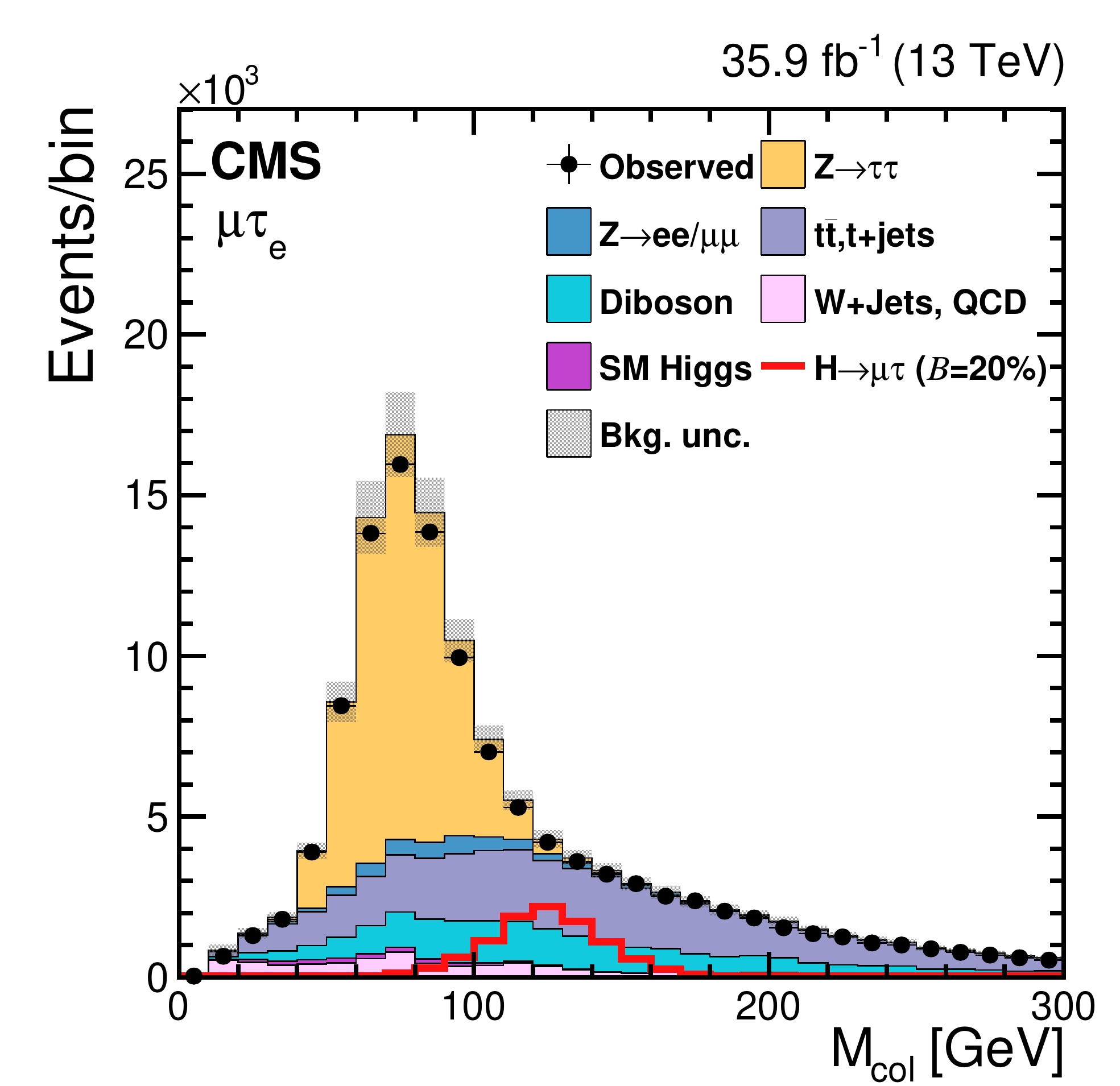}
 \includegraphics[width=0.315\textwidth]{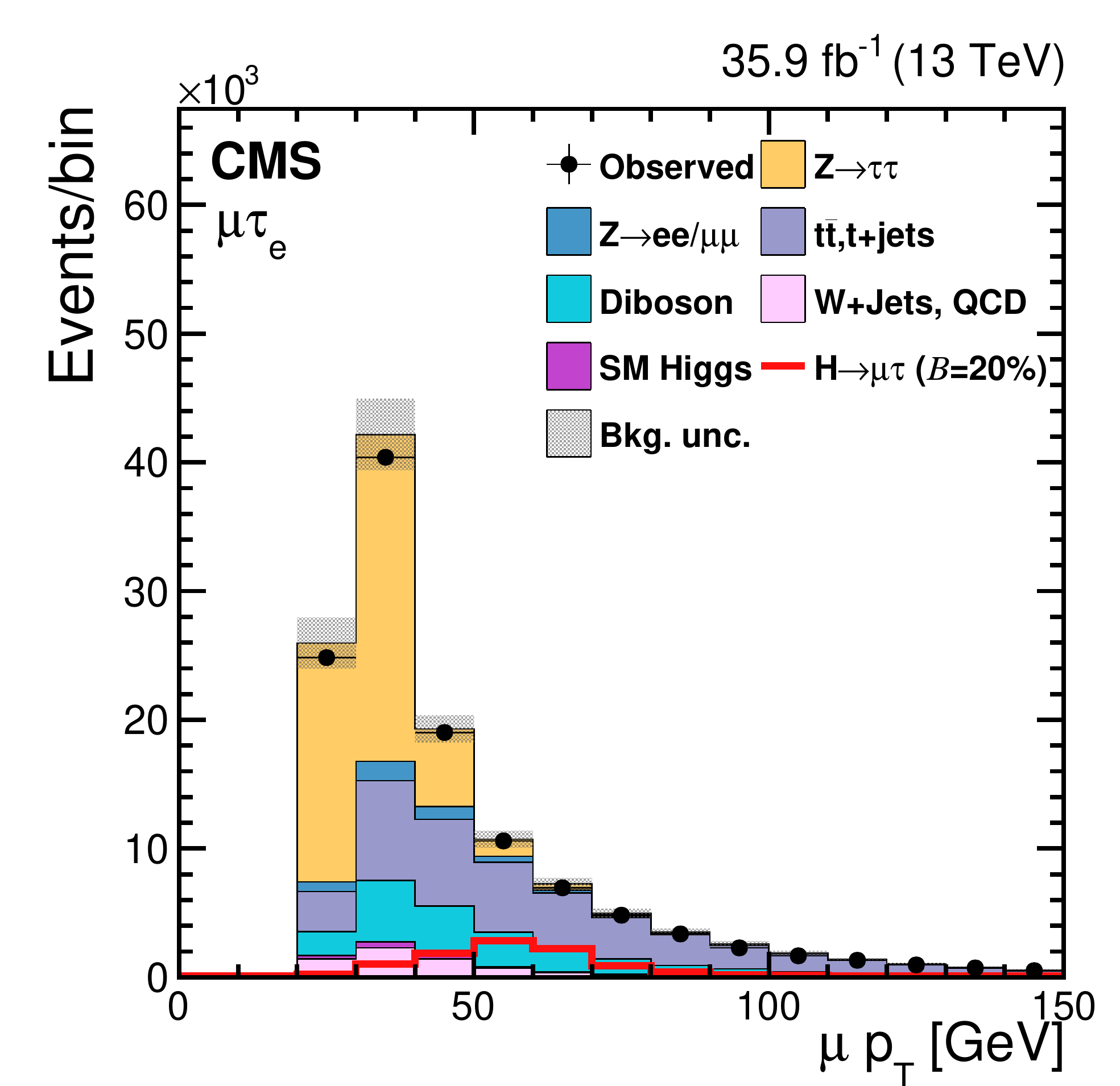} \\
 \includegraphics[width=0.315\textwidth]{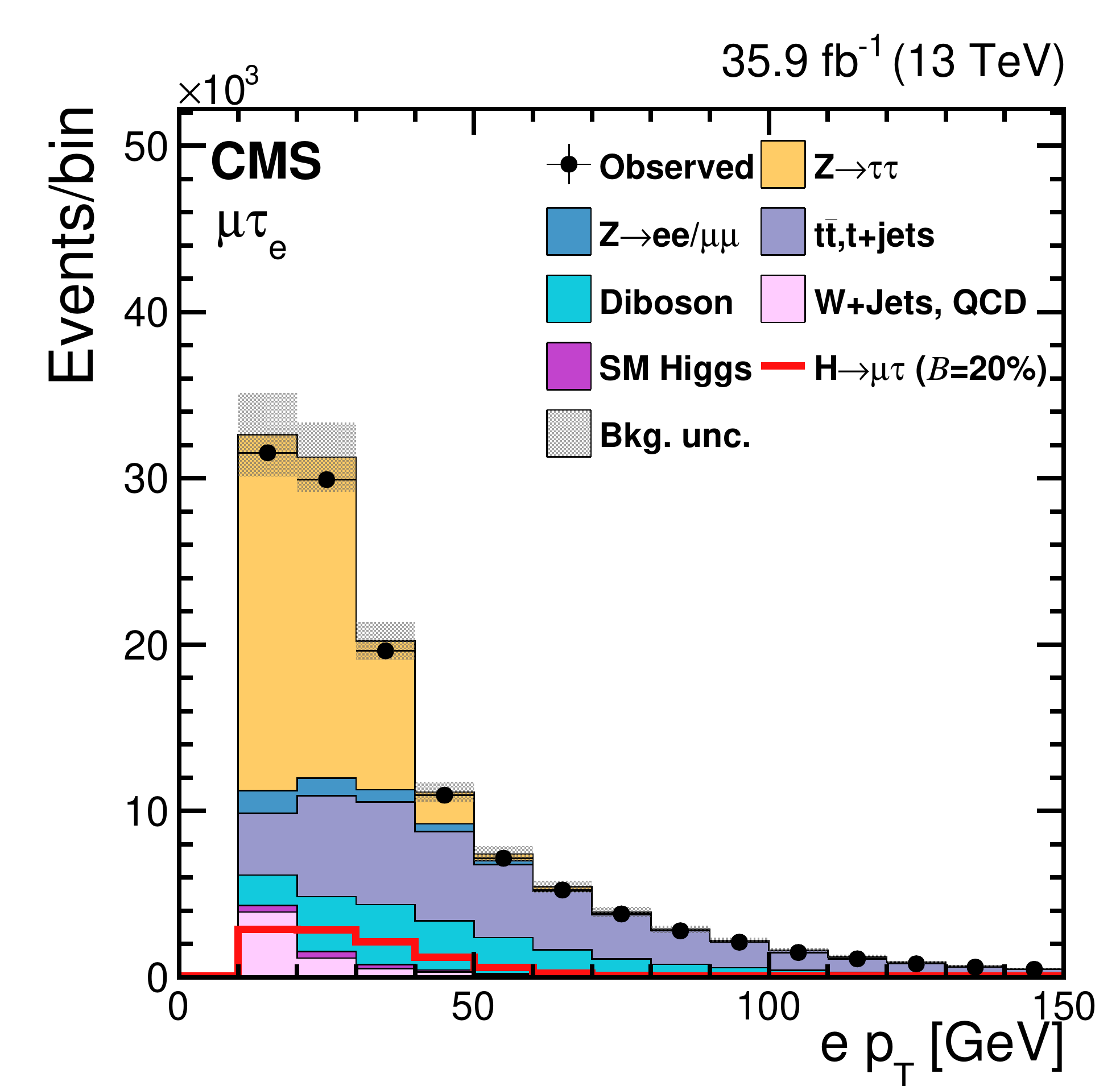}
 \includegraphics[width=0.315\textwidth]{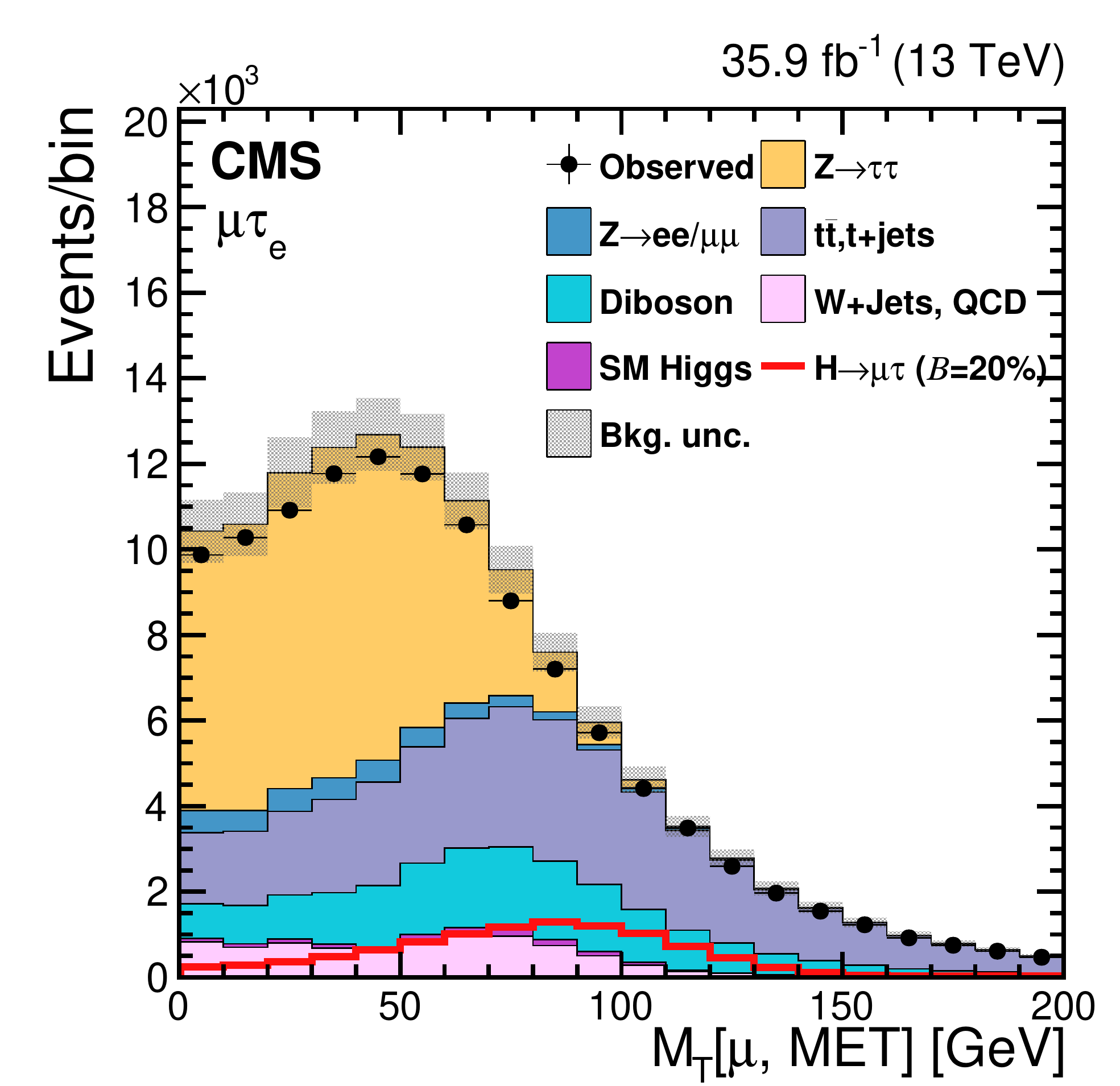}  \\
 \includegraphics[width=0.315\textwidth]{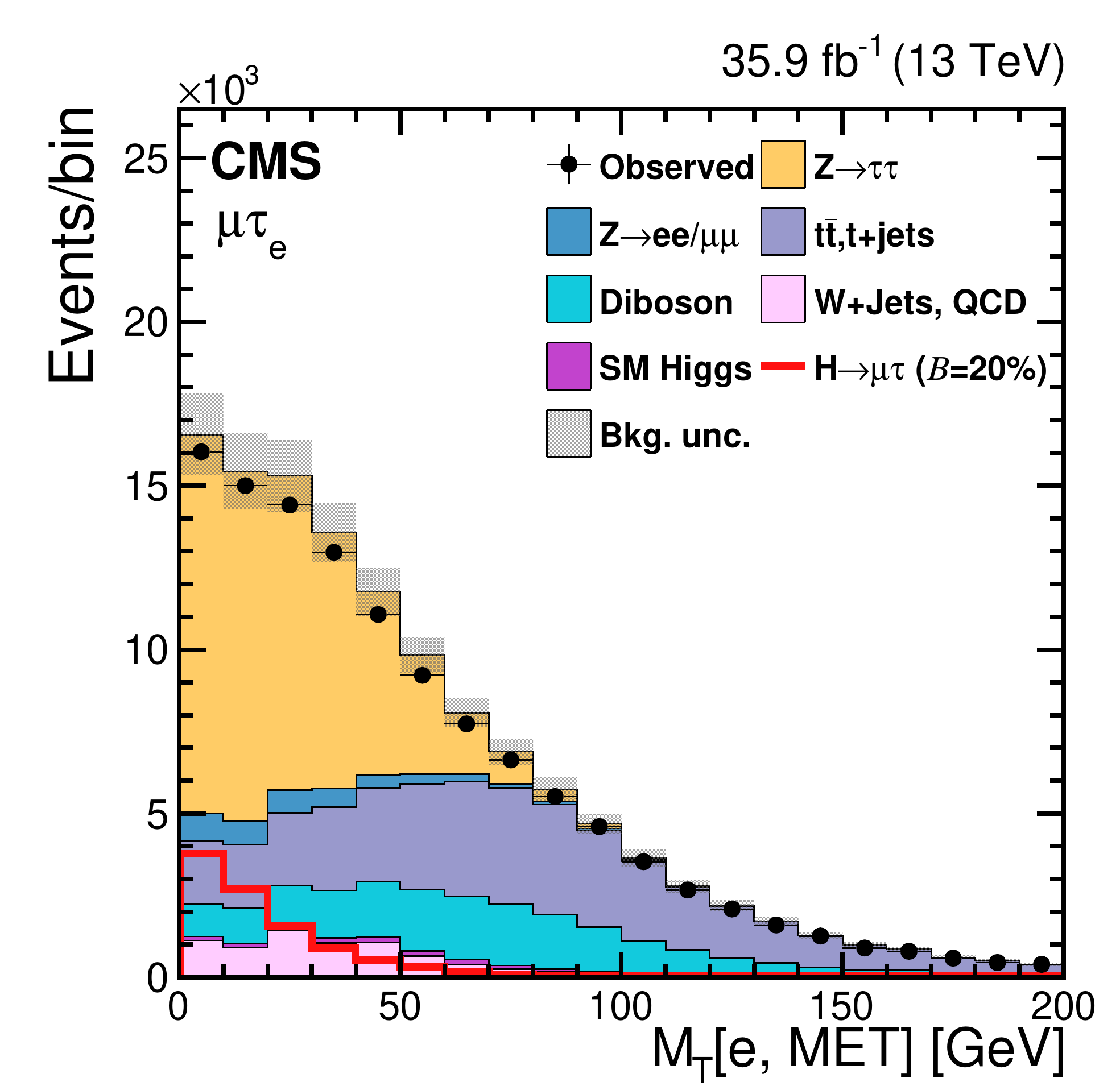}
 \includegraphics[width=0.315\textwidth]{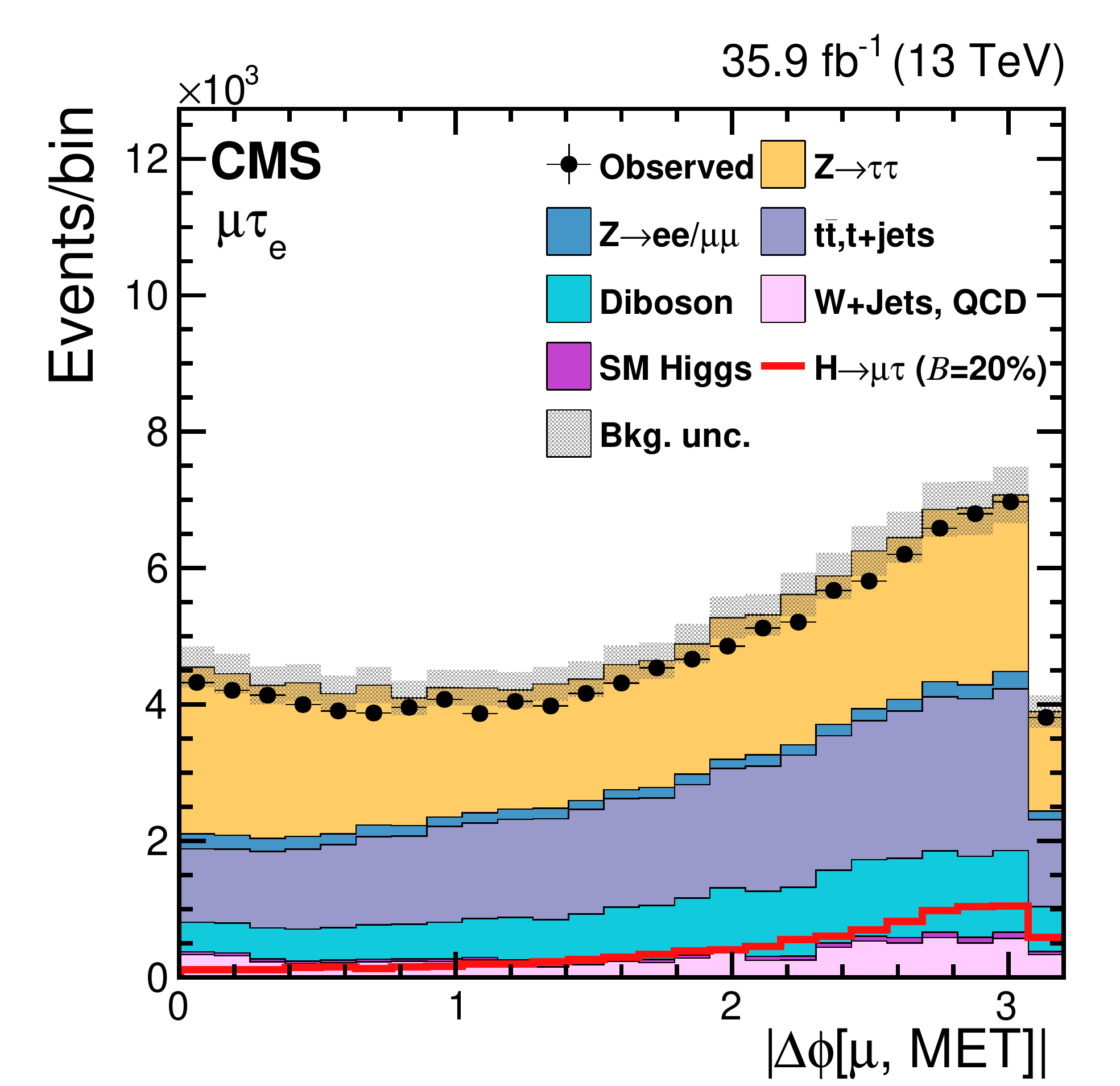} \\
 \includegraphics[width=0.315\textwidth]{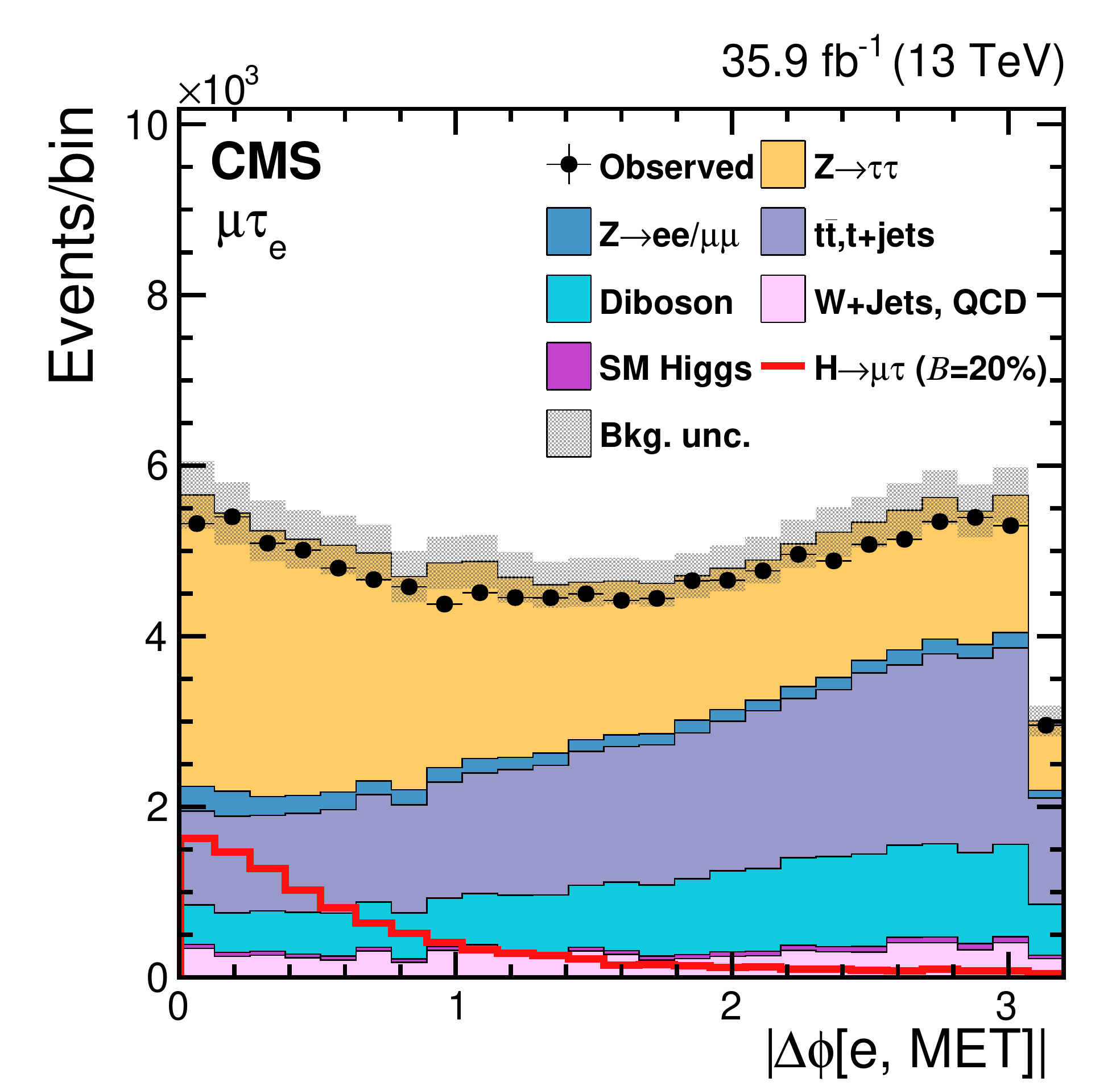}
 \includegraphics[width=0.315\textwidth]{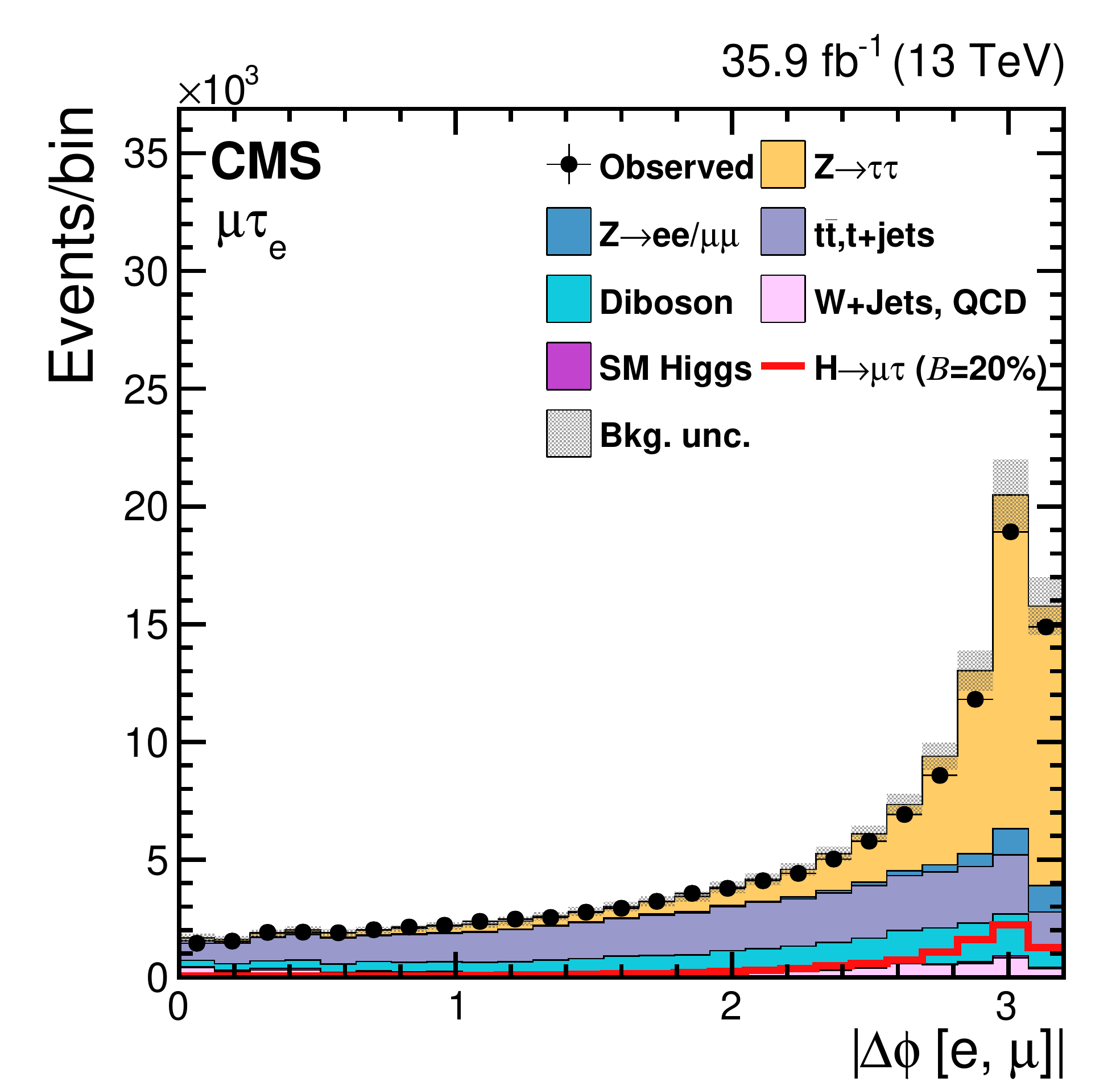}
\caption{Distributions of the input variables to the BDT for the \Hmue channel.}
 \label{fig:BDT_input_var_mutaue}
\end{figure*}

A BDT is trained after the loose selection, combining all categories. The background is
a mixed sample of \ttbar\ and $\cPZ \to \ell\ell$ $(\ell=\Pe,\Pgm,\Pgt)$  events weighted by their production cross-sections. The \ttbar\ background is the dominant background in
this channel for the 2-jet category and also very significant in the 1-jet category. It has many kinematic characteristics in common
with the other backgrounds, such as diboson and single top. The $\cPZ \to \ell\ell$ background is the dominant background in 0- and 1-jet category.
The input variables to the BDT are: $\pt^{\Pgm}$, $\pt^{\Pe}$, $\mcol$, $\mt(\Pgm)$, $\mt(\Pe)$, $\Delta\phi(\Pe, \Pgm)$, $\Delta\phi(\Pe, \ptvecmiss)$, and
$\Delta\phi(\Pgm, \ptvecmiss)$. The distributions of these variables are shown in Fig.~\ref{fig:BDT_input_var_mutaue}.

\begin{table}[htpb]
 \centering
 \topcaption{Event selection criteria for the kinematic variables for the $\PH \to \Pgm\Pgt$ channels.}
  \label{mutauSel}
  \begin{tabular}{lc*{4}{c}{c}@{\hspace*{5pt}}*{4}{c}} \hline
  Variable         &           &     \multicolumn{4}{c}{$ \PH \to \Pgm \tauh$}               && \multicolumn{4}{c}{$\PH \to \Pgm \Pgt_\Pe$}                  \\ \cline{3-6}\cline{8-11}
                   &           &  0 jet          & 1 jet       & \multicolumn{2}{c}{2 jet}  &&     0 jet       &     1 jet  &  \multicolumn{2}{c}{2 jet}     \\
                   &           &         &       &  ggH  & VBF                              &&        &     &  ggH  &  VBF                                    \\ \hline

$M_{jj}$  &[\GeVns{}]      &    \NA          &   \NA       & $<$550  & $\geq$550         &&    \NA    &   \NA    & $<$550 & $\geq$550                      \\       \hline
$\pt^{\Pe} $       & [\GeVns{}]     &  \multicolumn{4}{c}{---}                                    &&  \multicolumn{4}{c}{$>$10}                                    \\
$\pt^{\Pgm}$       & [\GeVns{}]     &  \multicolumn{4}{c}{$>$26}                                  &&  \multicolumn{4}{c}{$>$26}                                     \\
$\pt^{\tauh}$      &[\GeVns{}]      &  \multicolumn{4}{c}{$>$30}                                  &&  \multicolumn{4}{c}{---}                                       \\\hline
$|\eta^{\Pe}|$     &           &  \multicolumn{4}{c}{---}                                    &&  \multicolumn{4}{c}{$<$2.4}                                     \\
$|\eta^{\Pgm}|$    &           &  \multicolumn{4}{c}{$<$2.4}                                 &&  \multicolumn{4}{c}{$<$2.4}                                     \\
$|\eta^{\tauh}|$   &           &  \multicolumn{4}{c}{$<$2.3 }                                &&  \multicolumn{4}{c}{---}                                        \\\hline
$I_{\text{rel}}^{\Pe}$      &           &  \multicolumn{4}{c}{---}                                    &&  \multicolumn{4}{c}{$<$0.1}                                     \\
$I_{\text{rel}}^\Pgm$     &           &  \multicolumn{4}{c}{$<$0.15}                                &&  \multicolumn{4}{c}{$<$0.15}                                    \\[\cmsTabSkip]
                   &           &  \multicolumn{9}{c}{$M_\text{col}$ fit selection} \\\hline
$\pt^{\Pgm}$      & [\GeVns{}]     &  \multicolumn{4}{c}{---}            &&  $>$30 & \NA       & ----      & \NA                          \\
$\mt(\Pgm)$      & [\GeVns{}]     & \multicolumn{4}{c}{---}             &&  $>$60 & $>$40 & $>$15 & $>$15                                     \\
$\mt(\tauh)$    &[\GeVns{}]       &  $<$105 & $<$105 & $<$105 & $<$85    && \multicolumn{4}{c}{---}                                           \\
$\Delta\phi(\Pe, \ptvecmiss)$&[radians]      & \multicolumn{4}{c}{---}              &&  $<$0.7 & $<$0.7 & $<$0.5 & $<$0.3               \\

$\Delta\phi(\Pe, \Pgm)$      &[radians]      & \multicolumn{4}{c}{---}               & &  $>$2.5 & $>$1.0 & \NA & \NA                    \\
  \end{tabular}
\end{table}

\subsection{\texorpdfstring{\Hehad}{H to e tau[h]}}
The loose selection begins by requiring an isolated \Pe\ and an isolated \tauh  candidate of  opposite charge, separated  by $\Delta R>0.5$.
The \Pe\ candidate is  required to have $\pt^{\Pe}>26$\GeV, $|\eta^{\Pe}|<2.1$, and $I_\text{rel}^{\Pe}<0.1$.
The \tauh candidate is  required to have $\pt^{\tauh}>30$\GeV and $|\eta^{\tauh}|<2.3$.   Events with additional \Pe, \Pgm\ or \tauh candidates are vetoed. No veto is made on the number of b-tagged jets as the \ttbar\  contribution is small.
The additional selection used for the \mcol fit analysis further requires that $\mt(\tauh)< 60 $ \GeV. The selections are summarized in Table~\ref{etauSel}. A BDT is trained after the loose selection. The same training samples as for the \Hmuhad channel are used, except with an electron rather than a muon. The input variables to the BDT are also the same except for the addition of the visible mass, \mvis, and the removal of $\ptmiss$. The relative composition of the backgrounds in the \Hehad channel is different from the \Hmuhad channel, in particular the $\cPZ\to\Pe\Pe$ + jets background is larger in comparison to the $\cPZ\to\Pgm\Pgm$ + jets,  which leads to this change of variables.

\subsection{\texorpdfstring{\Hemu}{H to e tau[mu]}}
The loose selection begins by requiring an isolated \Pe\ and an  isolated \Pgm\ candidate  with  opposite  charge, separated by $\Delta R>0.4$.
The \Pe\  candidate is required to have $\pt^{\Pe}>24$ \GeV, $|\eta^{\Pe}|<2.1$, and  $I_\text{rel}^{\Pe} < 0.1$. The \Pgm\ candidate is required to have $\pt^{\Pgm}>10$\GeV, $|\eta^{\Pgm}|<2.4$, and $I_\text{rel}^{\Pgm} < 0.15$.
Events with additional \Pe, \Pgm\ or \tauh candidates, or  with at least one  b-tagged jet are vetoed.

The tighter selection used in the \mcol fit analysis further requires $\Delta\phi(\Pe, \ptvecmiss) < 1.0$ and $\mt(\Pe)>60$\GeV.
The large \ttbar\  background is further reduced by requiring $p_{\zeta} - 0.85 \, p_{\zeta}^{\rm{vis}} > -60$\GeV. This topological selection is based on the projections
\begin{equation*}
\label{eq:PZetaDefinition}
\begin{split}
p_{\zeta} = (\vec{\pt}^{\Pe} + \vec{\pt}^{\Pgm} + \vec{\pt}^{\rm{miss}})\, \frac{\vec{\zeta}}{|\vec{\zeta}|}\qquad
\text{ and }\qquad p_{\zeta}^{\rm{vis}} = (\vec{\pt}^{\Pe} + \vec{\pt}^{\Pgm})\, \frac{\vec{\zeta}}{|\vec{\zeta}|}
\end{split}
\end{equation*}
on the axis  $\vec{\zeta}$  bisecting the directions of the electron, $\vec{\pt}^{\Pe}$, and of the muon, $\vec{\pt}^{\Pgm}$. This selection criterion is highly efficient in rejecting background as the \ptvecmiss is oriented in the direction of the visible \Pgt\ decay products in signal events. The selection criteria are summarized in Table~\ref{etauSel}.

A BDT is trained after the loose selection. It uses the same input variables as for the \Hmue channel with the addition of the visible mass, \mvis, and the removal of $\mt(\Pe)$. The background  used for the training  is a sample of simulated \ttbar\  events.

\begin{table*}[hbtp]
 \centering
 \topcaption{Event selection criteria for the kinematic variables for the $\PH \to \Pe\Pgt$ channels.}
  \label{etauSel}
  \begin{tabular}{lc*{4}{c}{c}@{\hspace*{5pt}}*{4}{c}} \hline
  Variable         &           & \multicolumn{4}{c}{$\PH \to \Pe\tauh$}            &&     \multicolumn{4}{c}{$\PH \to \Pe \Pgt_\Pgm$}  \\ \cline{3-6}\cline{8-11}
                   &          &     0 jet       &     1 jet            &  \multicolumn{2}{c}{2 jet}      &&  0 jet          & 1 jet       & \multicolumn{2}{c}{2 jet} \\
                   &          &        &     &  ggH  &  VBF        &&            &       &  ggH  & VBF\\ \hline
$M_{jj}$            &  [\GeVns{}]       &    \NA          &   \NA       & $<$500 & $>$500    &&    \NA        &   \NA      & $<$500 & $>$500      \\       \hline
$\pt^{\Pe} $         &  [\GeVns{}]         &  \multicolumn{4}{c}{$>$26}                            &&  \multicolumn{4}{c}{$>$24} \\
$\pt^{\Pgm}$        &  [\GeVns{}]         &  \multicolumn{4}{c}{---}                            &&  \multicolumn{4}{c}{$>$10}  \\
$\pt^{\tauh}$     &  [\GeVns{}]         &  \multicolumn{4}{c}{$>$30}                                  &&  \multicolumn{4}{c}{---}     \\\hline
$|\eta^{\Pe}|$       &           &  \multicolumn{4}{c}{$<$2.1}                               &&  \multicolumn{4}{c}{$<$2.1}               \\
$|\eta^{\Pgm}|$     &           &  \multicolumn{4}{c}{---}                               &&  \multicolumn{4}{c}{$<$2.4}\\
$|\eta^{\tauh}|$   &           &  \multicolumn{4}{c}{$<$2.3}                                  &&  \multicolumn{4}{c}{---} \\\hline
$I_{\text{rel}}^{\Pe}$        &           &  \multicolumn{4}{c}{$<$0.15}                               &&  \multicolumn{4}{c}{$<$0.1}  \\
$I_{\text{rel}}^\Pgm$      &           &  \multicolumn{4}{c}{---}                               &&  \multicolumn{4}{c}{$<$0.1}\\[\cmsTabSkip]
                   &           &  \multicolumn{8}{c}{$M_\text{col}$ fit selection} \\\hline
$\mt(\tauh)$      &    [\GeVns{}]       & \multicolumn{4}{c}{$<$60}                             && \multicolumn{4}{c}{---} \\
$\mt(\Pe)$      &    [\GeVns{}]       & \multicolumn{4}{c}{---}                             && \multicolumn{4}{c}{$>$60} \\
$\Delta\phi(\Pe, \ptvecmiss)$&  [radians]       & \multicolumn{4}{c}{---}                        && \multicolumn{4}{c}{$<$1.0 }\\
$p_\zeta - 0.85 \, p_\zeta^{vis}$& [\GeVns{}] & \multicolumn{4}{c}{---}                     && \multicolumn{4}{c}{${>}-60$}\\\hline
  \end{tabular}
\end{table*}

\section{Background estimation}
\label{backgrounds}

\begin{figure}[htpb]
\centering
\includegraphics[width=0.325\textwidth]{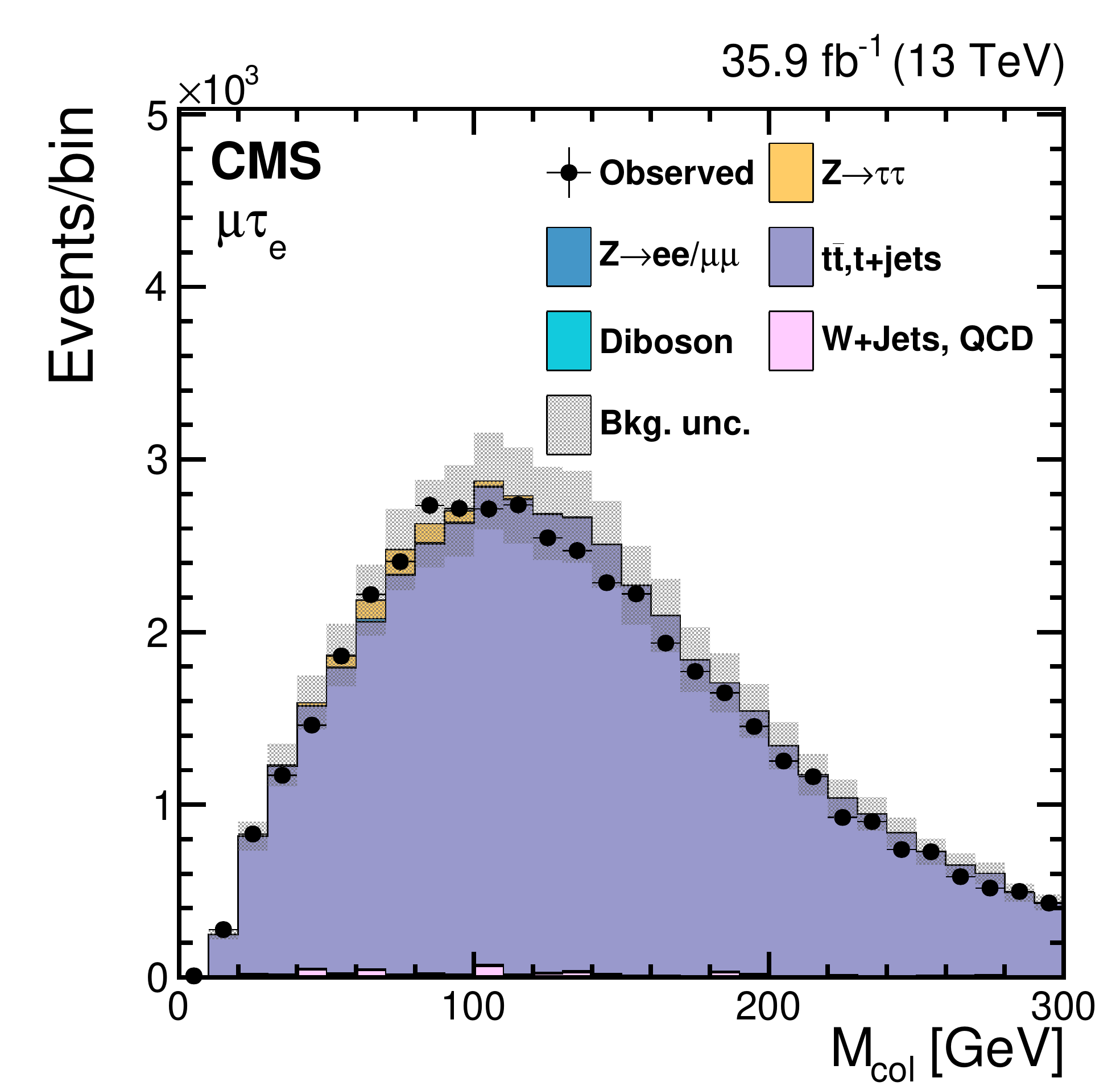}
\includegraphics[width=0.325\textwidth]{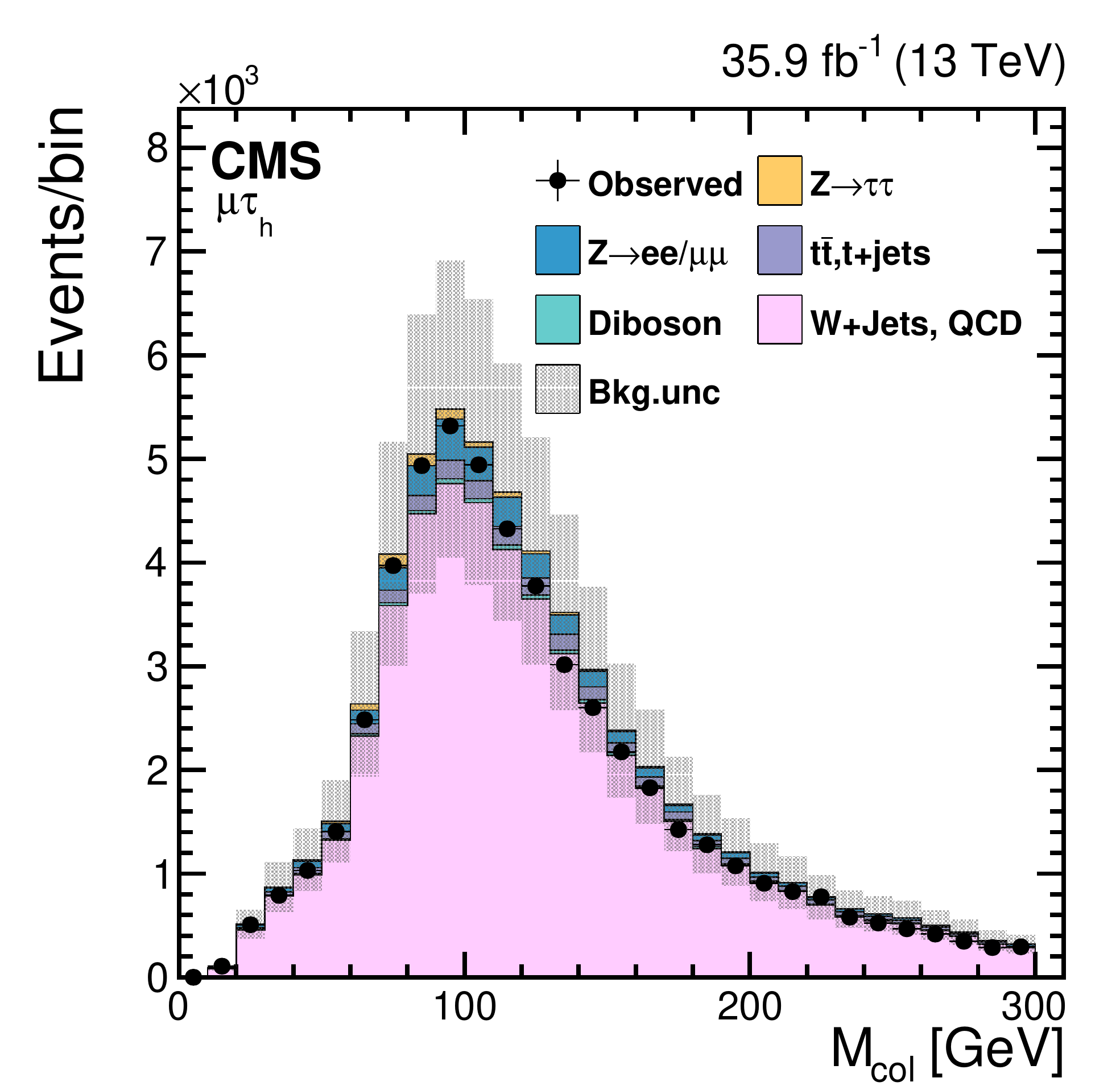}
\includegraphics[width=0.325\textwidth]{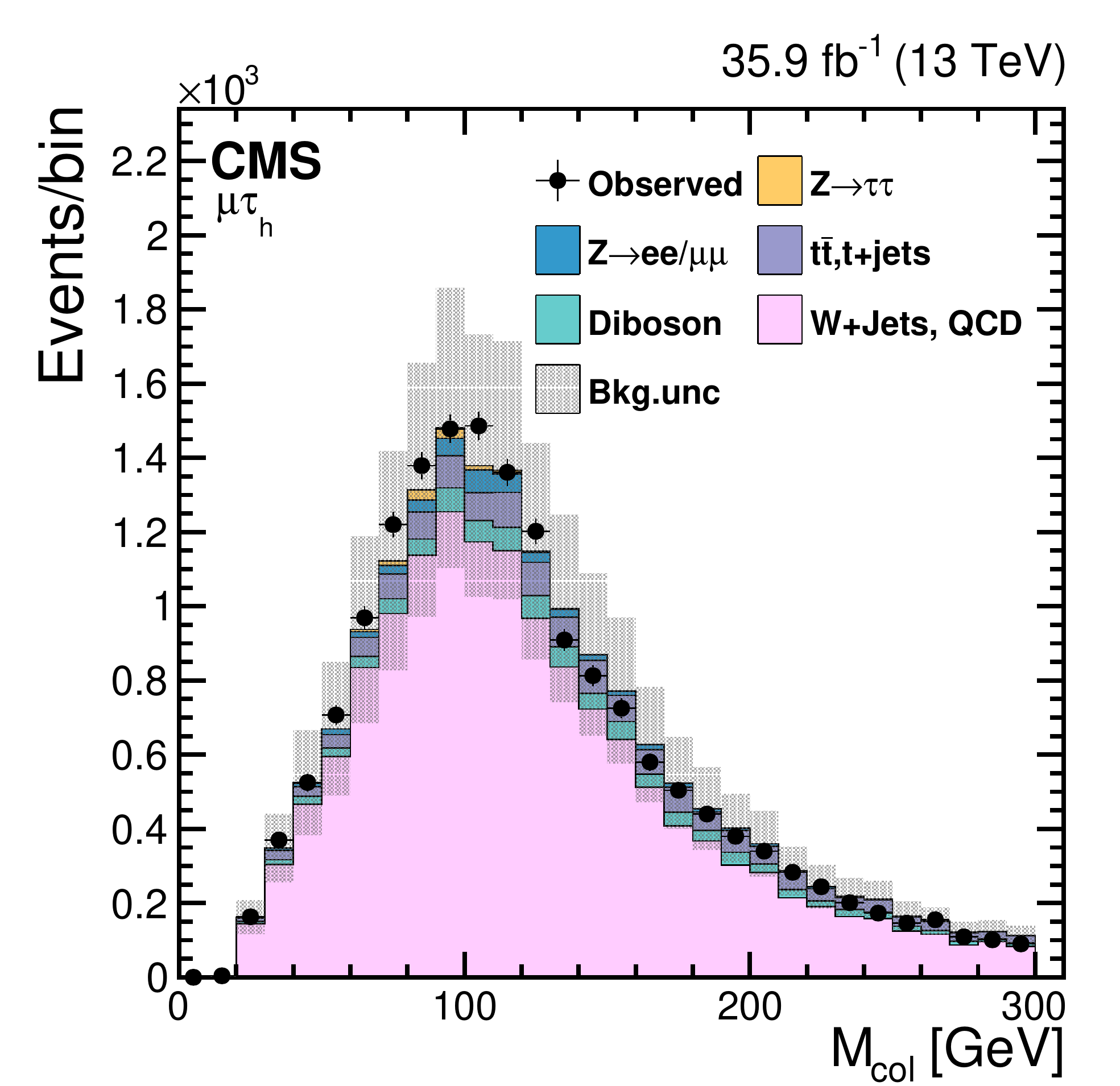}

\caption{\mcol distribution in \ttbar\ enriched (left),  like-sign lepton  (central), and $\PW+\text{jets}$ enriched (right) control samples defined in the text. The distributions include both statistical and systematic uncertainties.}
\label{fig:SScontrolregion}
\end{figure}

The main background processes are $\cPZ \to \Pgt \Pgt$, in which the \Pgm\ or \Pe\ arises from a \Pgt\ decay, and $\PW\mathrm{+jets}$ and QCD multijet production  where one or more of the  jets are misidentified as leptons. Other backgrounds come from processes in which the lepton pair is produced from the weak decays of quarks and vector bosons.  These include $\ttbar$ pairs, Higgs boson production ($\PH \to \Pgt\Pgt, \PW\PW$), $\PW\PW$, $\PW\cPZ$, and $\cPZ\cPZ$. There are also smaller contributions from  $\PW\gamma^{(*)}+\text{jets}$ processes, single top quark production,  and $Z\to\ell\ell$ ($\ell = \Pe, \Pgm)$.
All the backgrounds are estimated from simulated samples with the exception of the misidentified-lepton backgrounds that are estimated from data with either fully data-driven or semi data-driven methods.
These techniques are described in detail below. The background estimate is validated with control regions designed to have  enhanced contributions from the dominant backgrounds.

The $Z\to\ell\ell$   background is estimated from simulation.  A reweighting is applied to correct the generator-level \PZ $\pt$ and $m_{\ell\ell}$ distributions in LO  \aMCATNLO  samples to reduce the shape discrepancy between collision data and simulation. The reweighting factors are extracted from a $Z\to\Pgm\Pgm$ control region and are applied to both $Z\to\Pgm\Pgm$ and $Z\to\Pe\Pe$ simulated samples in bins of \PZ $\pt$ and $m_{\ell\ell}$.
Additional corrections for $\Pgm\to\tauh$ and $\Pe\to\tauh$ misidentification rates are applied when the reconstructed
\tauh\  candidate is matched to a muon or an electron, respectively, at the generator level.
These corrections are measured in  $Z\to\ell\ell$ events and  depend on the lepton $\eta$.  The $\ttbar+\text{jets}$ background is particularly important in the $\Pe\Pgm$ final state. A correction based on the generated \pt of the top quark and antiquark is applied to the  simulation  to match the  \pt distribution  observed in a \ttbar\ sample from collision data. The background estimation for this contribution is validated in a  \ttbar\ enriched control sample. It is defined by requiring the
loose selection for these channels but with the additional requirement that
at least one of the jets is b-tagged. Figure~\ref{fig:SScontrolregion} (left) shows the data compared
to the background estimation for this control sample in the \Hmue channel. The same samples are used in the \Hemu channel and show similar agreement.

The Higgs boson production contributes a small but non-negligible background.  It arises predominantly from $\PH \to  \Pgt\Pgt$ but also from $\PH \to \PW\PW$ decays and peaks at
lower values of \mcol than the signal, because of  additional neutrinos in the decays. The
event selection described in Section~\ref{eventsel} uses a BDT discriminator that combines  \mcol
with a set of other kinematic variables. The Higgs boson background also peaks below
the signal in the distribution of the BDT discriminator output.

Jets misidentified as leptons are a source of background arising from two sources, \wjets  and QCD multijet events. In \wjets background events, one lepton candidate is a real lepton from the $\PW$ boson decay and the
other is a jet misidentified as a lepton. In QCD multijet events, both lepton candidates
are misidentified jets.
In each of the four channels for this analysis (\muhad, \ehad, \mue, \emu), the misidentified-lepton background has been estimated using purely data-driven  methods. In  the \mue and \emu channels it is also estimated using a technique, called semi data-driven, partially based on control samples in data and partially on simulation. It has been used previously in the SM \Htt analysis~\cite{CMS-PAS-HIG-16-043}. The misidentified \wjets background is estimated from simulation and the QCD background
with data. The two techniques give consistent results; the semi data-driven technique is chosen for the leptonically decaying tau
channels as the fully data-driven technique is limited by the reduced size of the sample.

\subsection*{Fully data-driven technique}

The misidentified lepton background is estimated from collision data samples. The misidentification rates are evaluated with independent \zjets data sets and then applied to a control region, orthogonal to the signal region, to estimate the misidentified background in the signal region. This control region is obtained by relaxing the signal selection requirements, typically isolation, and excluding events passing the final selection.
The probabilities with which jets are misidentified as \Pe\ ($f_{\Pe}$), \Pgm\ ($f_{\Pgm}$), or \tauh ($f_{\Pgt}$), are estimated using events with a \PZ boson candidate plus one jet that can be misidentified as a lepton. The \PZ boson candidate is formed from two muons with  $\pt>26$ \GeV, $\abs{\eta}<2.4$, and $I_\text{rel}^{\ell}<0.15$ $(0.25)$ for the $\text{jet}\to\tauh, \Pgm$ ($\text{jet}\to \Pe $) misidentification rate. The muons are required to have opposite charge and their invariant mass ($M_{\Pgm\Pgm}$) must satisfy  $70<M_{\Pgm\Pgm}<110$\GeV. The contribution from  diboson events, where the third lepton candidate corresponds to a genuine lepton, is subtracted using
simulation. Two \zjets samples are defined: the signal-like one, in which the jet satisfies the same lepton selection criteria used in the $\PH \to \Pe\Pgt$ or $\PH \to \Pgm\Pgt$ selections,  and the background-enriched \zjets sample with relaxed lepton identification on the jet but excluding events selected in the signal-like sample. The requirements for the third lepton candidate vary depending on the lepton flavour. The two samples are used to estimate $f_{\Pe}$, $f_{\Pgm}$ and $f_{\Pgt}$ which are obtained as
\begin{equation*}
f_{i}=\frac{N_{i}(\text{\zjets signal-like})}{N_{i}(\text{\zjets background-enriched})+N_{i}(\text{\zjets signal-like})},
\end{equation*}
where  $N_{i}(\text{\zjets signal-like})$ is the number of events with a third lepton candidate that passes the signal-like sample selection, $N_{i}(\text{\zjets background-enriched})$ is the number of events in the background-enriched sample and $i=$~\Pe, \Pgm\ or \Pgt.
 The lepton selection criteria for the signal are given in Table~\ref{mutauSel} and \ref{etauSel}. The background-enriched lepton selection used to estimate  the misidentified $\Pgm$ and $\Pe$ contribution requires an isolation of $0.15 < I_\text{rel}^{\ell} < 0.25 $ and $0.1 < I_\text{rel}^{\ell} < 0.5$, respectively. In both cases the misidentification rate is computed as a function of the lepton \pt. The lepton selection for the \tauh\ background-enriched sample requires that the tau candidates are identified using a loose HPS working point but are not identified by the tight working point used for the signal selection. The loose and the tight working points have an efficiency of 75\% and 60\% for genuine \tauh\ candidates, respectively. The misidentification rates show a \pt dependence that varies with the \Pgt\ decay mode and $\abs{\eta}$. The misidentification rates are thus obtained as a function of \pt for the different decay modes and $\abs{\eta}$ regions ($\abs{\eta}<1.5$ or $\abs{\eta}>1.5$).

The final misidentified lepton background in the signal region for the two analyses (BDT and \mcol fit) is obtained from  background-enriched signal-like samples (LFV background-enriched, type $i$), where the lepton $i$ ($i$ = \Pe, \Pgm\ or \Pgt) passes  the identification and isolation criteria used for the \zjets background-enriched sample but not those defining the \zjets signal-like sample,  but otherwise  uses the same selection as the signal. To estimate  the misidentified lepton background in the signal sample, each event in  this LFV background-enriched sample of type $i$ is weighted by a factor ${f_{i}}/{(1-f_{i})}$, depending on the lepton \pt for electrons and muons or on \pt,
$\eta$, and decay mode for the \Pgt\ lepton candidates.
Both background yield and
shape distributions are thus estimated.
Double-counted events with two misidentified leptons are subtracted. For example, events
with a misidentified  \Pgm\ (\Pe) and a misidentified \tauh are subtracted in the \Hmuhad (\Hehad) channel using a weight $ {f_\Pgt\, f_{\ell}}/{[(1-f_{\Pgt})\,(1-f_{\ell})]}$  (where $\ell$ = \Pgm\ or \Pe) applied to the
events of a LFV background-enriched sample defined requiring both leptons to pass the identification and isolation criteria used for the \zjets background-enriched sample but not those defining the \zjets signal-like sample.

The background estimation is validated in a like-sign sample applying the misidentification rate $f_{i}$ to events selected inverting the charge requirement of the lepton pair in both the back\-ground-\-en\-riched and the signal-like samples. It is performed after the loose selection described in Section~\ref{eventsel}.  Figure~\ref{fig:SScontrolregion} (central) shows the data compared to the background estimation in the like-sign control region for the \Hmuhad channel. The like-sign selection enhances the misidentified lepton background and there is
good agreement in the control sample. The background estimation can also be validated in a \PW\ boson enriched control sample. This data sample is obtained by applying the signal  sample requirements and  $\mt(\ell)>60$\GeV ($\ell = \Pe$  or \Pgm)
and $\mt(\tauh)>80$\GeV.
Figure~\ref{fig:SScontrolregion} (right) shows the data compared to the background estimation in the \PW\ enriched sample for the \Hmuhad channel.
The same samples are used in the \Hehad channel with similar agreement.

\subsection*{Semi data-driven technique}
The \wjets background contribution to the misidentified-lepton background is estimated with simulated samples. The QCD multijet contribution is estimated with like-sign collision data
events that pass the signal requirement. The expected yield from non-QCD processes
is subtracted using simulation. The resulting sample is then rescaled to account for the differences
between the composition in the like- and opposite-sign samples. The scaling factors are extracted from QCD multijet enriched control samples, composed of events with the lepton candidates satisfying inverted isolation requirements as illustrated in Ref.~\cite{CMS-PAS-HIG-16-043}. This technique is chosen for the leptonically decaying tau channels as the size of the samples allows a more precise background description.

\section{Systematic uncertainties}\label{sec:systematics}
The systematic uncertainties affect the normalization and the shape of the distributions of the
different processes, and arise from either experimental or theoretical sources.
They are summarized in Table~\ref{tab:systematics}.
\begin{table}[htpb]
\topcaption{Systematic uncertainties in the expected event yields. All uncertainties are treated as correlated between the categories, except those that have two values separated by the $\oplus$ sign. In this case, the first value is the correlated uncertainty and the second value is the  uncorrelated uncertainty for each individual category. Theoretical uncertainties on VBF Higgs boson production~\cite{YR4}  are also applied to VH production. Uncertainties on acceptance lead to migration of events between the categories, and can be
correlated or anticorrelated between categories. Ranges of uncertainties for the Higgs boson
production indicate the variation in size, from negative (anticorrelated) to positive (correlated).
}
\label{tab:systematics}
\centering
\begin{tabular}{l*{4}{c}} \hline
Systematic  uncertainty            & $\PH\to\Pgm\tauh$ & $\PH\to\Pgm\Pgt_{\Pe}$ & $\PH\to\Pe\tauh $ & $\PH\to\Pe\Pgt_{\mu}$ \\ \hline
Muon  trigger/identification/isolation         &       2\%             &       2\%          &          \NA          &  2\%   \\
Electron trigger/identification/isolation      &       \NA               &       2\%          &          2\%          &    2\% \\
Hadronic tau lepton efficiency     &       5\%              &       \NA          &          5\%           & \NA\\
b tagging veto                     &      2.0--4.5\%        &      2.0--4.5\%     &      \NA       &    2.0--4.5\% \\[\cmsTabSkip]
$\cPZ\to\Pgm\Pgm, \Pe\Pe$ + jets background&   \NA    &  10\%$\oplus$5\%           &     \NA   &   10\%$\oplus$5\%  \\
$\cPZ\to\Pgt\Pgt$ + jets background &     10\%$\oplus$5\%   &   10\%$\oplus$5\%  &     10\%$\oplus$5\%   & 10\%$\oplus$5\%  \\
$\PW+\text{jets}$ background     &           \NA         &        10\%          &         \NA          & 10\%  \\
QCD multijet background            &           \NA         &        30\%          &         \NA          & 30\%  \\
$\PW\PW, \cPZ\cPZ$ background      &     5\%$\oplus$5\%    &   5\%$\oplus$5\%   &     5\%$\oplus$5\%    & 5\%$\oplus$5\%    \\
\ttbar\  background                &     10\%$\oplus$5\%   &   10\%$\oplus$5\%  &     10\%$\oplus$5\%   & 10\%$\oplus$5\%     \\
$\PW\gamma$ background           &   \NA  &       10\%$\oplus$5\%        &    \NA   &   10\%$\oplus$5\%    \\
Single top quark background   &     5\%$\oplus$5\%    &   5\%$\oplus$5\%   &     5\%$\oplus$5\%    &         5\%$\oplus$5\%      \\[\cmsTabSkip]
$\Pgm\to\tauh$ background         &         25\%           &         \NA      &           \NA           & \NA \\
$\Pe\to\tauh$ background          &           \NA           &          \NA         &           12\%           & \NA \\
$\text{Jet}\to\tauh, \Pgm, \Pe $ background &  30\%$\oplus$10\%   &  \NA  &            30\%$\oplus$10\%         &   \NA  \\
Jet energy scale                   &        3--20\%         &        3--20\%      &          3-20\%       &   3--20\%\\
\tauh energy scale                 &         1.2\%             &        \NA     &           1.2$\%$           &  \NA \\
$\Pgm,\Pe\to\tauh$ energy scale          &           1.5\%         &       \NA      &           3\%          & \NA \\
\Pe\ energy scale                   &       \NA    &        0.1 -- 0.5\%         &      0.1 -- 0.5\%      &  0.1 -- 0.5\% \\
\Pgm\ energy scale                  &        0.2\%          &        0.2\%       &           \NA         &  0.2\%  \\
Unclustered energy scale           &        $\pm 1 \sigma$    &  $\pm 1 \sigma$       &      $\pm 1 \sigma$      &  $\pm 1 \sigma$\\[\cmsTabSkip]
Renorm./fact. scales ({\cPg\cPg}H)   \cite{YR4}     &   \multicolumn{4}{c}{3.9\%}\\
Renorm./fact. scales (VBF and VH) \cite{YR4}             &   \multicolumn{4}{c}{0.4\%}\\
PDF + $\alpha_s$ ({\cPg\cPg}H)    \cite{YR4}        &   \multicolumn{4}{c}{ 3.2\%}\\
PDF + $\alpha_s$ (VBF and VH)   \cite{YR4}                  &   \multicolumn{4}{c}{ 2.1\%}\\
Renorm./fact. acceptance ({\cPg\cPg}H)     &   \multicolumn{4}{c}{$-3.0$\% -- $+2.0$\% } \\
Renorm./fact. acceptance (VBF and VH)           &   \multicolumn{4}{c}{$-0.3$\% -- $+1.0$\% } \\
PDF + $\alpha_s$ acceptance ({\cPg\cPg}H)       &   \multicolumn{4}{c}{ $-1.5$\% --  $+0.5$\%}\\
PDF + $\alpha_s$ acceptance (VBF and VH)             &   \multicolumn{4}{c}{ $-1.5$\% --  $+1.0$\%}\\[\cmsTabSkip]
Integrated luminosity               &   \multicolumn{4}{c}{ 2.5\%  } \\ \hline
\end{tabular}
\end{table}
The uncertainties in the lepton (\Pe, \Pgm, \tauh) selection including the  trigger, identification, and isolation efficiencies are
estimated using tag-and-probe measurements in collision data sets of $\cPZ$ bosons decaying  to
$\Pe\Pe, \Pgm\Pgm, \Pgt_{\Pgm}\tauh$~\cite{Khachatryan2011,Chatrchyan:2012xi,Khachatryan:2015hwa,Khachatryan:2015dfa,CMS:2016gvn}. The b tagging efficiency in the simulation is adjusted to match the efficiency measured in data. The uncertainty in this measurement is taken as the systematic uncertainty. The uncertainties on the $\cPZ \to \Pe\Pe, \cPZ \to\Pgm\Pgm, \cPZ \to\Pgt\Pgt$, $\PW\PW$, $\cPZ\cPZ$, $\PW\gamma$, \ttbar,  and single top production
background contributions arise predominantly from the uncertainties in  the measured cross sections of these processes.
The uncertainties in the estimate of the misidentified-lepton backgrounds ($\Pgm\to\tauh$, $\Pe\to\tauh$, $\text{jet}\to\tauh, \Pgm, \Pe$)  are extracted from the validation tests in control samples, described in Section~\ref{backgrounds}.

Shape and normalization uncertainties arising from the uncertainty in the  jet energy scale are computed
by propagating the effect of altering each source of jet energy scale uncertainty by one standard deviation to the fit templates of each process.
This takes  into account differences in yield and shape.
The uncertainties on the \Pe, \Pgm, $\tauh$ energy scale are propagated to the \mcol  and BDT distributions. For $\tauh$, the energy scale uncertainty is treated independently for each reconstructed hadronic decay mode of the \Pgt\ lepton.
The systematic uncertainties in the energy resolutions of lepton candidates have negligible effect. The energy scale of  muons (electrons) misidentified as hadronically decaying tau candidates ($\Pgm,\Pe\to\tauh$ energy scale)  is considered independently from true hadronic tau leptons. There is also an uncertainty in the
unclustered energy scale. The unclustered energy comes from jets having $\pt < 10$\GeV and PF candidates not within jets. It is propagated to \ptmiss.
The unclustered energy scale is considered independently for charged particles, photons, neutral hadrons, and very forward
particles which are not contained in jets. The effect of varying the energy of each particle by its uncertainty leads to  changes in both shape of the distribution and yield.  The four different systematic uncertainties are uncorrelated.

The uncertainties in the Higgs boson production cross sections due to the factorization and the renormalization scales, as well as the parton distribution functions (PDF) and the strong coupling constant ($\alpha_s$),  result  in changes in normalization and they are taken from Ref.~\cite{YR4}. They also affect the acceptance and lead to the migration of events between the categories. They are listed as acceptance uncertainties in Table~\ref{tab:systematics} and depend on the production process, Higgs boson decay channel, and category. For the {\cPg\cPg}H production this variation on the acceptance varies from $-3$\% (anticorrelated between the categories)  to 2\% (correlated) for  the factorization and the renormalization scales, and from $-1.5$\%  to 0.5\% for PDF and $\alpha_s$. For the VBF and associated production (VH) the ranges go from $-0.3$\% to 1.0\% for the factorization and the renormalization scales, and from $-1.5$\% to 1.0\% for PDF and $\alpha_s$.

The bin-by-bin uncertainties account for the statistical uncertainties in every bin of the template distributions of every process. They are uncorrelated between
bins, processes, and categories.
The uncertainty of 2.5\% on the integrated luminosity~\cite{CMS-PAS-LUM-17-001} affects all processes with the normalization taken directly from simulation.
Shape uncertainties related to the pileup have been considered by varying the weights applied to simulation. The weight variation is obtained by a  5\% change of the total inelastic cross section used to estimate the number of pileup events in data. The new values are then used to compute the weights for the simulation samples and these are applied, event by event, to produce alternate collinear mass and BDT distributions used as shape uncertainties in the fit. Other minimum bias event modelling and simulation uncertainties are estimated to be much smaller than those on the rate and are therefore neglected.

\section{Results}\label{results}

After applying the selection criteria, a maximum likelihood fit is performed to derive the expected and observed limits. Each systematic uncertainty is used as a nuisance parameter in the fit. The fits are performed simultaneously in all channels and categories.
A profile likelihood ratio is used as test statistic. The upper limits on the signal branching fraction  are calculated with the asymptotic formula, using the CL$_{\mathrm{s}}$ criterion~\cite{Junk,Read2,CLs}.

\begin{figure}[!htpb]\centering
 \includegraphics[width=0.33\textwidth]{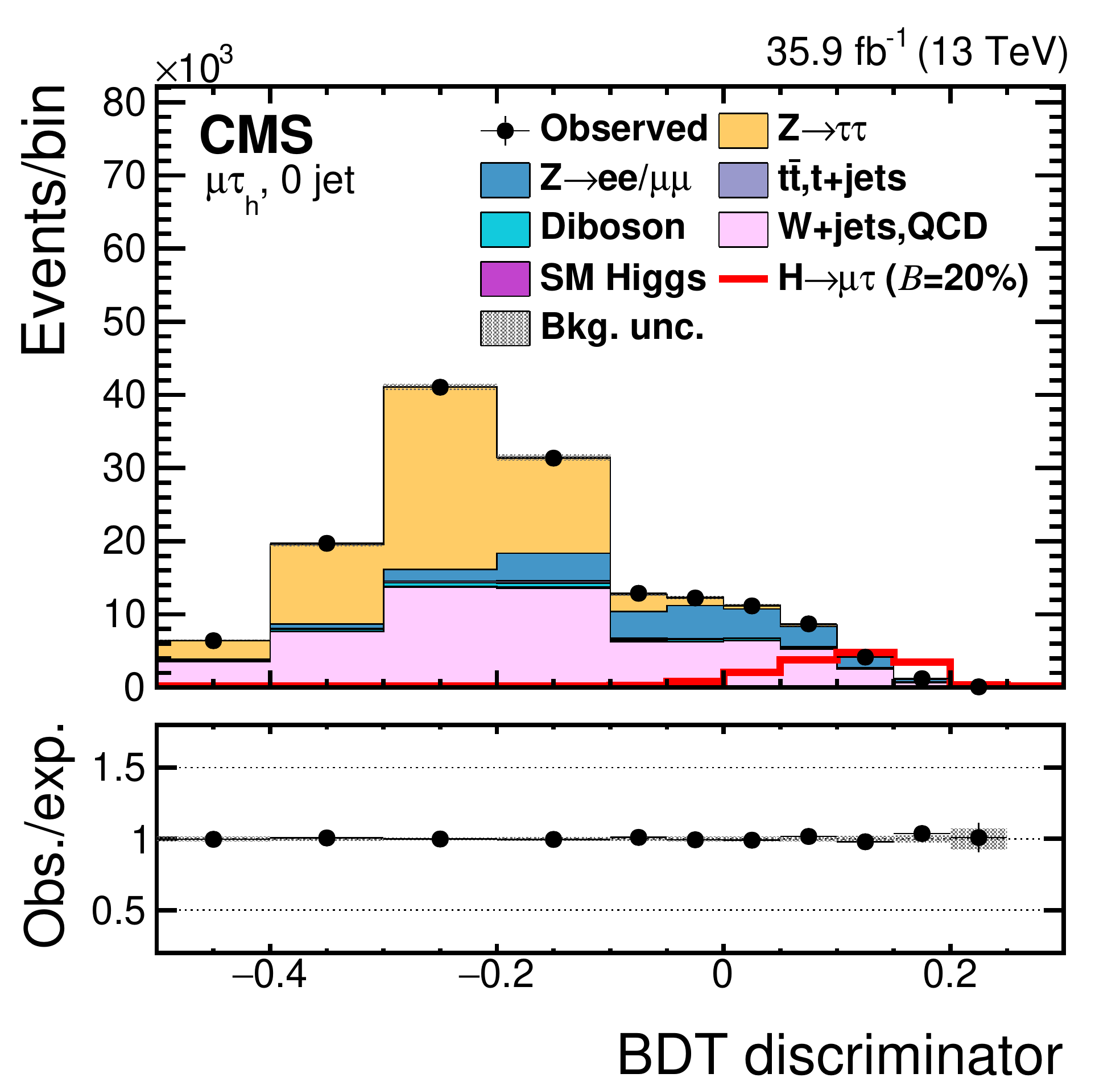}
 \includegraphics[width=0.33\textwidth]{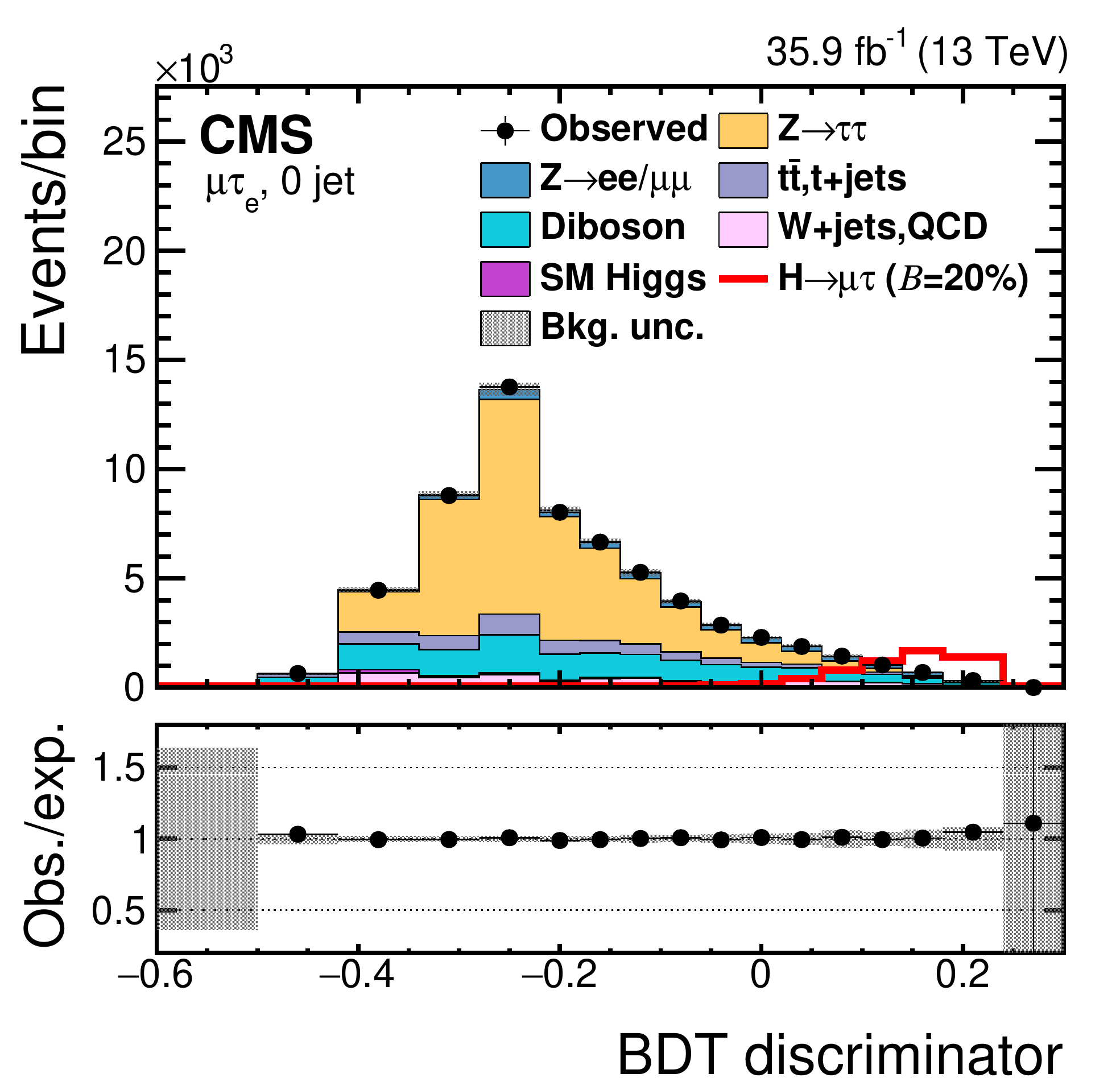} \\
 \includegraphics[width=0.33\textwidth]{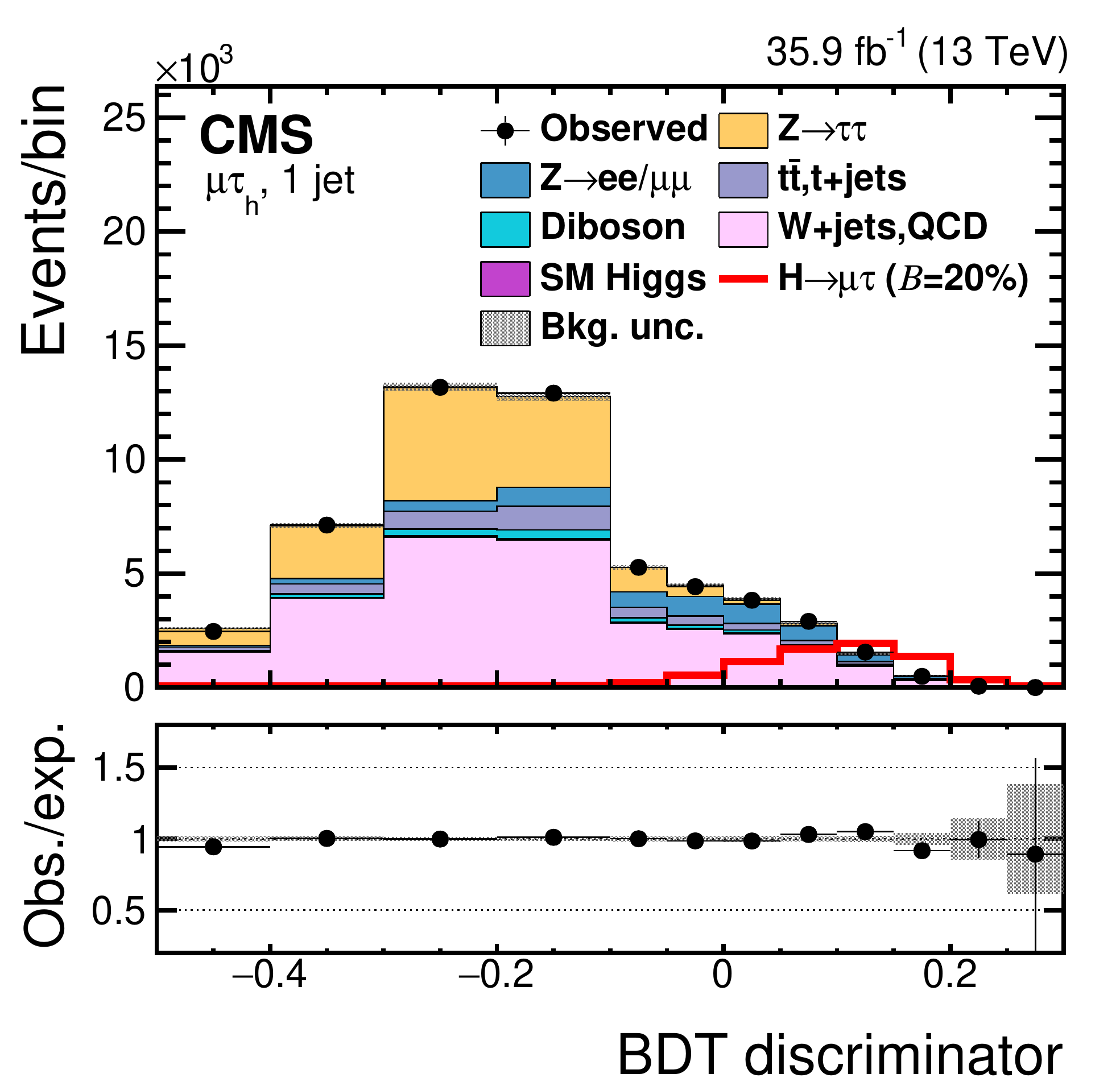}
 \includegraphics[width=0.33\textwidth]{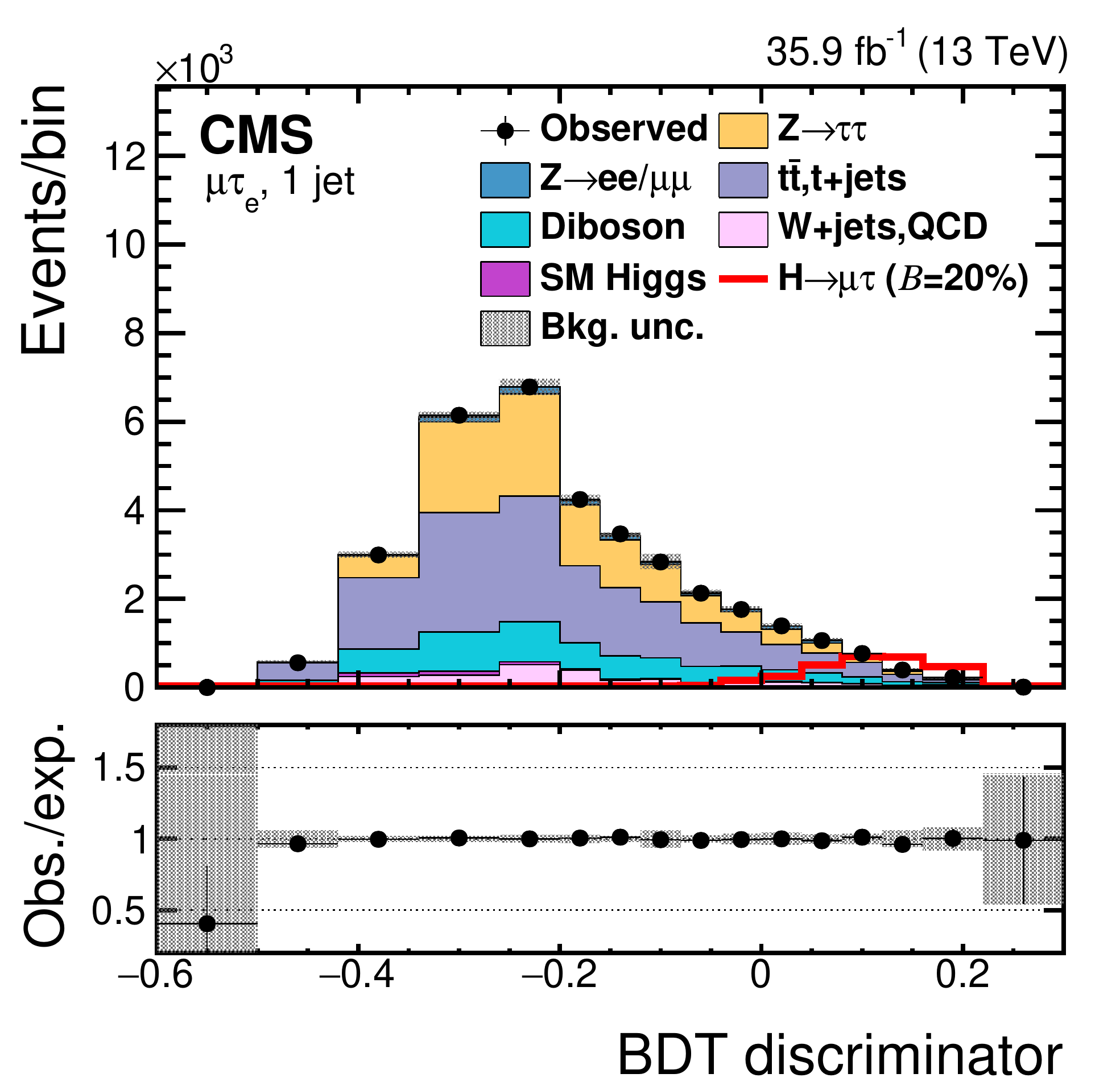}  \\
 \includegraphics[width=0.33\textwidth]{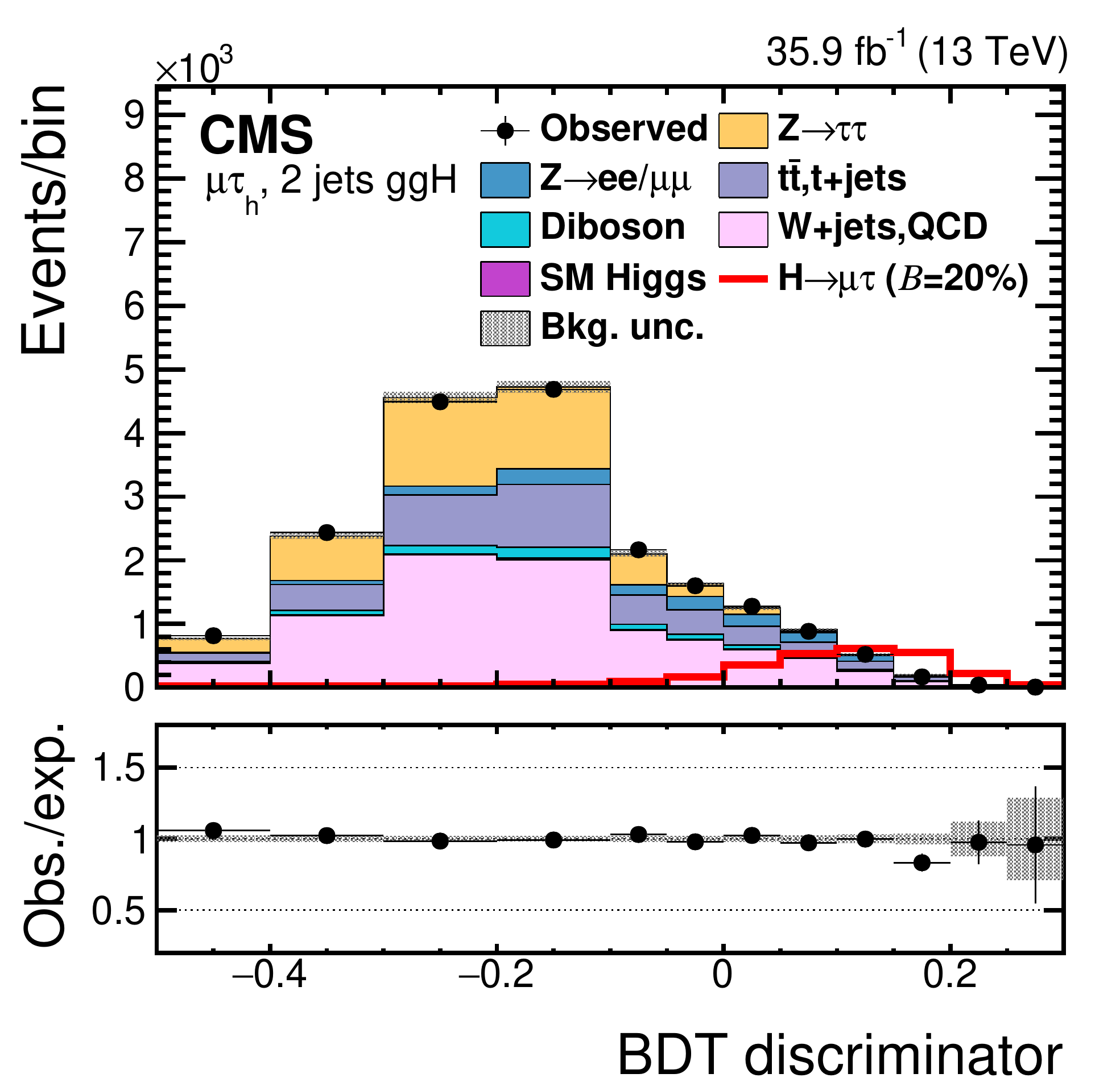}
 \includegraphics[width=0.33\textwidth]{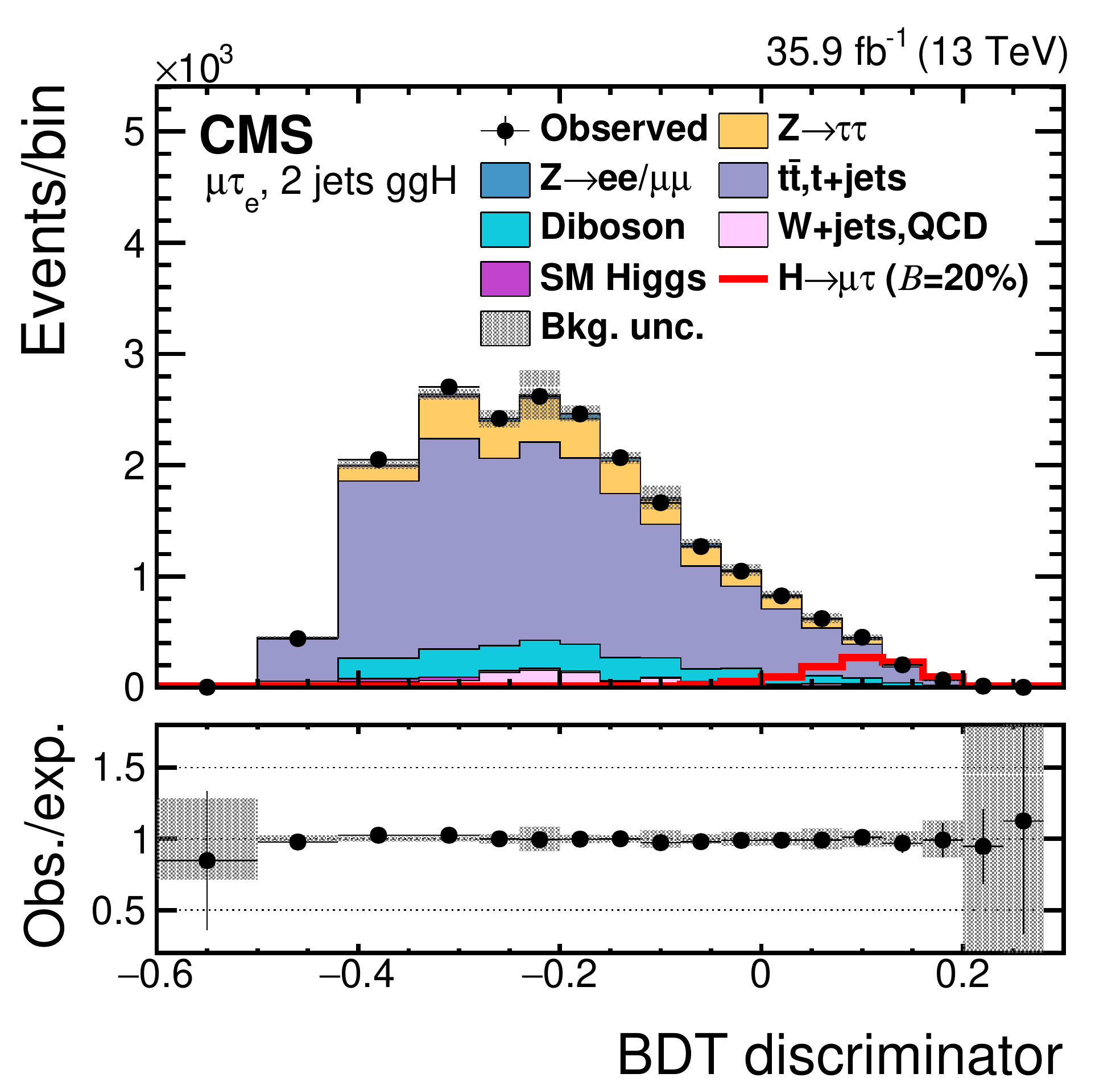} \\
 \includegraphics[width=0.33\textwidth]{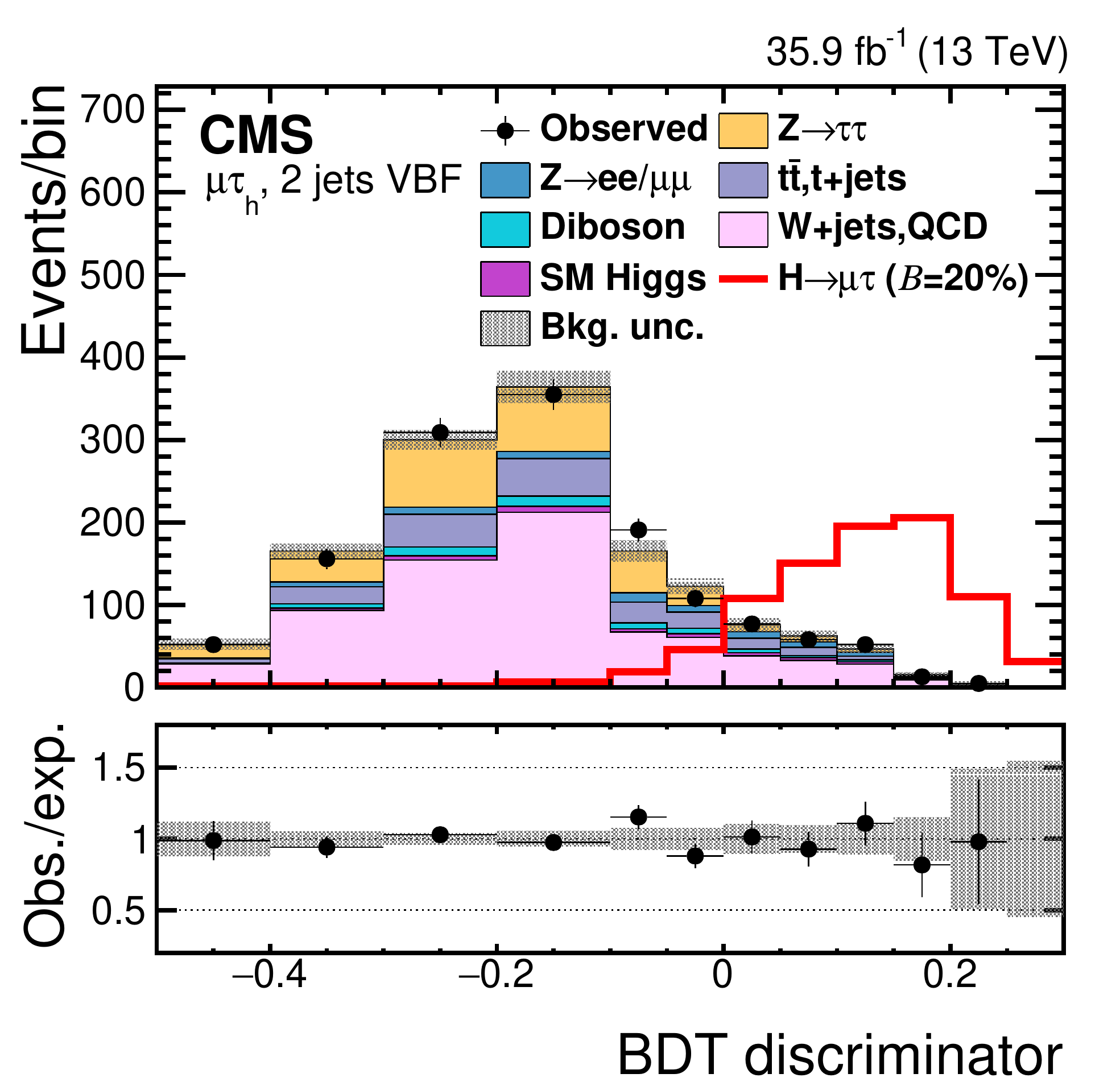}
 \includegraphics[width=0.33\textwidth]{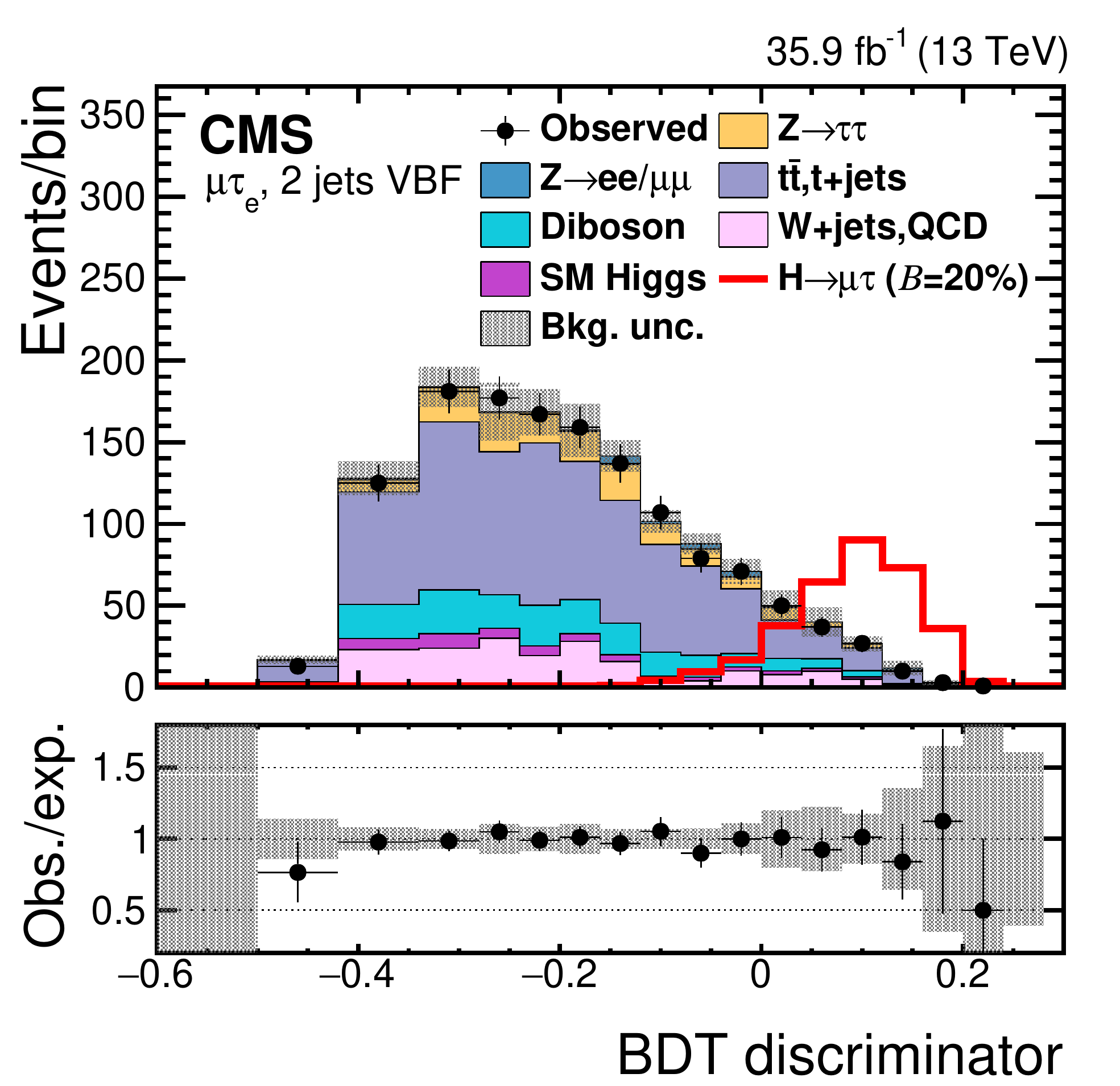}
\caption{Distribution of the BDT discriminator  for the $\PH \to \Pgm\Pgt$  process in the BDT fit analysis, in the individual channels and categories compared to the signal and background estimation.
The background is normalized to the best fit values from the signal plus background fit while the simulated signal corresponds to
 $\mathcal{B}(\PH \to \Pgm \Pgt )=5\%$.
The bottom  panel in each plot shows the fractional difference between the observed data and the fitted background.
The left column of plots corresponds to the $\PH \to \Pgm \tauh$ categories, from 0-jets (first row) to 2-jets VBF (fourth row). The right one to their $\PH \to \Pgm \Pgt_{\Pe}$ counterparts.}
 \label{fig:Mcol_SignalRegion_BDTMethod2_MuTau}
\end{figure}

\begin{figure}[!htpb]\centering
 \includegraphics[width=0.33\textwidth]{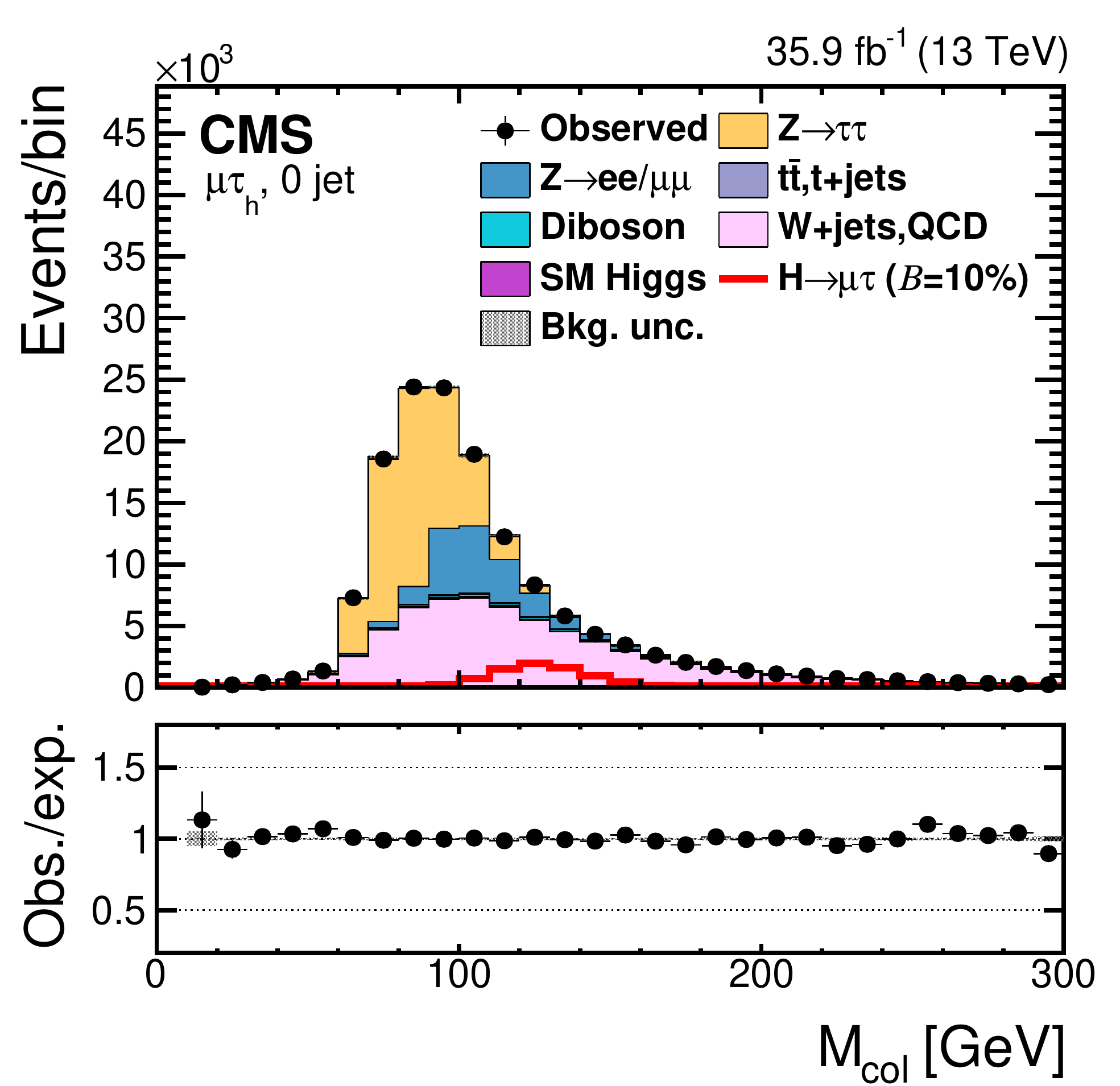}
 \includegraphics[width=0.33\textwidth]{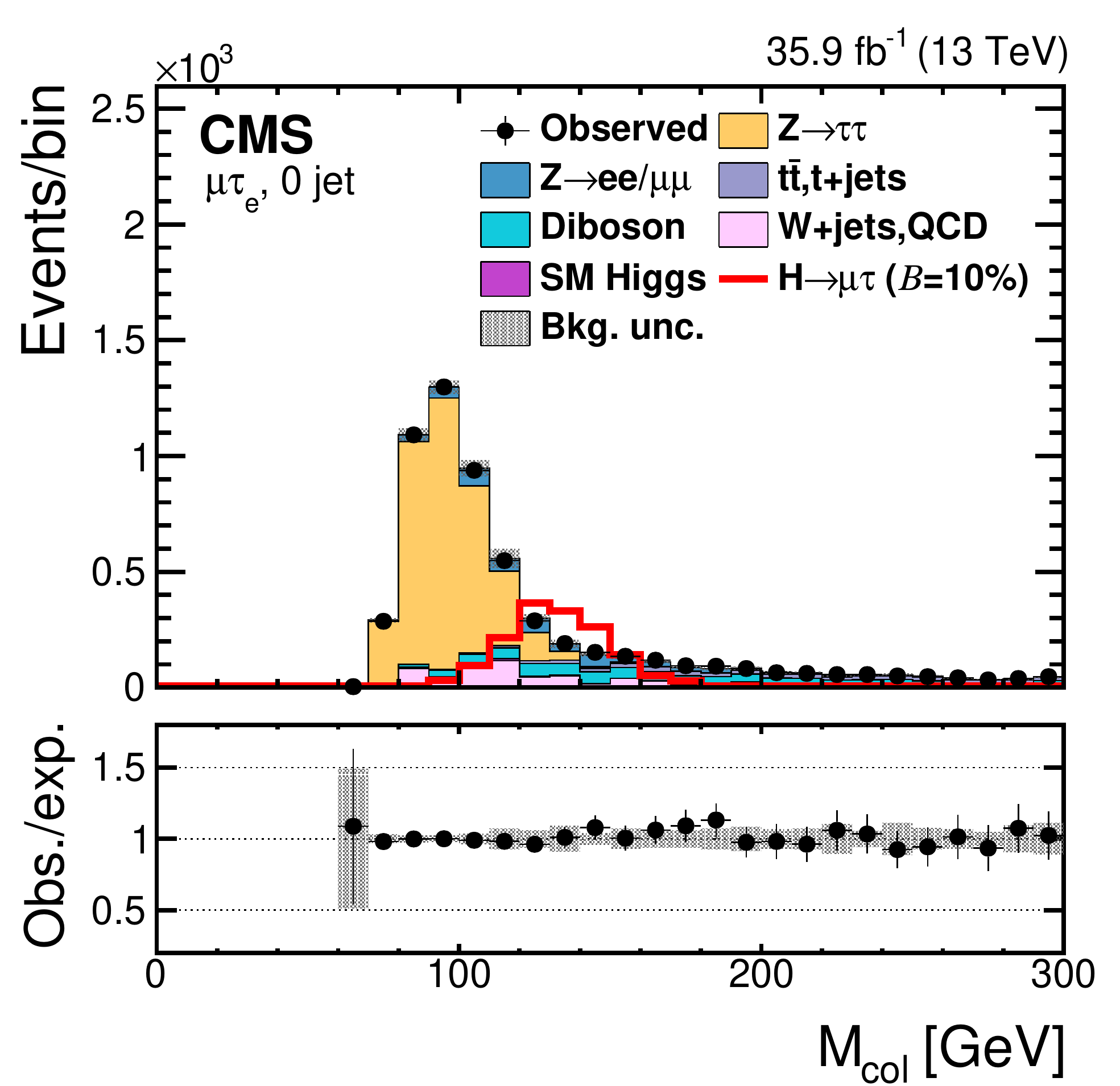} \\
 \includegraphics[width=0.33\textwidth]{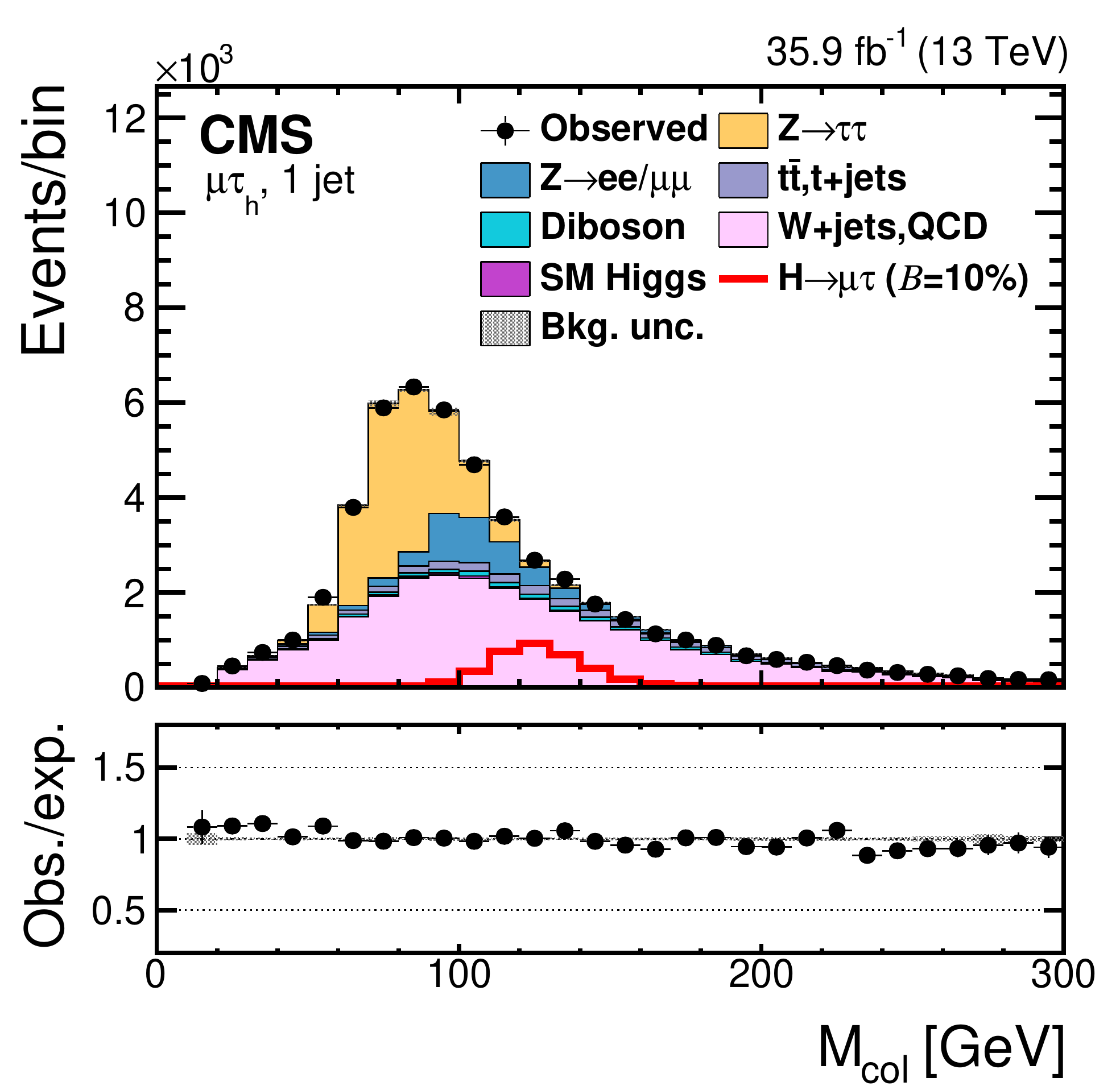}
 \includegraphics[width=0.33\textwidth]{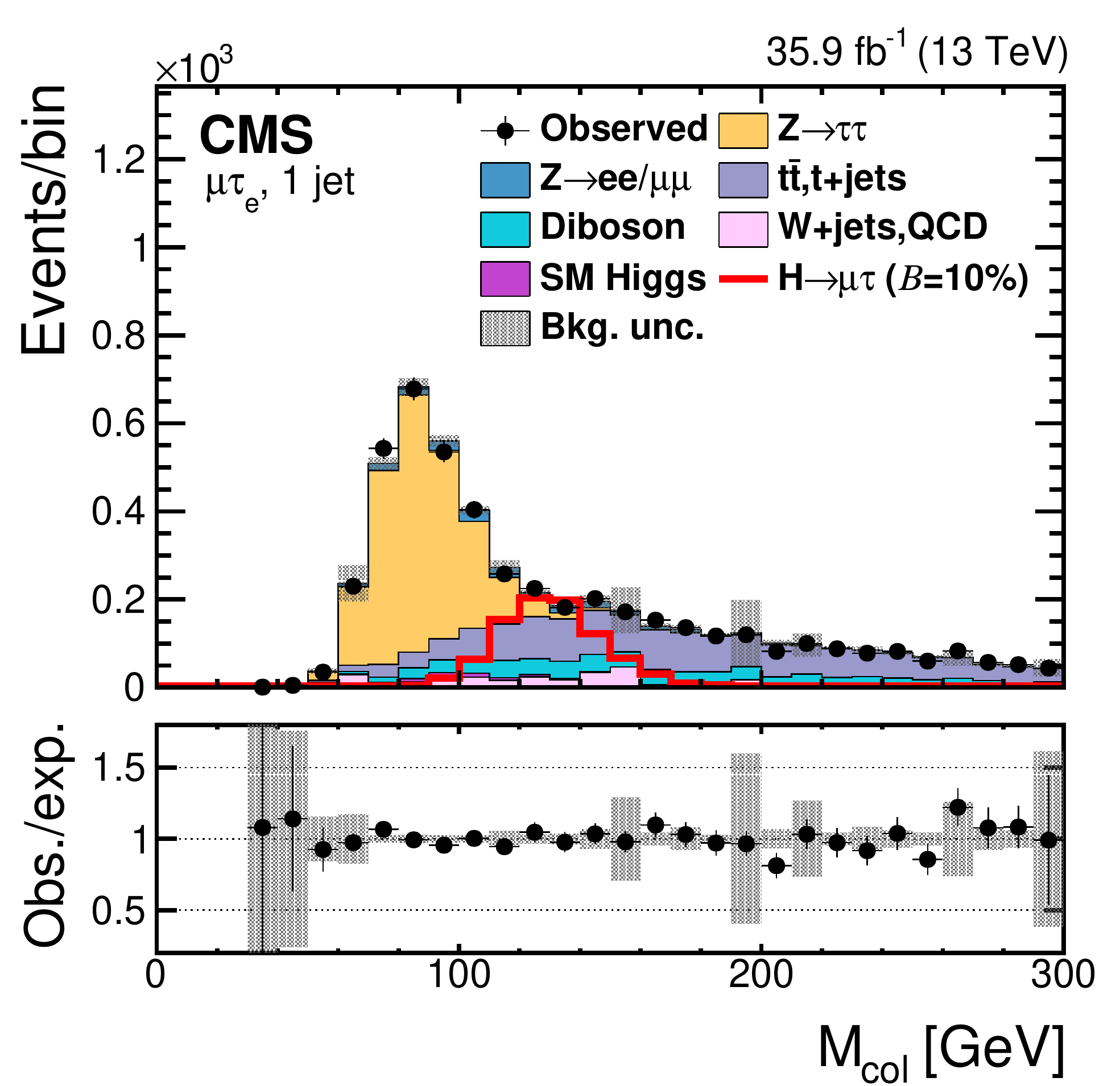}  \\
 \includegraphics[width=0.33\textwidth]{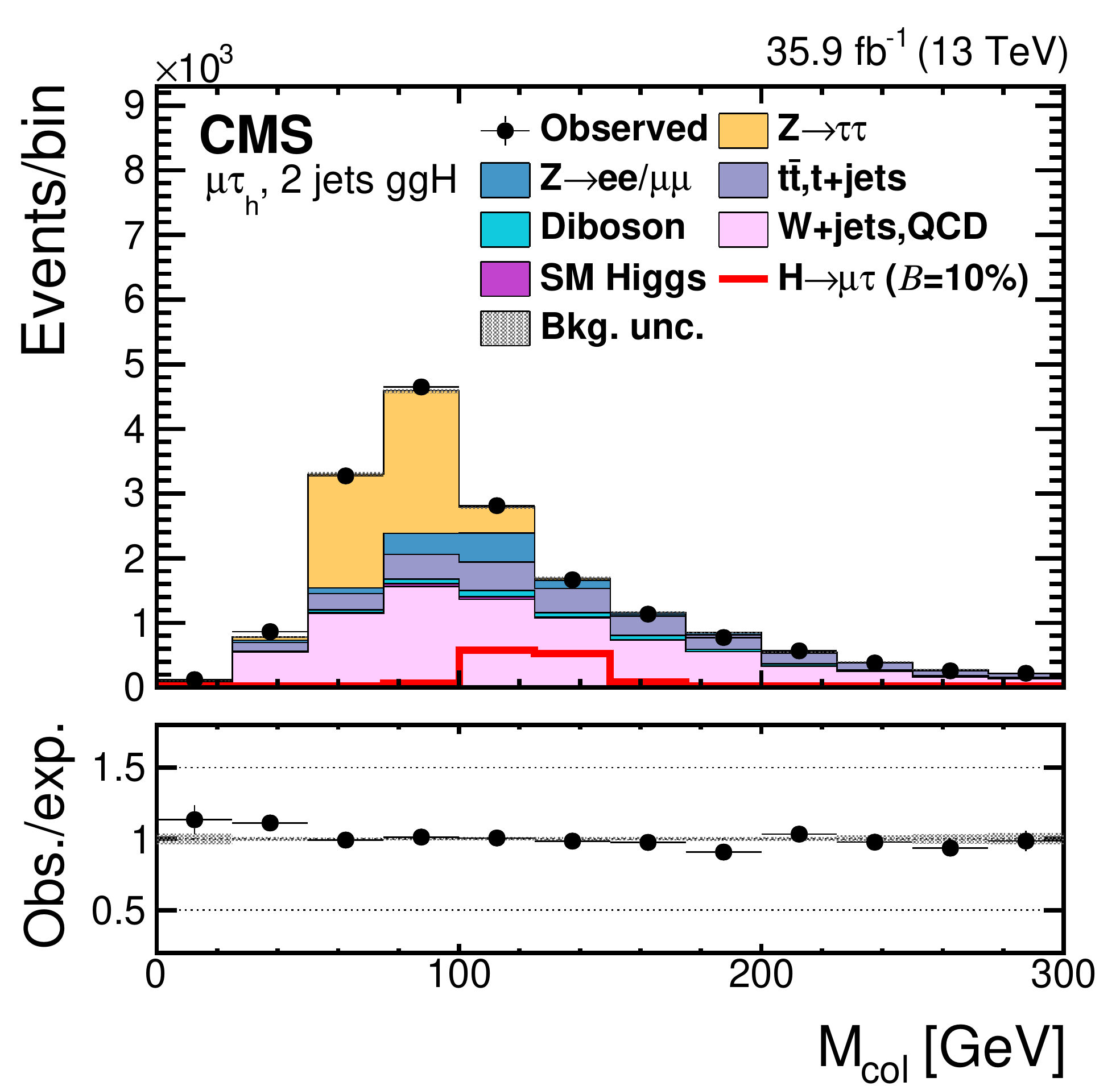}
 \includegraphics[width=0.33\textwidth]{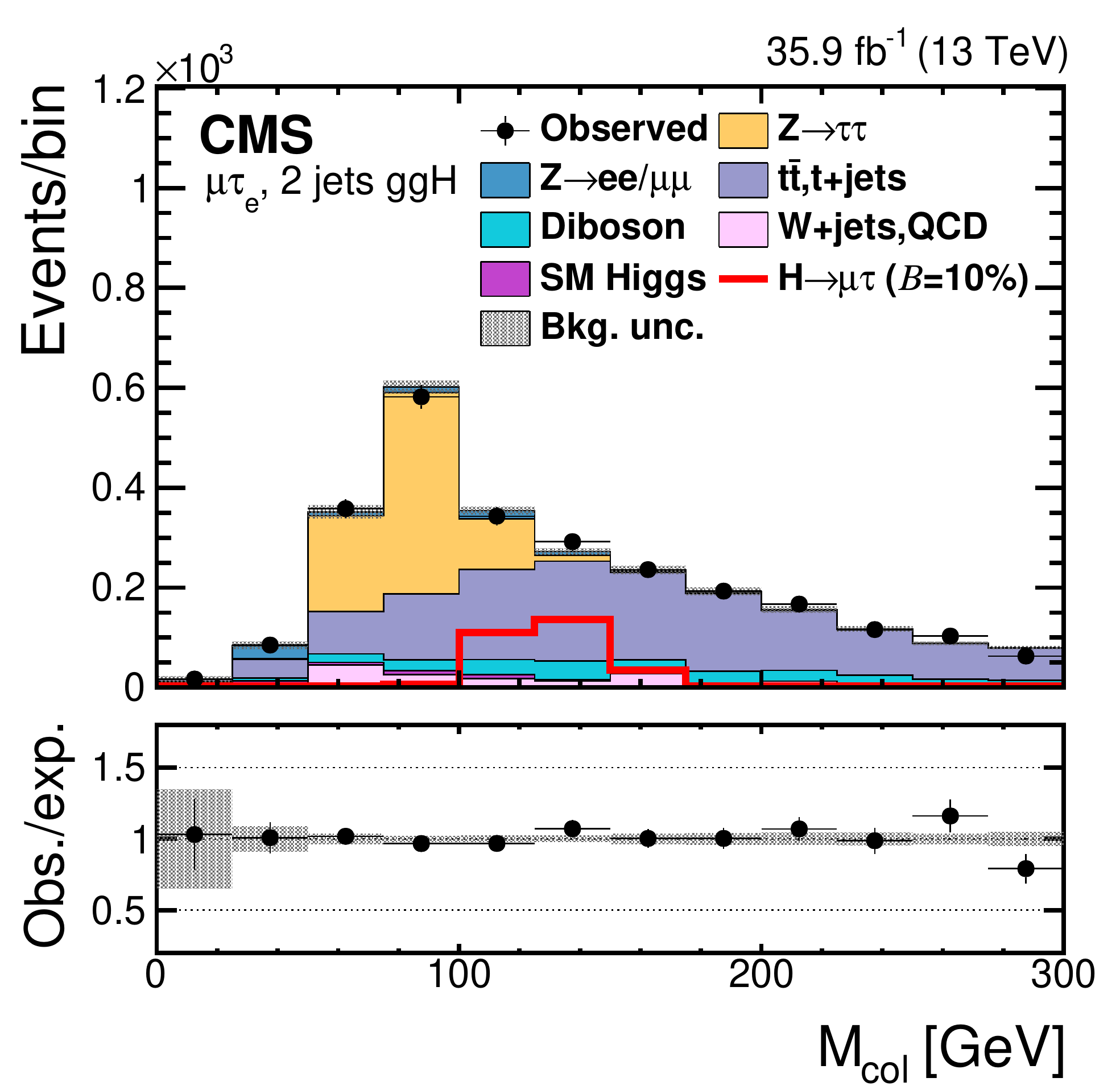} \\
 \includegraphics[width=0.33\textwidth]{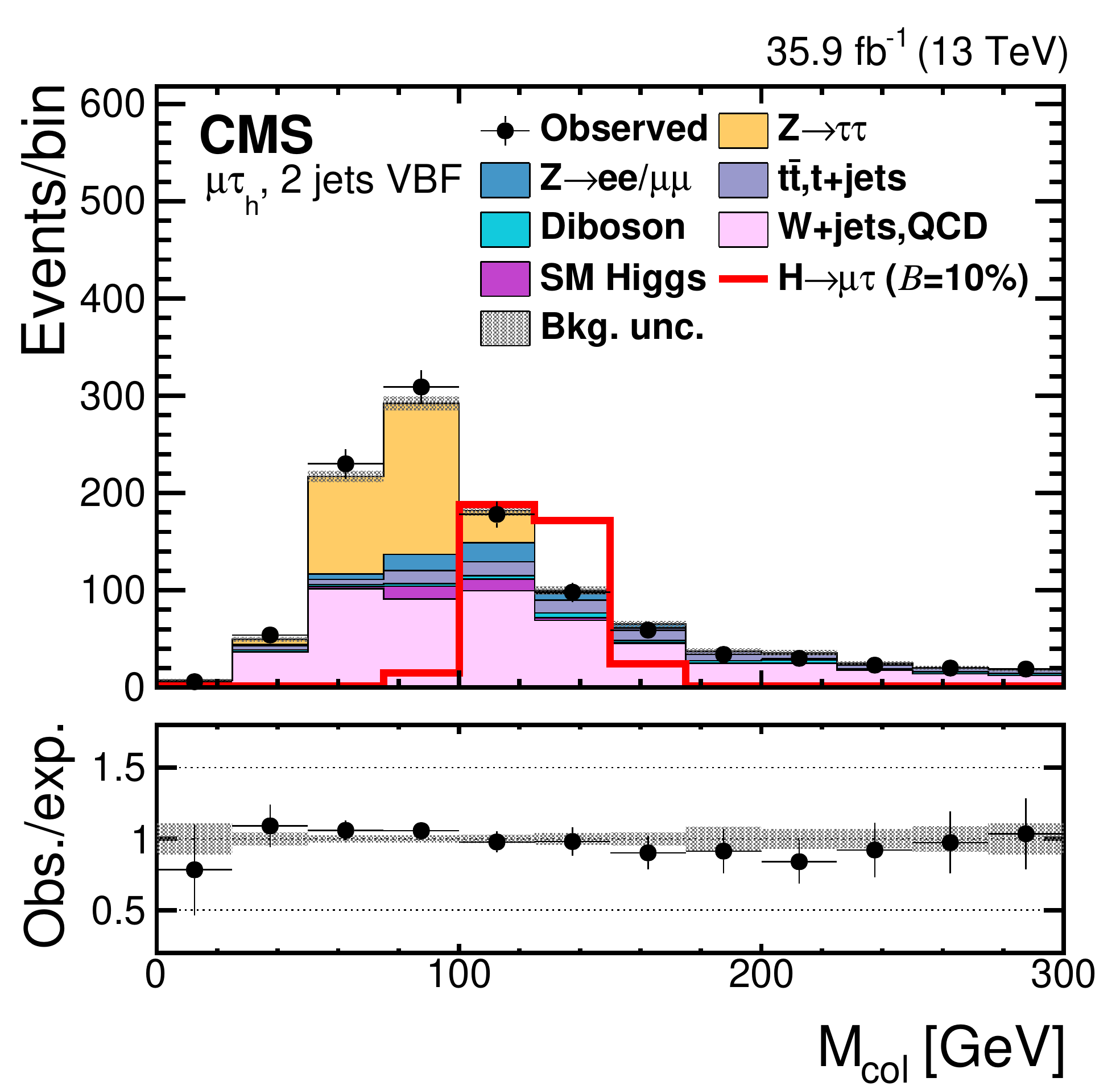}
 \includegraphics[width=0.33\textwidth]{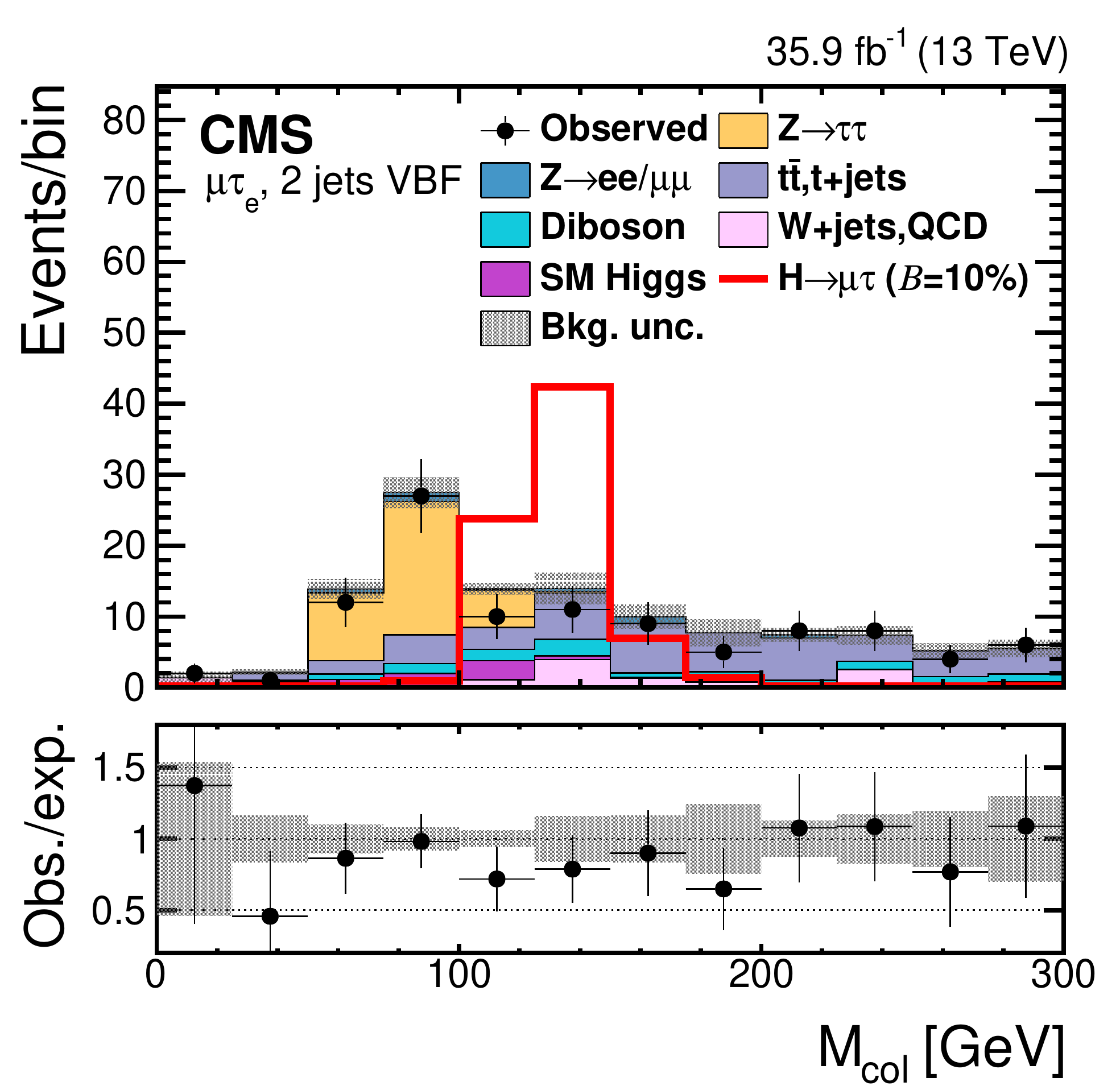}
\caption{Distribution of the collinear mass $\mcol$ for the $\PH \to \Pgm\Pgt$  process in \mcol fit analysis, in  different channels and categories compared to the signal and background estimation.
The background is normalized to the best fit values from the signal plus background fit while the overlaid simulated signal corresponds to $\mathcal{B}(\PH \to \Pgm \Pgt)=5\%$.
The bottom  panel in each plot shows the ratio between the observed data and the fitted background.
The left column of plots corresponds to the $\PH \to \Pgm \tauh$ categories, from 0-jets (first row) to 2-jets VBF (fourth row). The right one to their $\PH \to \Pgm \Pgt_{\Pe}$ counterparts.}
 \label{fig:Mcol_SignalRegion_CutBased_MuTau}
\end{figure}

\begin{figure}[!htpb]\centering
\includegraphics[width=0.45\textwidth]{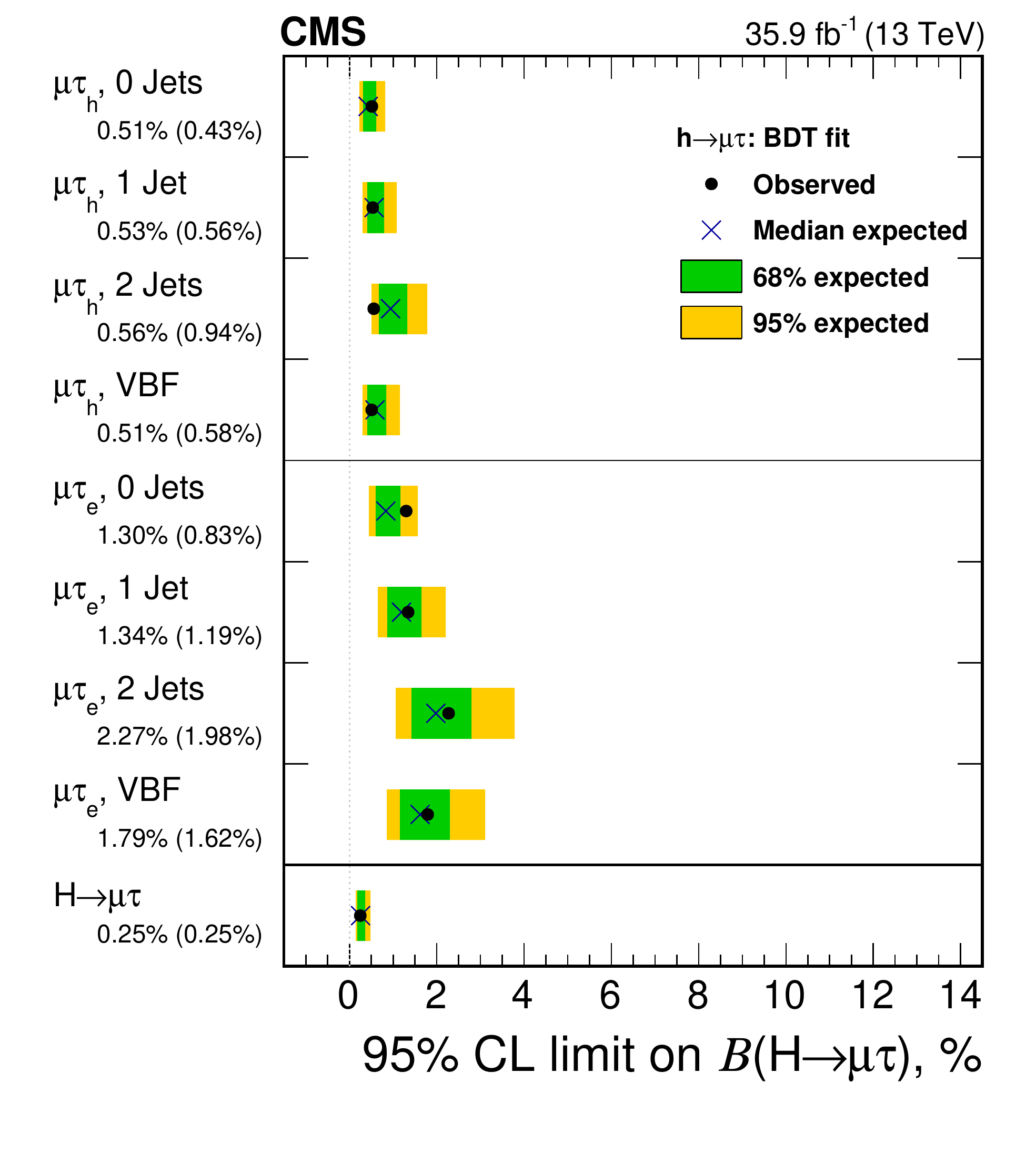}
\includegraphics[width=0.45\textwidth]{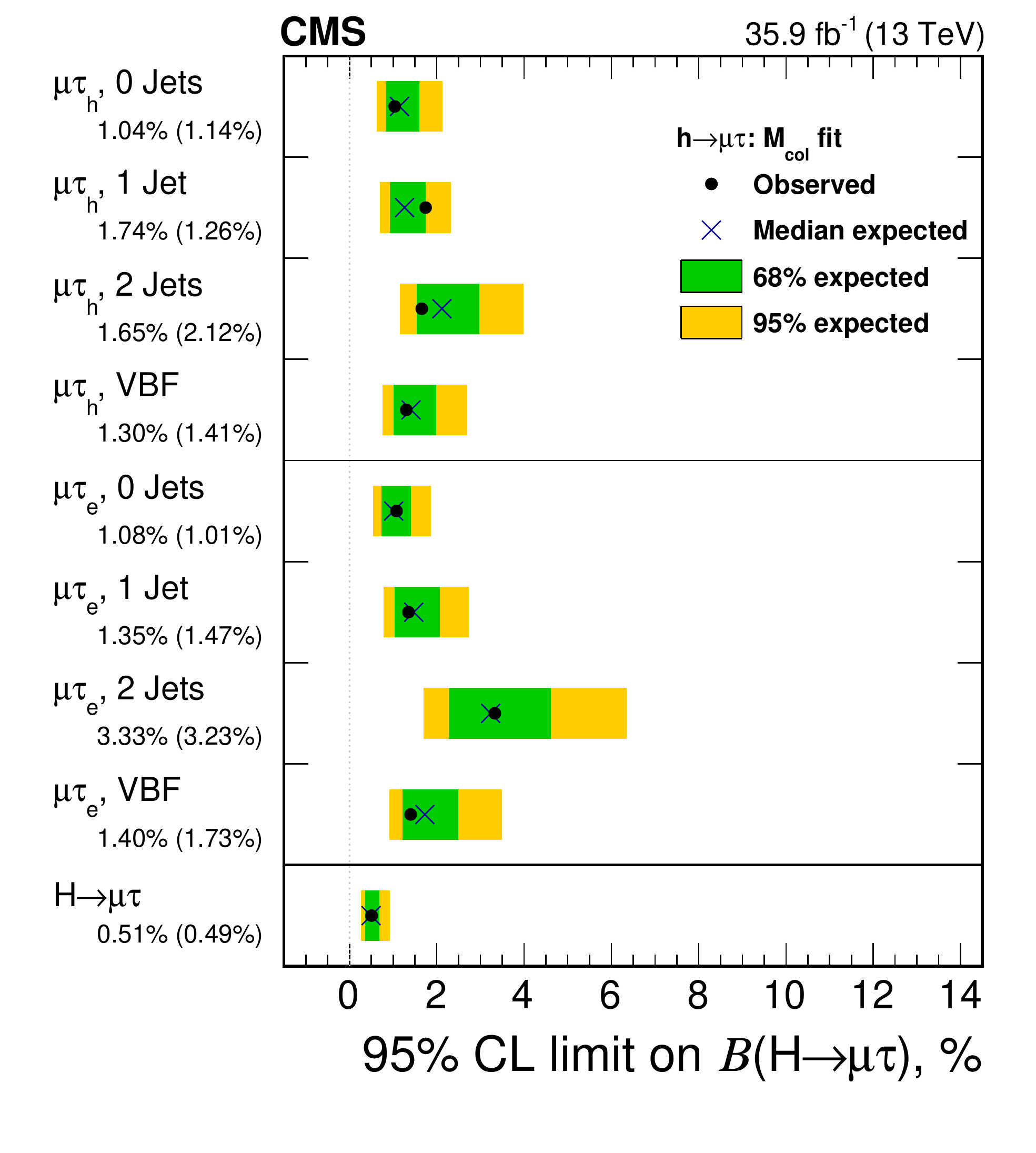}
\caption{Observed and expected 95\% CL upper limits on the $\mathcal{B}(\PH \to \Pgm \Pgt)$ for each individual category and combined. Left: BDT fit analysis. Right: $\mcol$ fit analysis.}
 \label{fig:limits_summary_MUTAU}
\end{figure}

\begin{figure}[!htpb]\centering
 \includegraphics[width=0.33\textwidth]{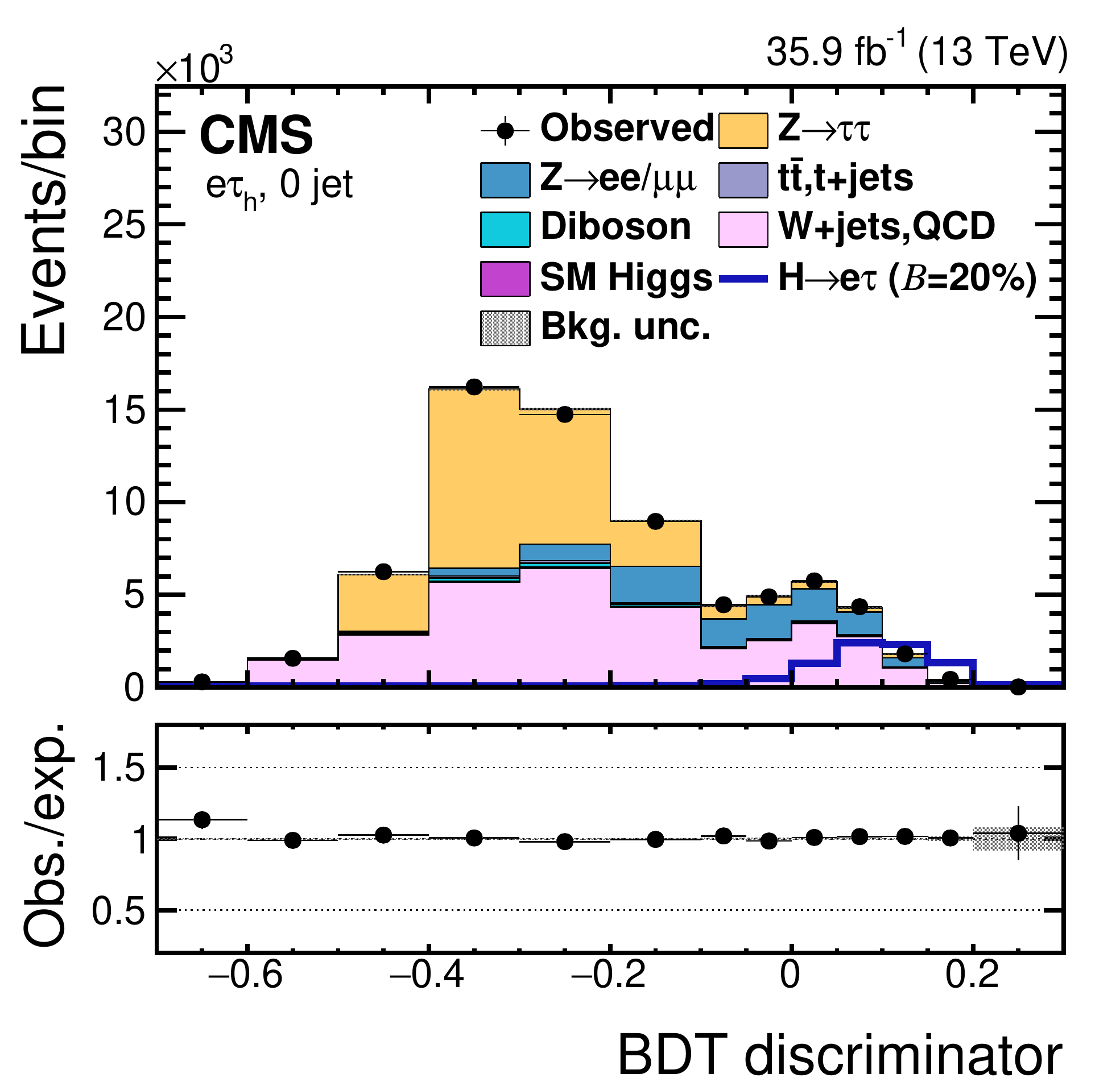}
 \includegraphics[width=0.33\textwidth]{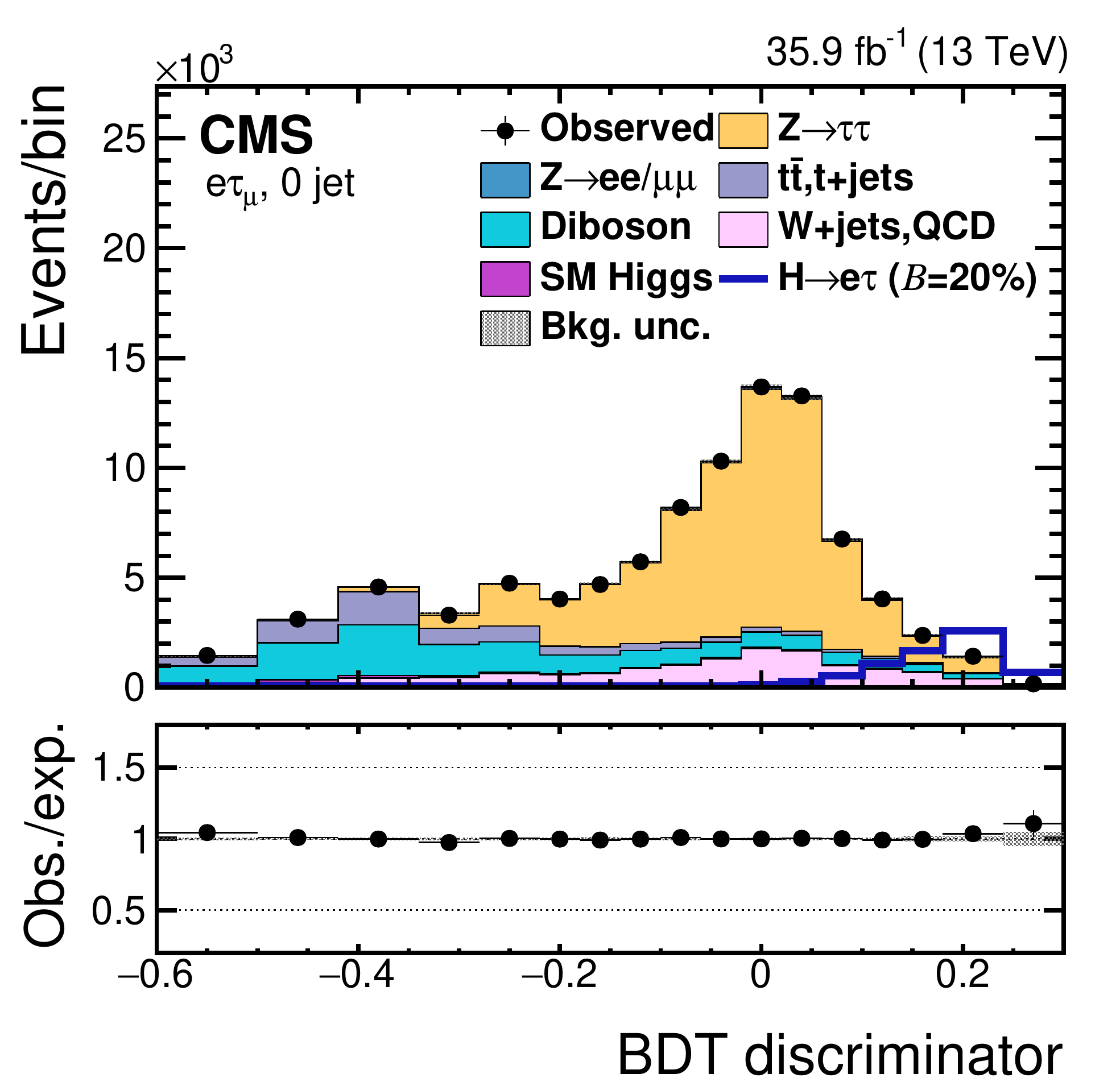} \\
 \includegraphics[width=0.33\textwidth]{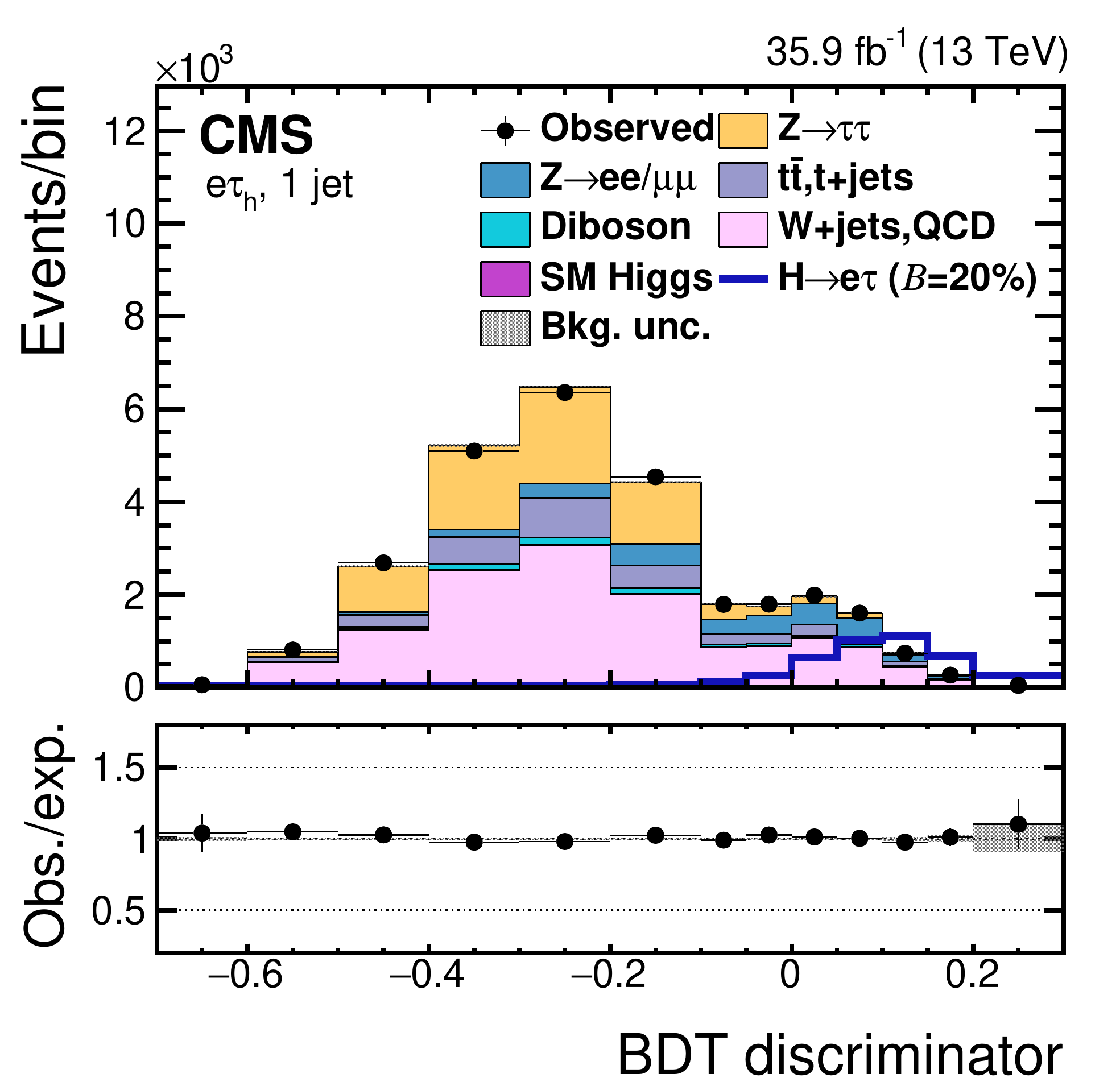}
 \includegraphics[width=0.33\textwidth]{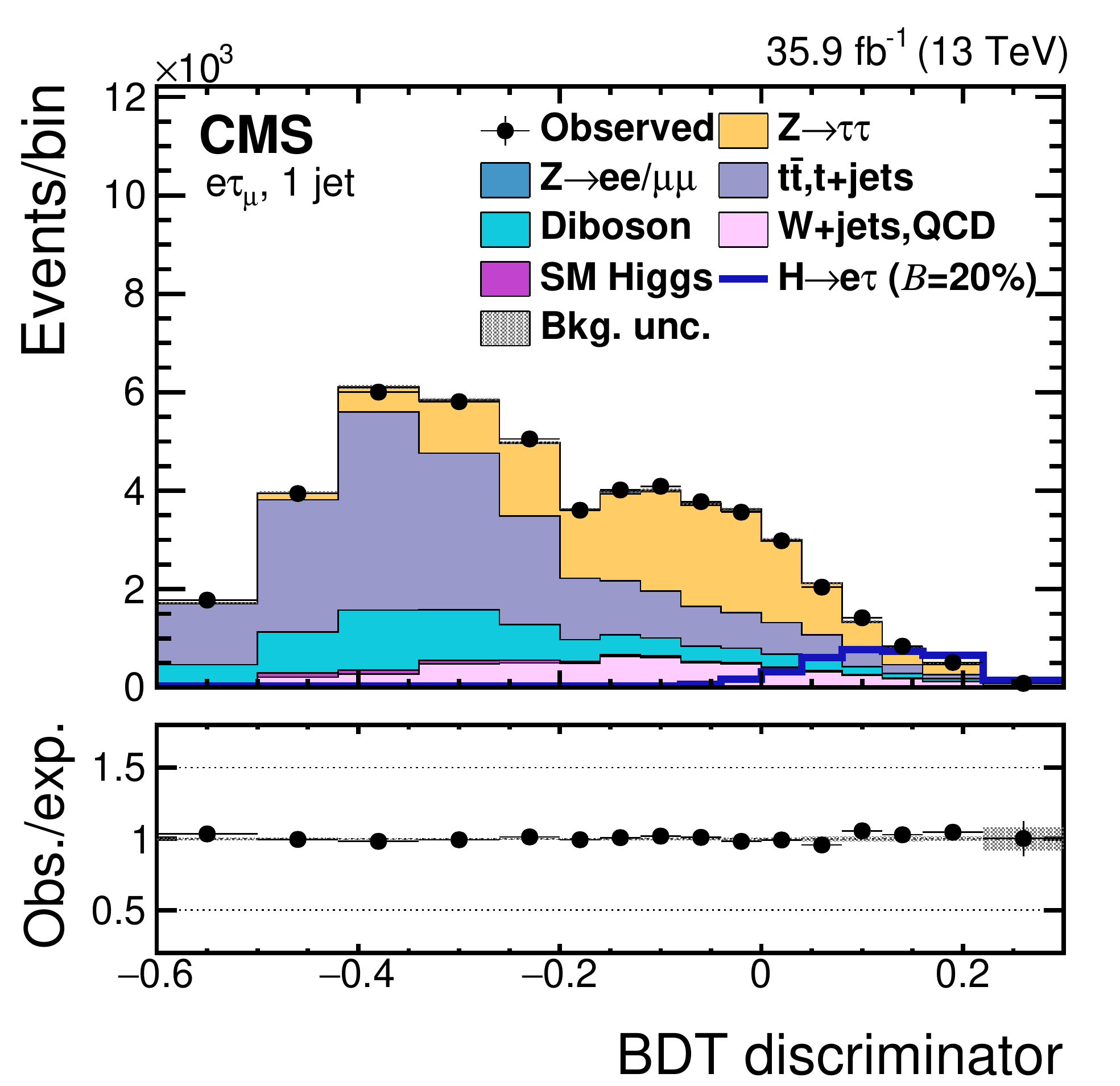}  \\
 \includegraphics[width=0.33\textwidth]{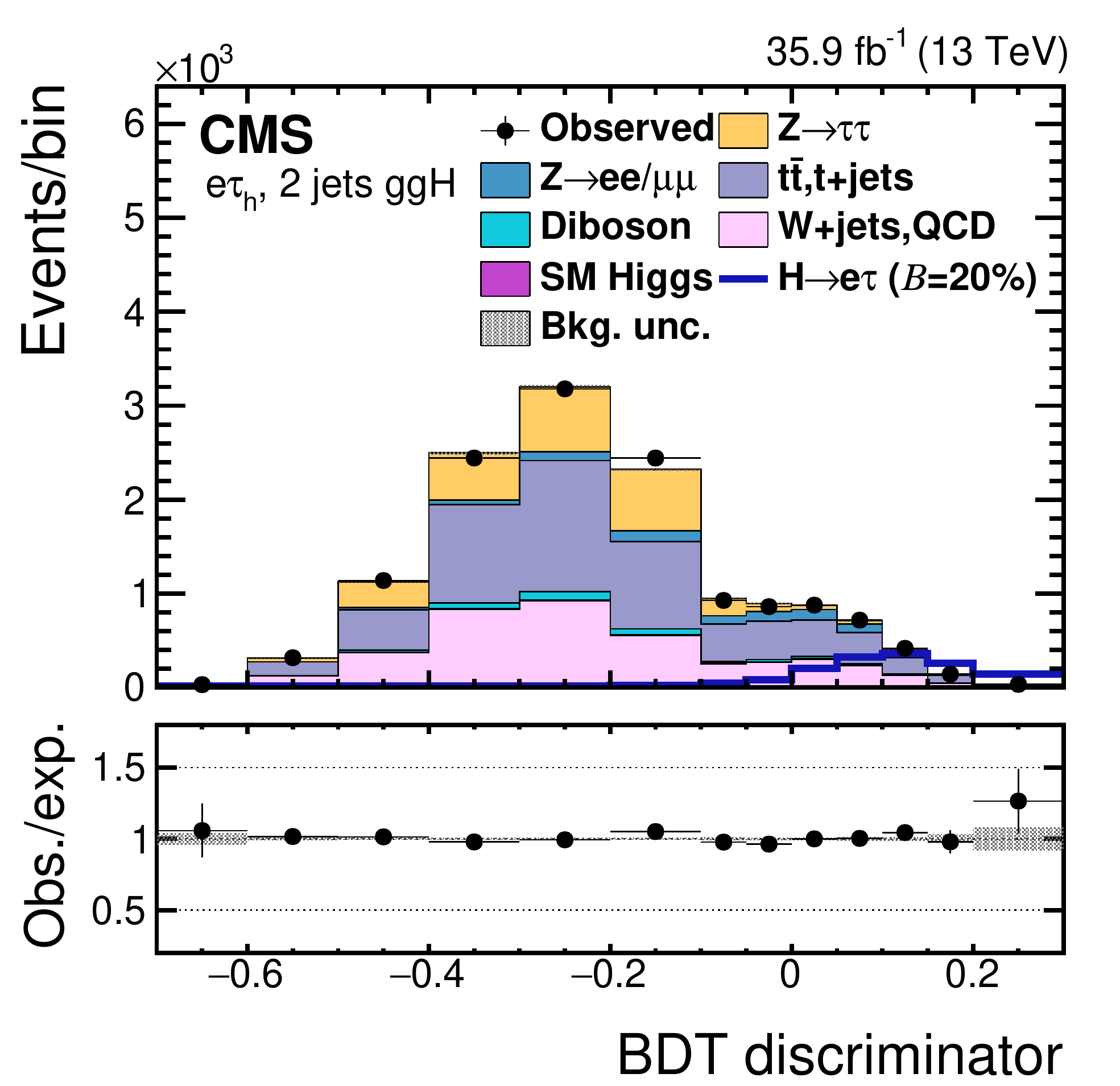}
 \includegraphics[width=0.33\textwidth]{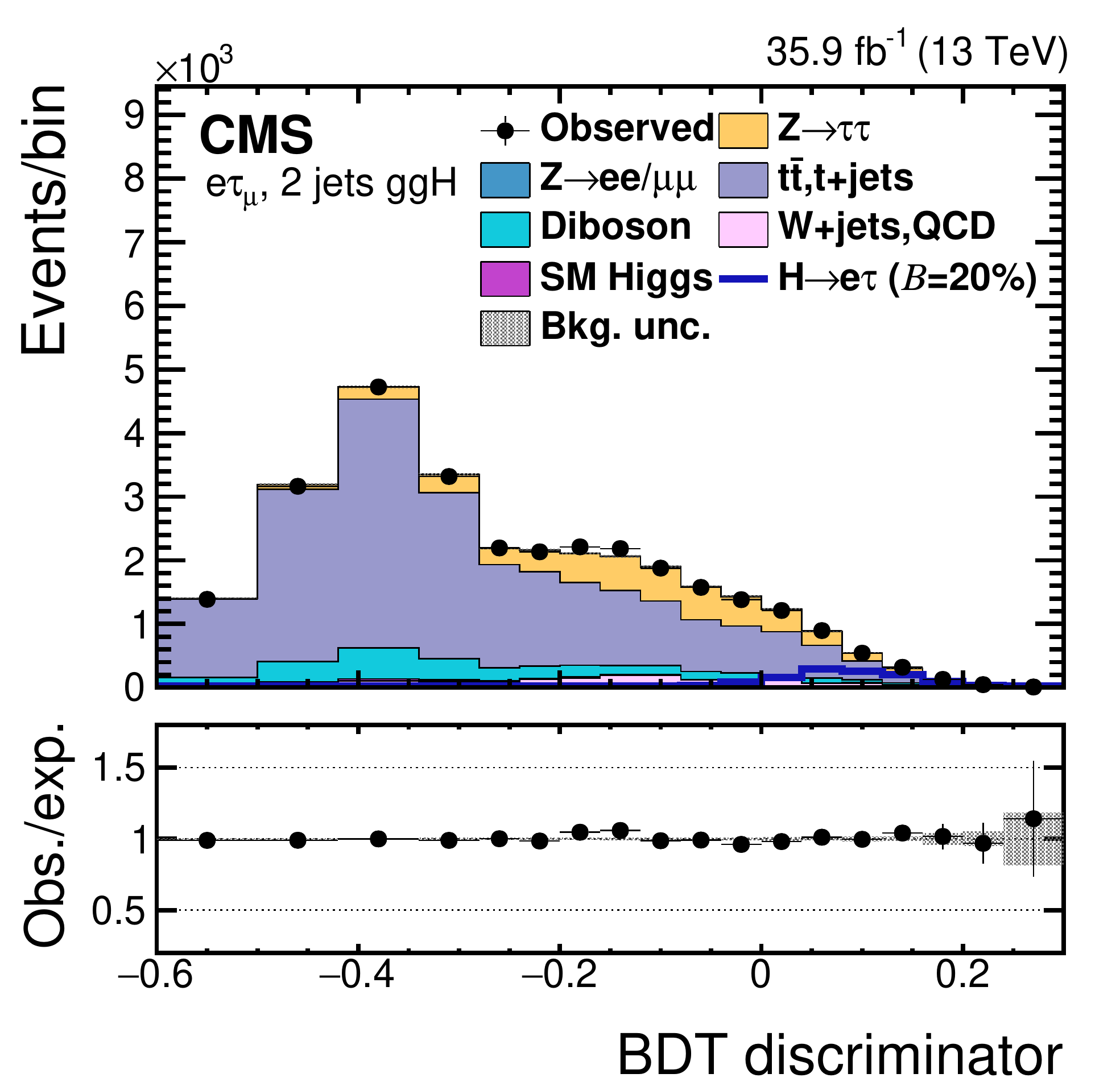} \\
 \includegraphics[width=0.33\textwidth]{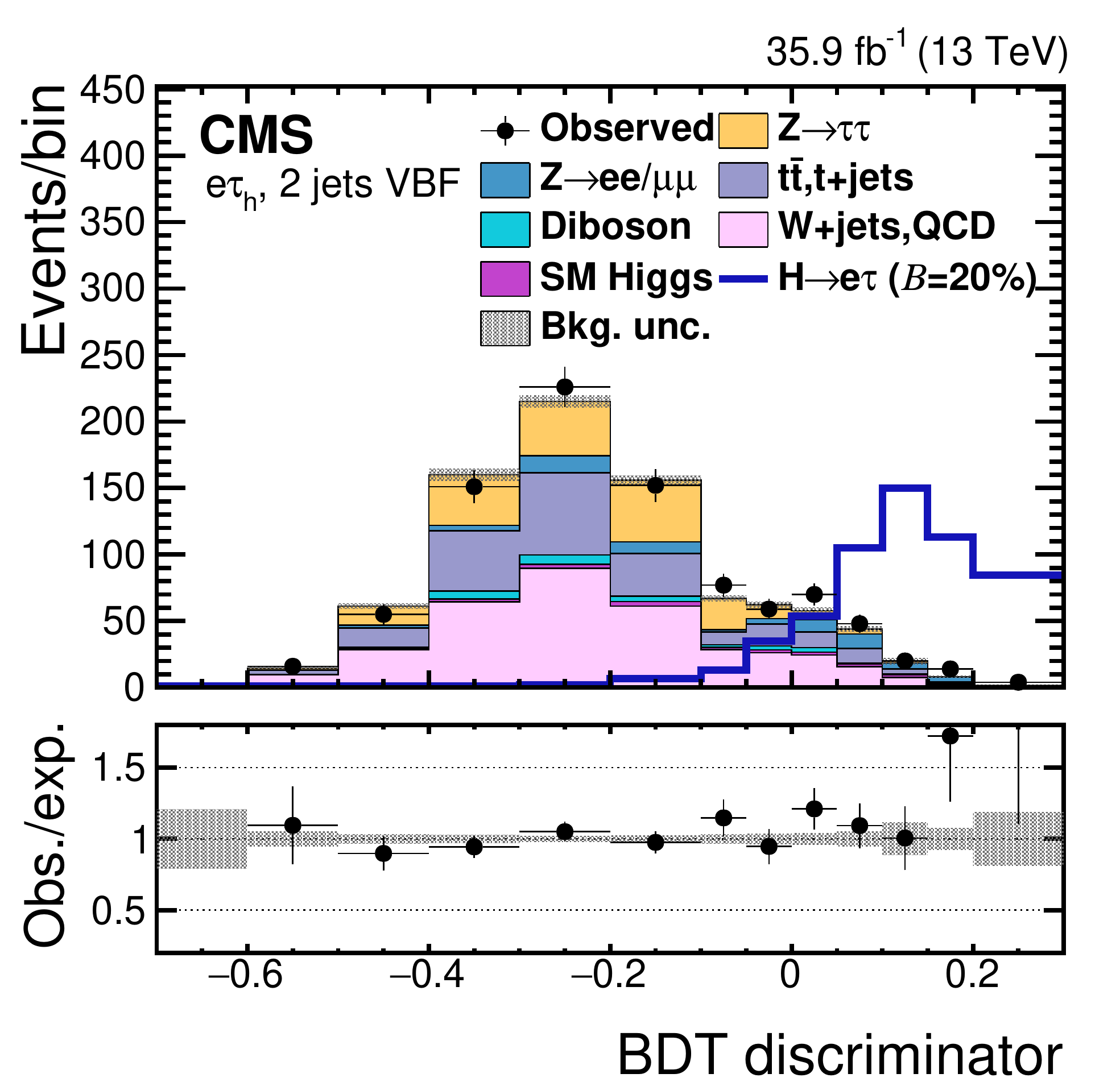}
 \includegraphics[width=0.33\textwidth]{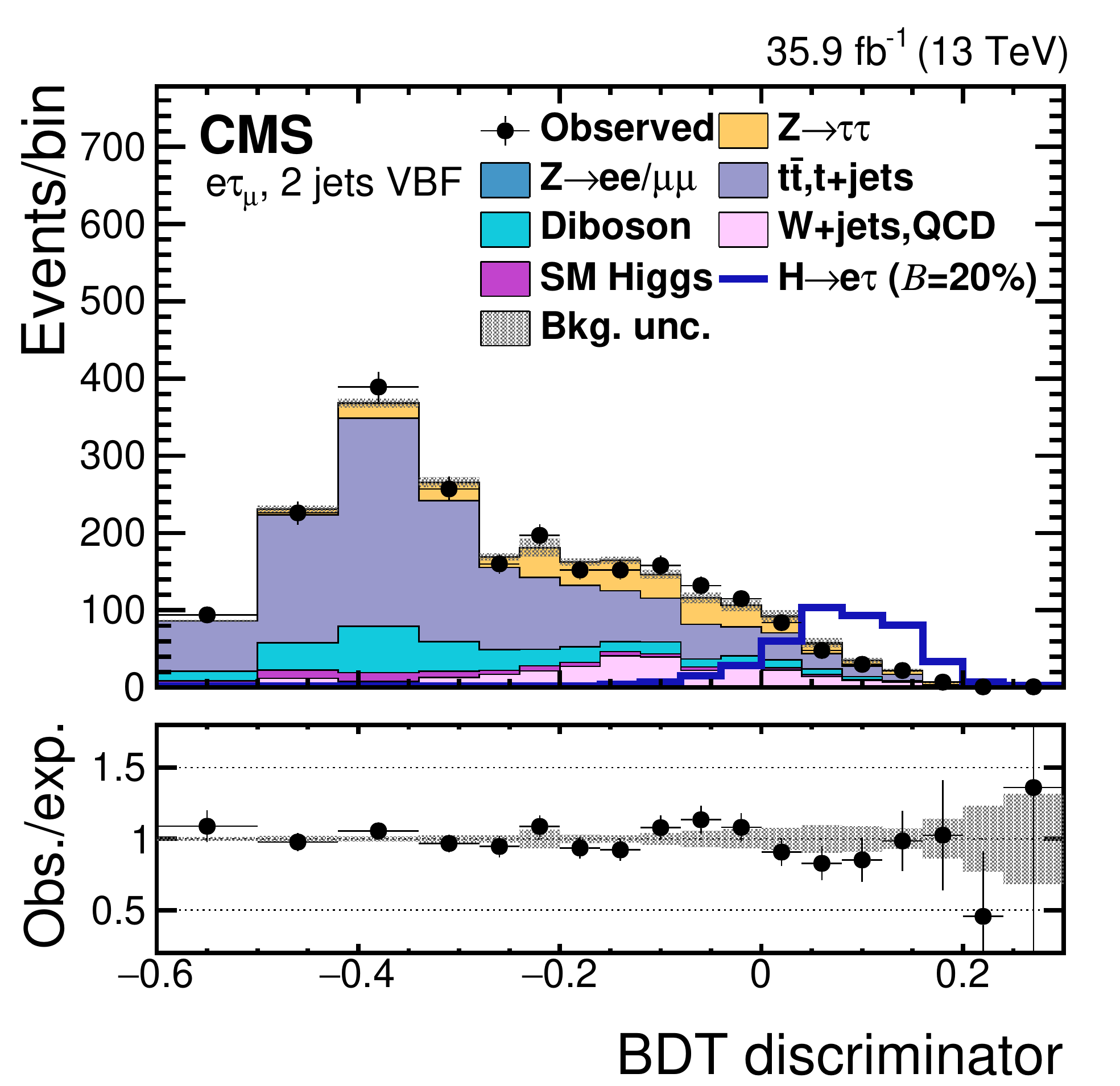}
\caption{Distribution of the BDT discriminator for the $\PH \to \Pe\Pgt$  process for the BDT fit analysis, in different channels and categories compared to the signal and background estimation.
The background is normalized to the best fit values from the signal plus background fit while the simulated signal corresponds to $\mathcal{B}(\PH \to \Pe \Pgt )=5\%$.
The bottom  panel in each plot shows the ratio  between the observed data and the fitted background.
The left column of plots corresponds to the $\PH \to \Pe \tauh$ categories, from 0-jets (first row) to 2-jets VBF (fourth row). The right one to their $\PH \to \Pe \Pgt_{\Pgm}$ counterparts.}
 \label{fig:Mcol_SignalRegion_BDTMethod2_ETau}
\end{figure}

\begin{figure}[!htpb]\centering
 \includegraphics[width=0.33\textwidth]{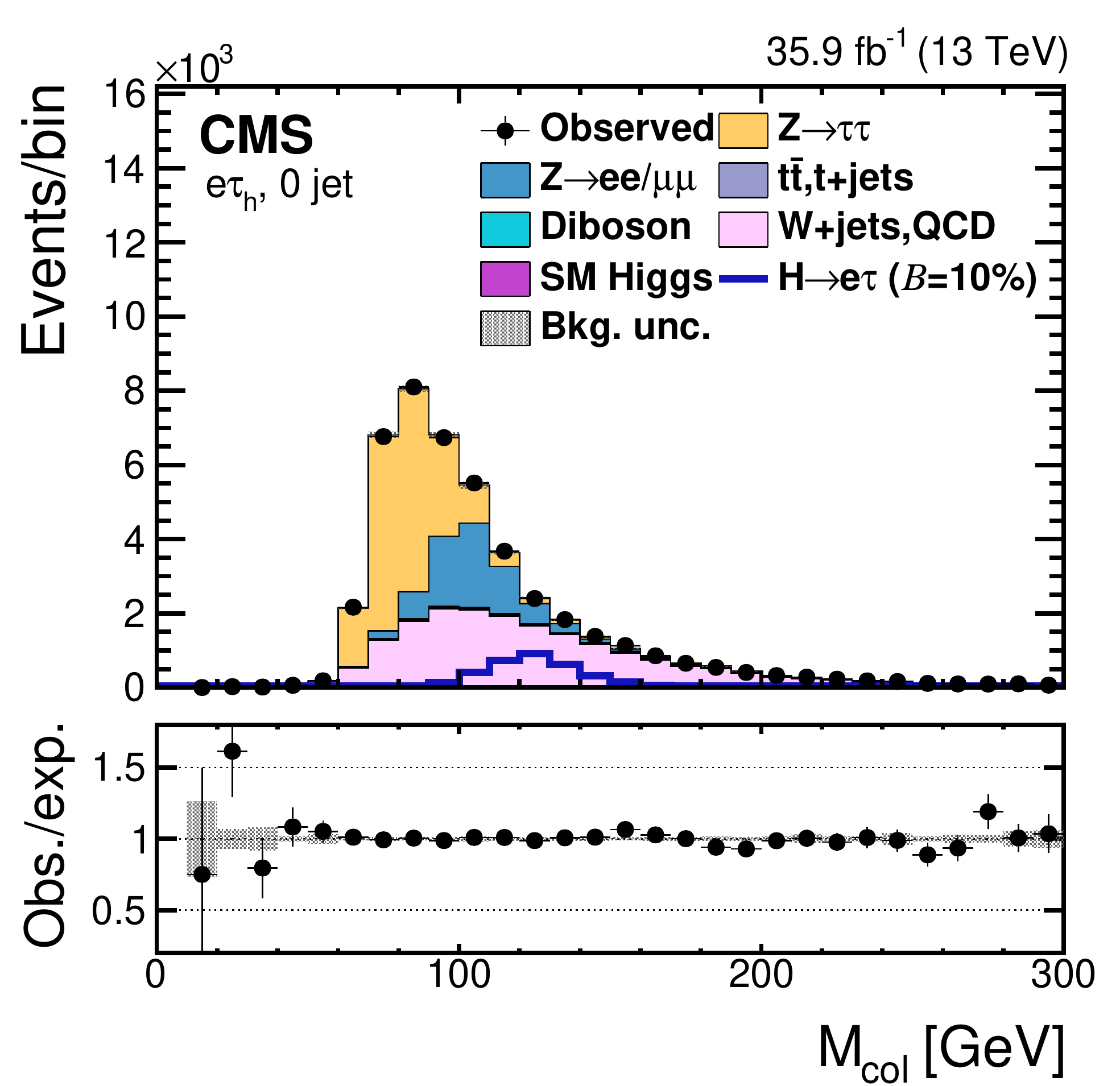}
 \includegraphics[width=0.33\textwidth]{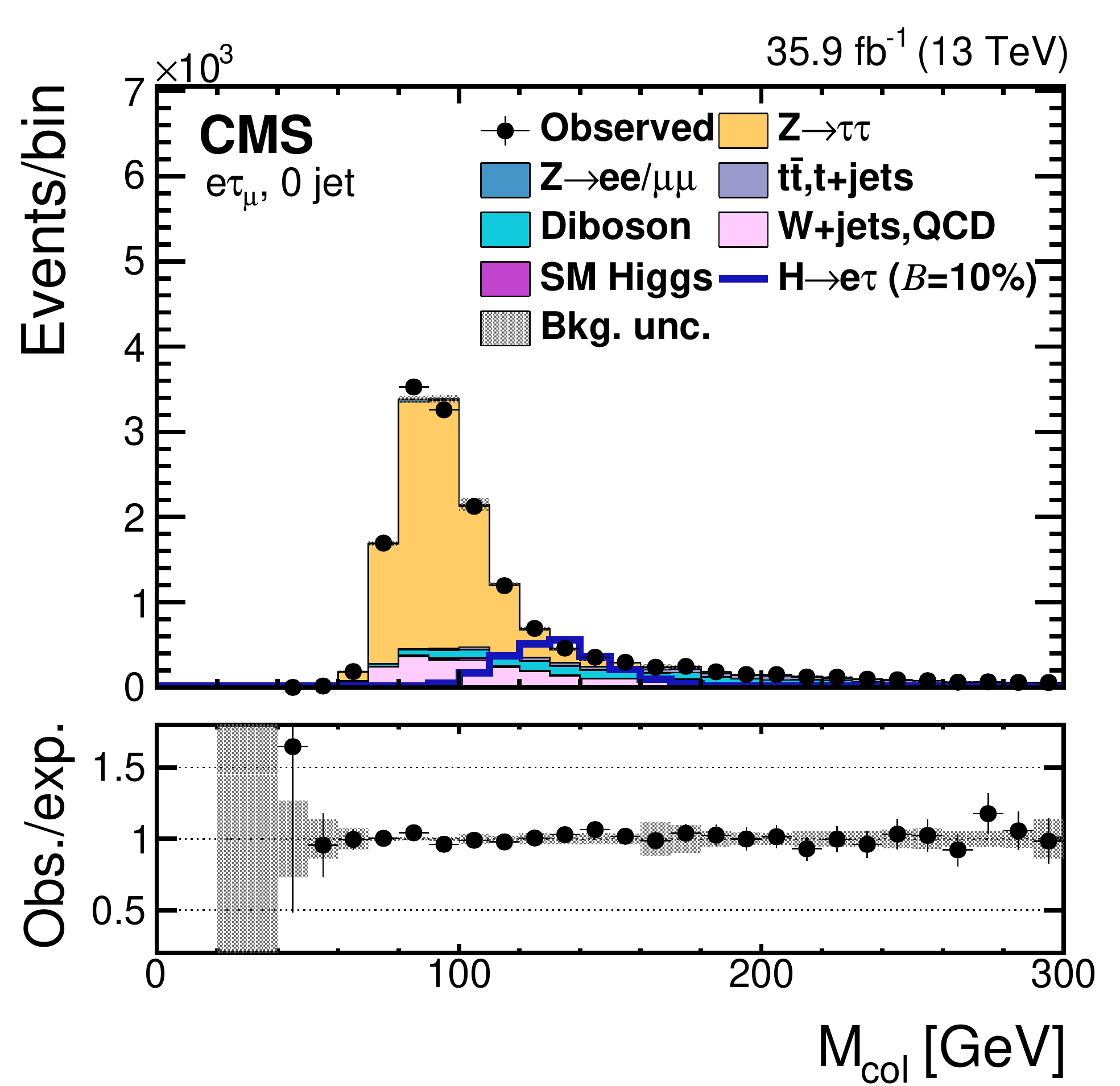} \\
 \includegraphics[width=0.33\textwidth]{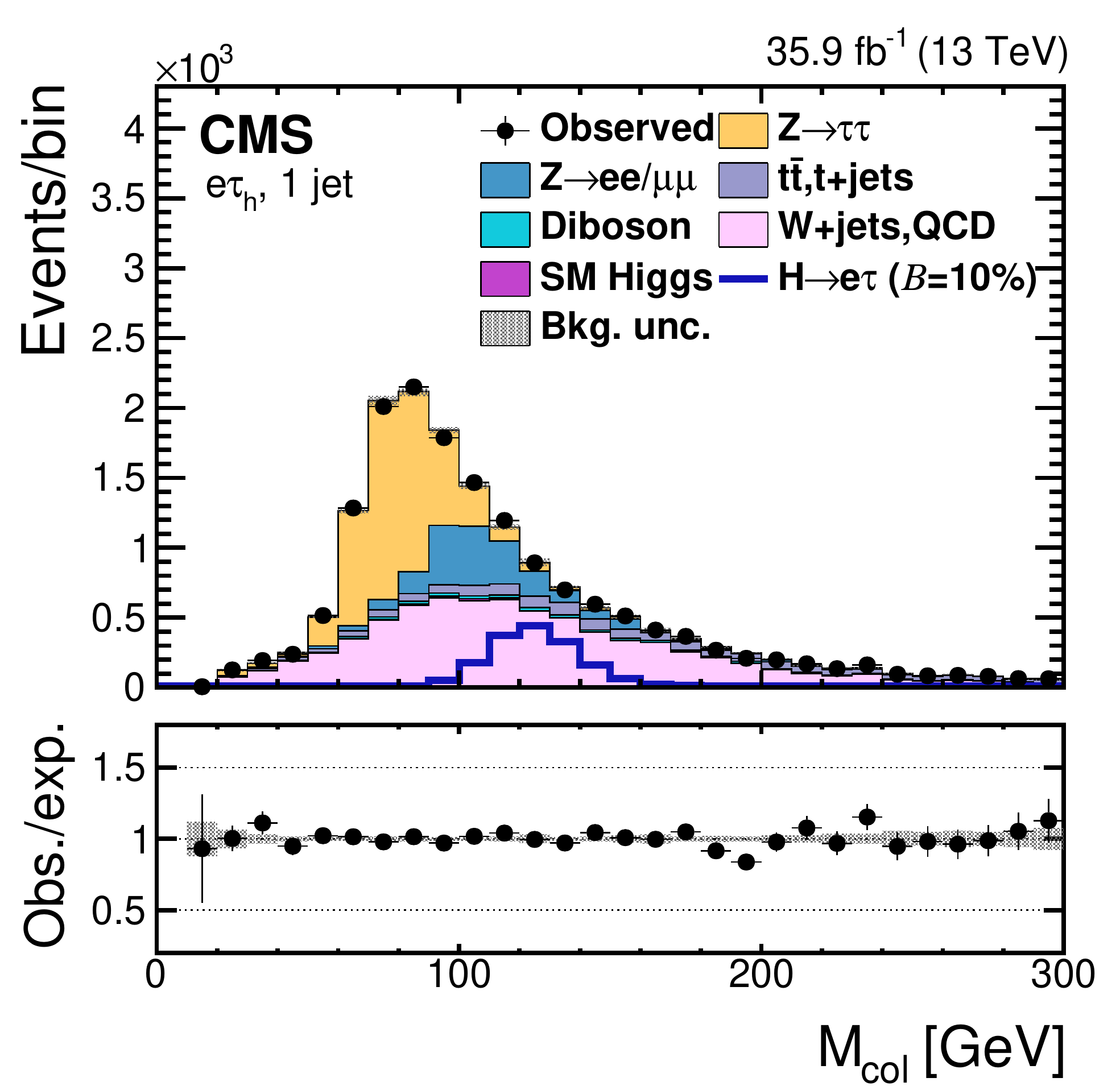}
 \includegraphics[width=0.33\textwidth]{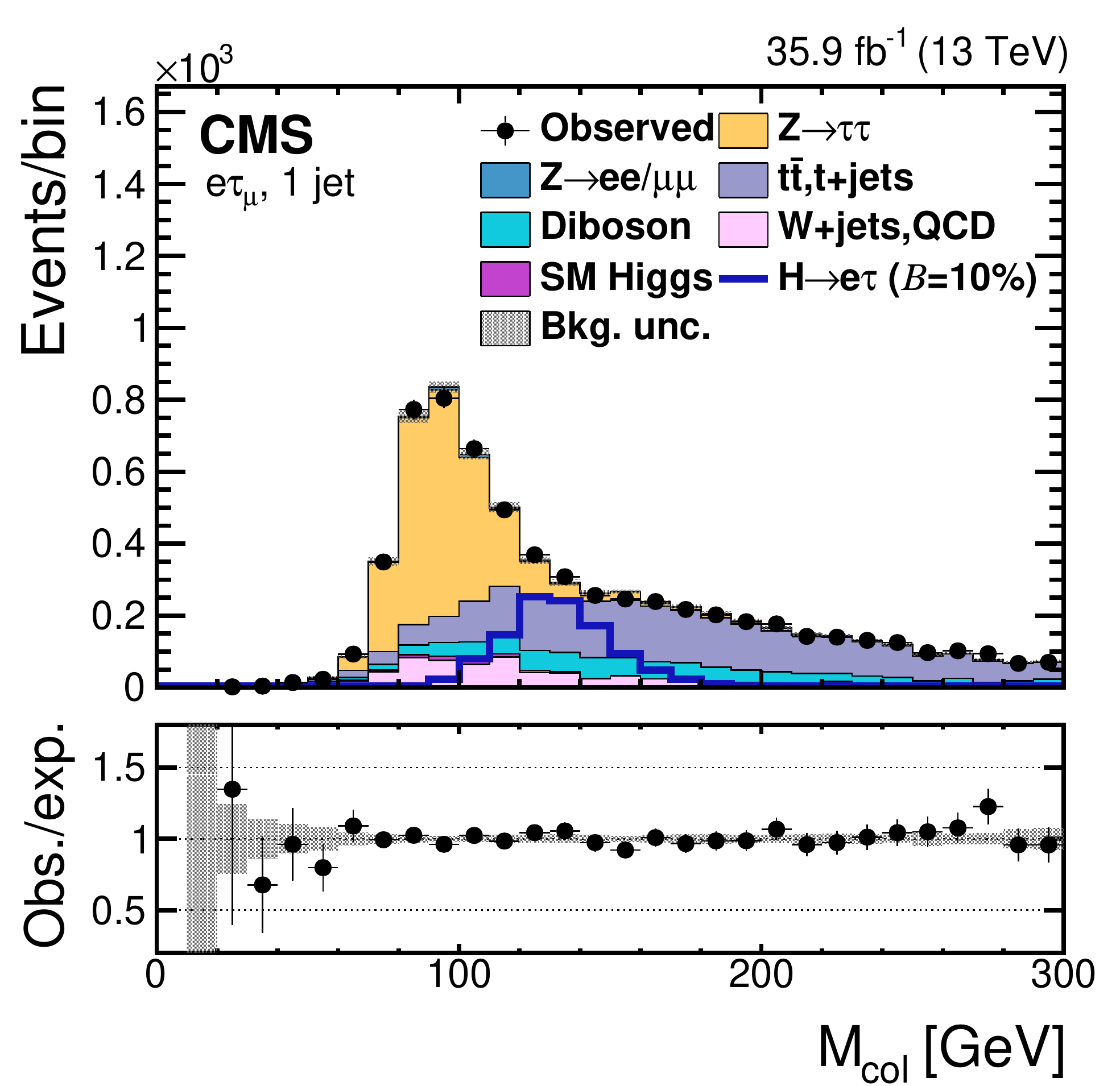}  \\
 \includegraphics[width=0.33\textwidth]{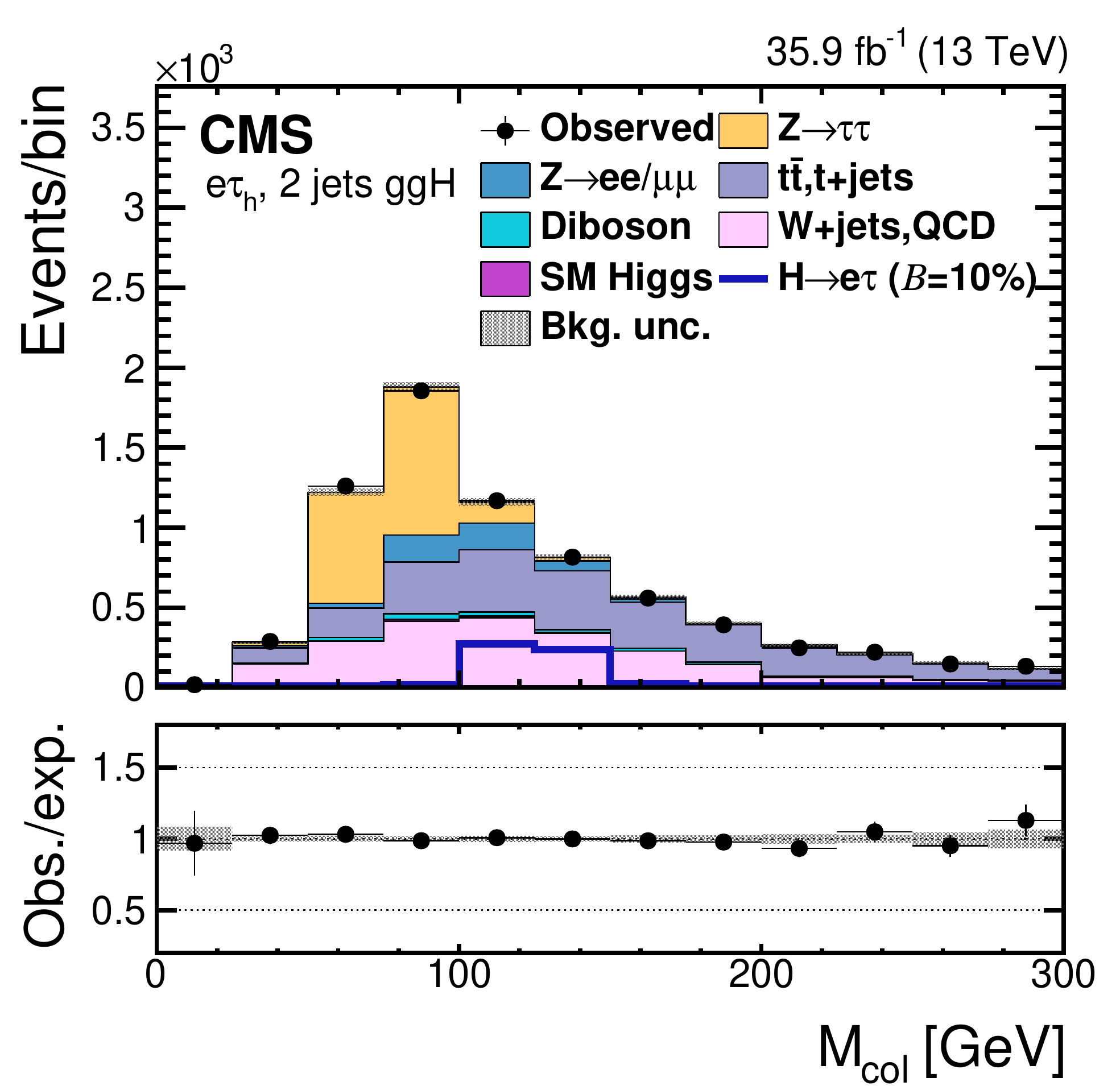}
 \includegraphics[width=0.33\textwidth]{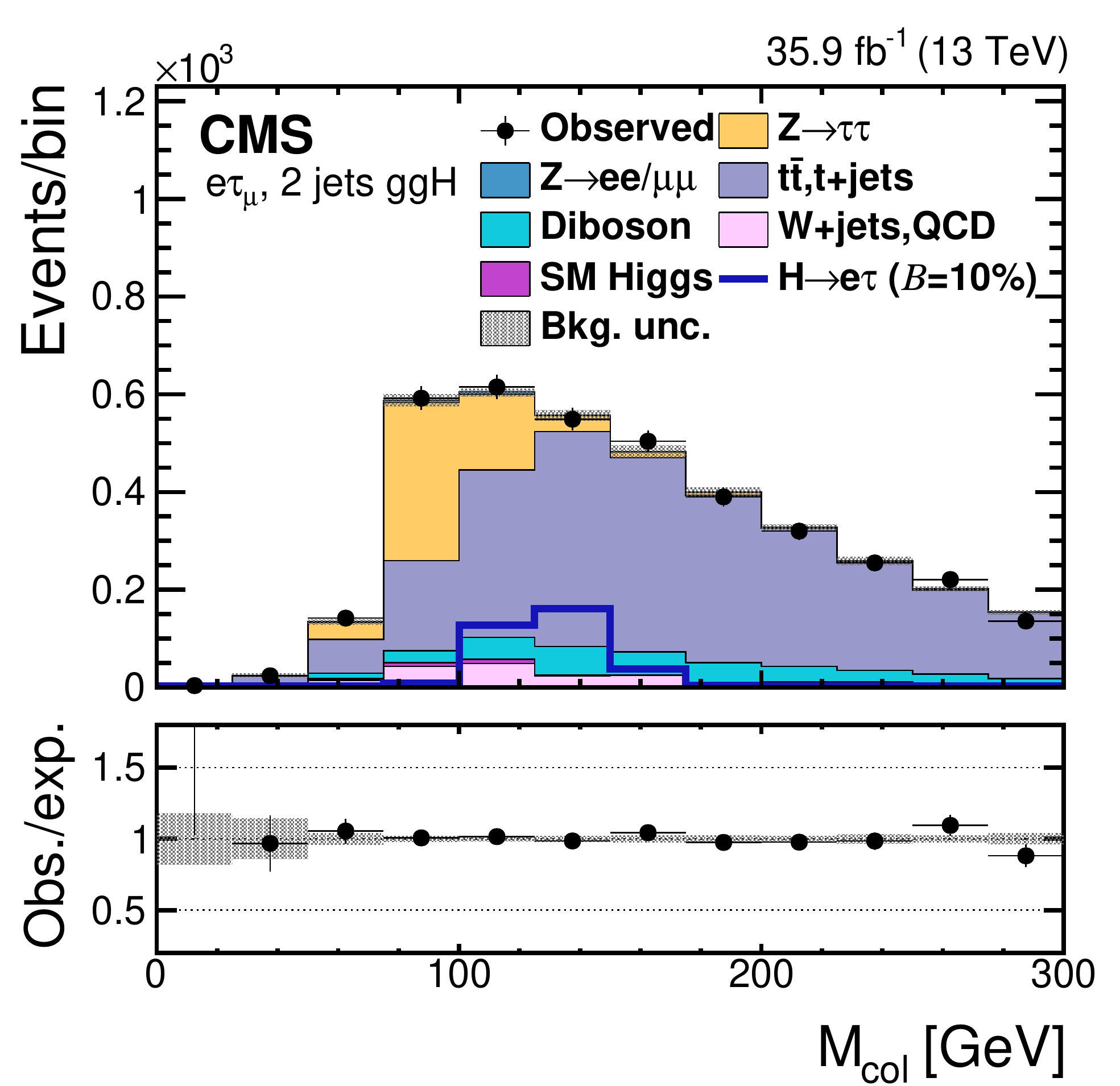} \\
 \includegraphics[width=0.33\textwidth]{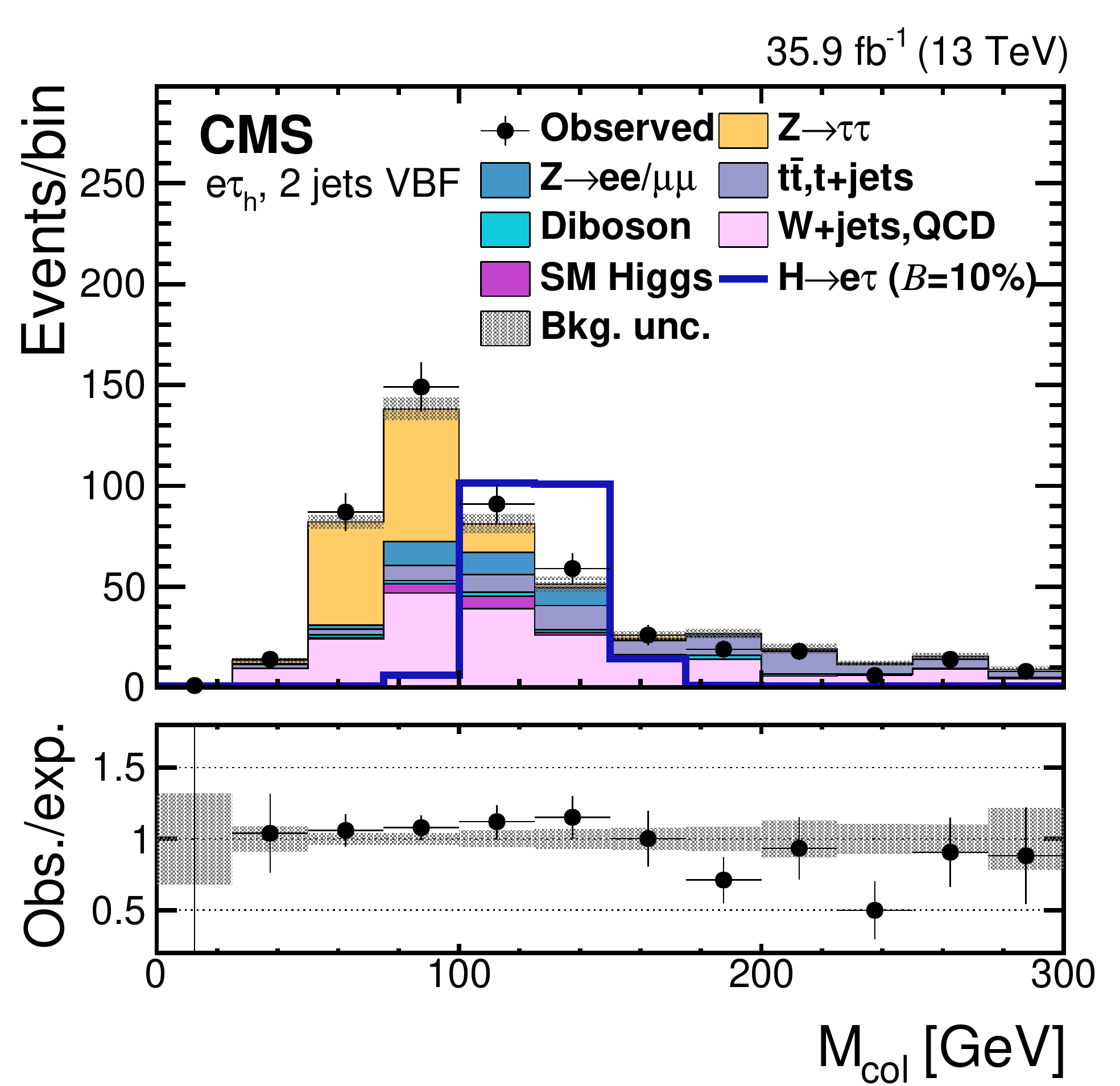}
 \includegraphics[width=0.33\textwidth]{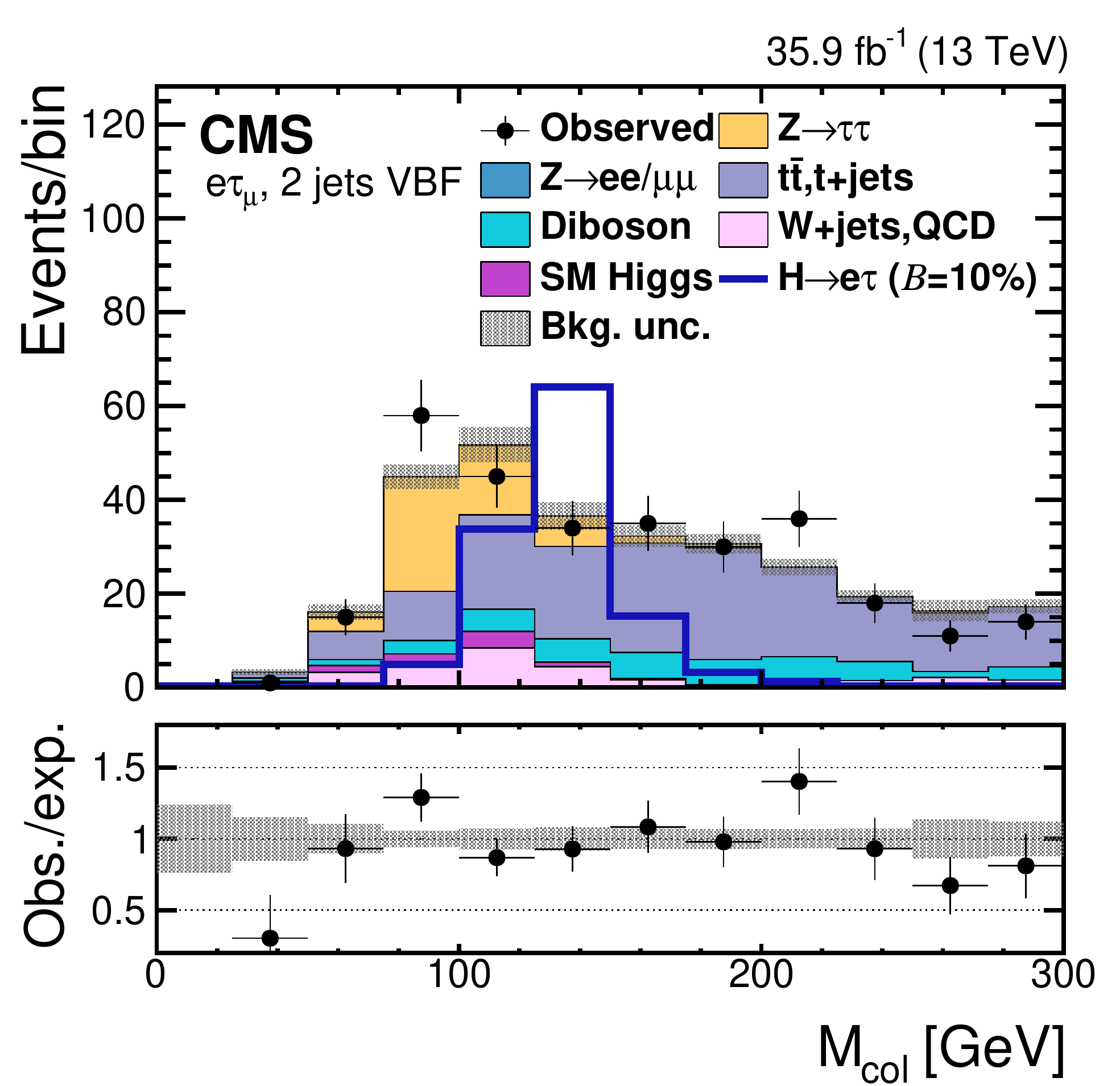}
\caption{Distribution of the collinear mass $M_\text{col}$ for the $\PH \to \Pe\Pgt$  process in the \mcol fit analysis, in  different channels and categories compared to the signal and background estimation.
The background is normalized to the best fit values from the signal plus background fit while the simulated signal
corresponds  to $\mathcal{B}(\PH \to \Pe \Pgt)=5\%$.
The lower panel in each plot shows the ratio  between the observed data and the fitted background.
The left column of plots correspond to the $\PH \to \Pe \tauh$ categories, from 0-jets (first row) to 2 jets VBF (fourth row). The right one to their $\PH \to \Pe \Pgt_{\Pgm}$ counterparts.}
 \label{fig:Mcol_SignalRegion_CutBased_ETau}
\end{figure}

\begin{figure}[!hbtp]\centering
\includegraphics[width=0.45\textwidth]{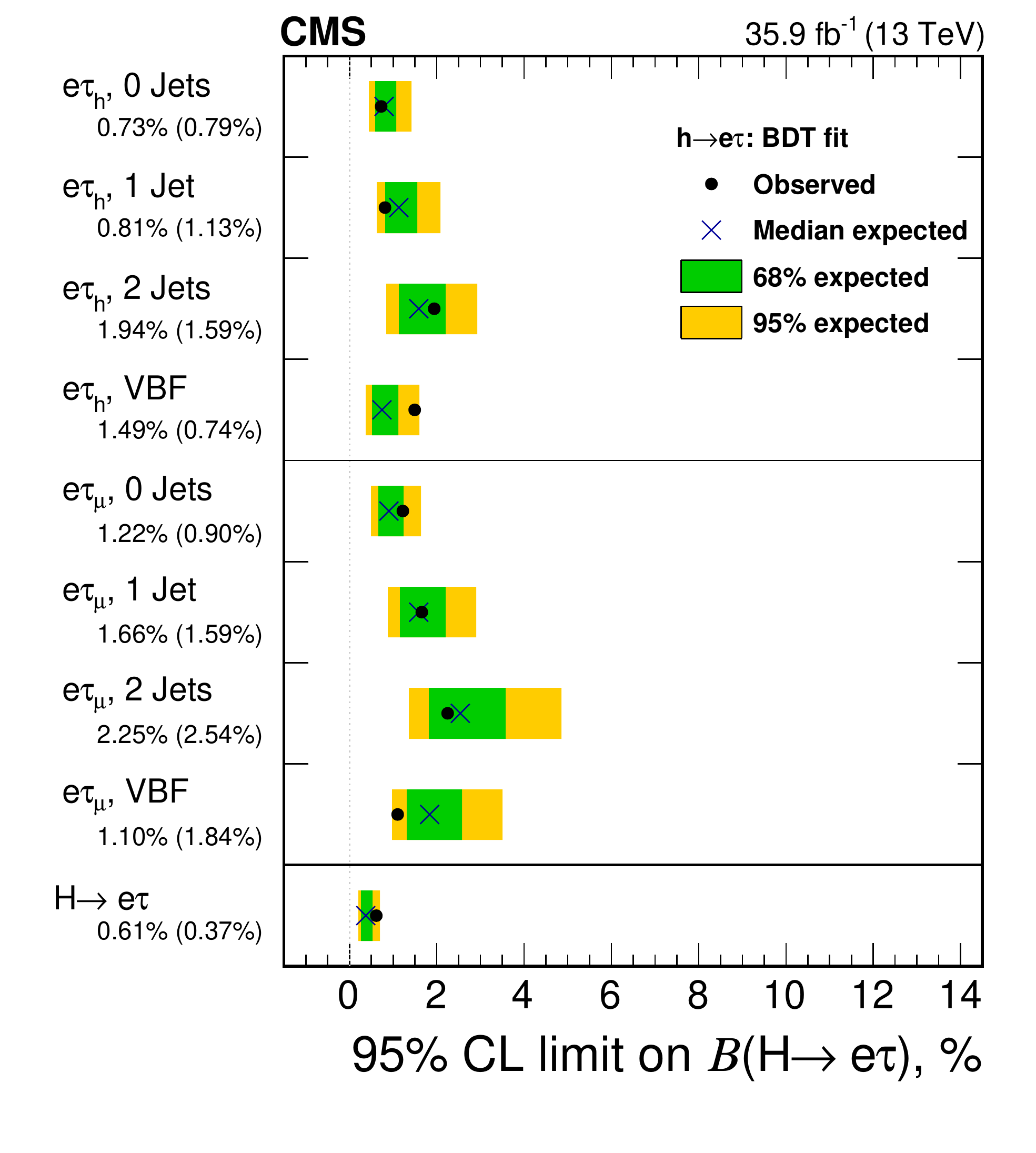}
\includegraphics[width=0.45\textwidth]{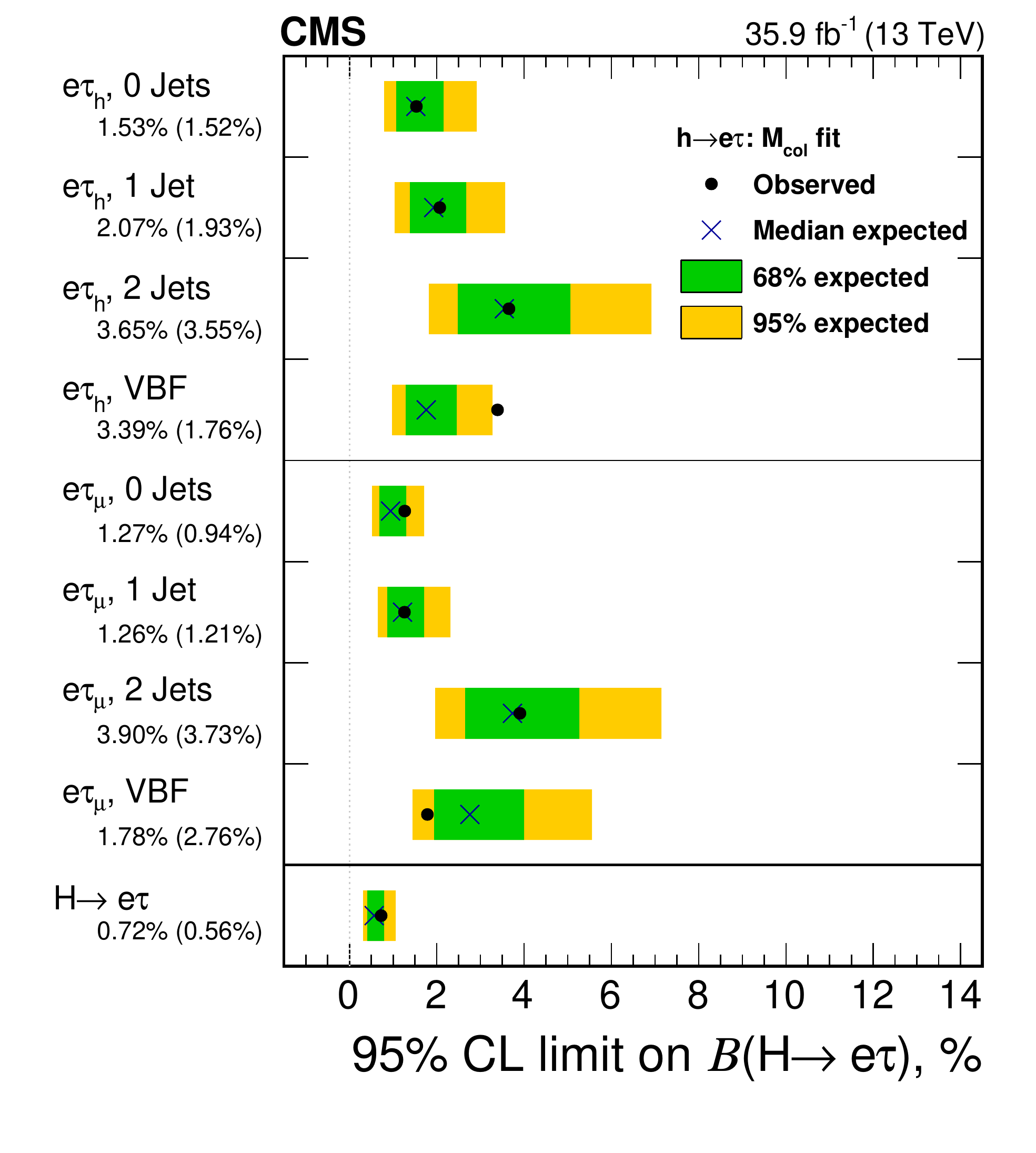}
 \caption{Observed and expected 95\% CL upper limits on the $\mathcal{B}(\PH \to \Pe \Pgt)$ for each individual category and combined. Left: BDT fit analysis. Right: \mcol fit analysis.}
 \label{fig:limits_summary_ETAU}\end{figure}

The BDT discriminator distributions of signal and background  for each category  are shown in Fig.~\ref{fig:Mcol_SignalRegion_BDTMethod2_MuTau} and~\ref{fig:Mcol_SignalRegion_BDTMethod2_ETau} in the $\PH \to \Pgm\Pgt$ and $\PH \to \Pe\Pgt$ channels respectively.
Figures~\ref{fig:Mcol_SignalRegion_CutBased_MuTau} and \ref{fig:Mcol_SignalRegion_CutBased_ETau} show the corresponding $\mcol$ distributions used as cross-check. All the distributions are shown after they have been adjusted  by the fit.
No excess over the background expectation is observed.
The observed and median expected $95\%$ CL upper limits, and best fit branching fractions, for $\mathcal{B}(\PH \to \Pgm \Pgt)$ and $\mathcal{B}(\PH \to \Pe \Pgt)$, assuming
$m_\PH$=125\GeV, are given for each category in Tables~\ref{tab:expected_limits_BDTMethod2_MuTau}-\ref{tab:expected_limits_CutBased_ETau}.
 The limits are also summarized graphically  in Figs.~\ref{fig:limits_summary_MUTAU} and ~\ref{fig:limits_summary_ETAU}.

\begin{table}[!htpb]
 \centering
  \topcaption{Expected and observed upper limits at 95\% CL, and best fit branching fractions in percent for each individual jet category, and combined,  in the $\PH \to \Pgm\Pgt$  process obtained with the BDT fit analysis.}
 \label{tab:expected_limits_BDTMethod2_MuTau}
\begin{tabular}{*{6}{c}}
\multicolumn{6}{c}{Expected limits~(\%) } \\ \hline
                       &  \multicolumn{1}{c}{0-jet}   & \multicolumn{1}{c}{1-jet}    &  \multicolumn{1}{c}{2-jets} & \multicolumn{1}{c}{VBF}  & \multicolumn{1}{c}{Combined}                 \\  \cline{2-6}
$\Pgm\Pgt_{\Pe}$ 	 &  $<$0.83   	 &  $<$1.19   	 &  $<$1.98   	 &  $<$1.62   	 &  $<$0.59    \\
$\Pgm\tauh$ 	 &  $<$0.43   	 &  $<$0.56   	 &  $<$0.94   	 &  $<$0.58   	 &  $<$0.29    \\
\cline{2-6}
 $\Pgm\Pgt$  & \multicolumn{5}{c}{ $<$0.25  }\\[\cmsTabSkip]
\multicolumn{6}{c}{Observed limits~(\%)} \\ \hline
                       &  \multicolumn{1}{c}{0-jet}   & \multicolumn{1}{c}{1-jet}    &  \multicolumn{1}{c}{2-jets} & \multicolumn{1}{c}{VBF} &\multicolumn{1}{c}{Combined}                 \\ \cline{2-6}
$\Pgm\Pgt_{\Pe}$   		 & $<$1.30   	 & $<$1.34   	 & $<$2.27   	 & $<$1.79   	 & $<$0.86    \\
$\Pgm\tauh$   		 & $<$0.51   	 & $<$0.53   	 & $<$0.56   	 & $<$0.51   	 & $<$0.27    \\
\cline{2-6}
  $\Pgm\Pgt$  & \multicolumn{5}{c}{ $<$0.25  } \\[\cmsTabSkip]
\multicolumn{6}{c}{Best fit branching fractions~(\%)} \\ \hline
                       &  \multicolumn{1}{c}{0-jet}   & \multicolumn{1}{c}{1-jet}    &  \multicolumn{1}{c}{2-jets} & \multicolumn{1}{c}{VBF} &\multicolumn{1}{c}{Combined}                 \\  \cline{2-6}
$\Pgm\Pgt_{\Pe}$    		 & 0.61 $\pm$ 0.36  	 & 0.22 $\pm$ 0.46  	 & 0.39 $\pm$ 0.83  	 & 0.10 $\pm$ 1.37  	 & 0.35 $\pm$ 0.26  \\
$\Pgm\tauh$    		 & 0.12 $\pm$ 0.20  	 & $-0.05$ $\pm$ 0.25  	 & $-0.72$ $\pm$ 0.43  	 & $-0.22$ $\pm$ 0.31  	 & $-0.04$ $\pm$ 0.14  \\
\cline{2-6}
 $\Pgm\Pgt$  & \multicolumn{5}{c}{ 0.00 $\pm$ 0.12 } \\ \hline
  \end{tabular}
\end{table}

\begin{table}[!htpb]
 \centering
  \topcaption{Expected and observed upper limits at 95\% CL, and best fit branching fractions in percent for each individual jet category, and combined, in the $\PH \to \Pgm\Pgt$  process obtained with  the  $\mcol$ fit analysis.}
 \label{tab:expected_limits_CutBased_MuTau}
\begin{tabular}{*{6}{c}}
\multicolumn{6}{c}{Expected limits~(\%) } \\ \hline
                       &  \multicolumn{1}{c}{0-jet}   & \multicolumn{1}{c}{1-jet}    &  \multicolumn{1}{c}{2-jets} & \multicolumn{1}{c}{VBF}  & \multicolumn{1}{c}{Combined}                 \\  \cline{2-6}
$\Pgm\Pgt_{\Pe}$ 	 &  $<$1.01   	 &  $<$1.47   	 &  $<$3.23   	 &  $<$1.73   	 &  $<$0.75    \\
$\Pgm\tauh$ 	 &  $<$1.14   	 &  $<$1.26   	 &  $<$2.12   	 &  $<$1.41   	 &  $<$0.71    \\
\cline{2-6}
 $\Pgm\Pgt$  & \multicolumn{5}{c}{ $<$0.49  } \\[\cmsTabSkip]

\multicolumn{6}{c}{Observed limits~(\%)} \\ \hline
                       &  \multicolumn{1}{c}{0-jet}   & \multicolumn{1}{c}{1-jet}    &  \multicolumn{1}{c}{2-jets} & \multicolumn{1}{c}{VBF}  & \multicolumn{1}{c}{Combined}                 \\  \cline{2-6}
$\Pgm\Pgt_\Pe$   		 & $<$1.08   	 & $<$1.35   	 & $<$3.33   	 & $<$1.40   	 & $<$0.71    \\
$\Pgm\tauh$   		 & $<$1.04   	 & $<$1.74   	 & $<$1.65   	 & $<$1.30   	 & $<$0.66    \\
\cline{2-6}
  $\Pgm\Pgt$  & \multicolumn{5}{c}{ $<$0.51  } \\[\cmsTabSkip]
\multicolumn{6}{c}{Best fit branching fractions~(\%)} \\ \hline
                       &  \multicolumn{1}{c}{0-jet}   & \multicolumn{1}{c}{1-jet}    &  \multicolumn{1}{c}{2-jets} & \multicolumn{1}{c}{VBF} &\multicolumn{1}{c}{Combined}                 \\  \cline{2-6}
$\Pgm\Pgt_\Pe$    		 & 0.13 $\pm$ 0.43  	 & $-0.22$ $\pm$ 0.75  	 & 0.22 $\pm$ 1.39  	 & $-1.73$ $\pm$ 1.05  	 & $-0.04$ $\pm$ 0.33  \\
$\Pgm\tauh$    		 & $-0.30$ $\pm$ 0.45  	 & 0.68 $\pm$ 0.56  	 & $-1.23$ $\pm$ 1.04  	 & $-0.23$ $\pm$ 0.66  	 & $-0.08$ $\pm$ 0.34  \\
\cline{2-6}
 $\Pgm\Pgt$  & \multicolumn{5}{c}{ 0.02 $\pm$ 0.20 } \\ \hline
\end{tabular}
\end{table}

\begin{table}[!hbtp]
 \centering
  \topcaption{Expected and observed upper limits at 95\% CL and best fit branching fractions in percent for each individual jet category, and combined, in the $\PH \to \Pe\Pgt$  process obtained  with the BDT fit analysis.}
 \label{tab:expected_limits_BDTMethod2_ETau}
\begin{tabular}{*{6}{c}}
\multicolumn{6}{c}{Expected limits~(\%) } \\ \hline
                       &  \multicolumn{1}{c}{0-jet}   & \multicolumn{1}{c}{1-jet}    &  \multicolumn{1}{c}{2-jets} & \multicolumn{1}{c}{VBF}  & \multicolumn{1}{c}{Combined}                 \\ \cline{2-6}
$\Pe\Pgt_{\Pgm}$   	 & $<$0.90  	 & $<$1.59  	 & $<$2.54  	 & $<$1.84  	 & $<$0.64   \\
$\Pe\tauh$   	 & $<$0.79  	 & $<$1.13  	 & $<$1.59  	 & $<$0.74  	 & $<$0.49   \\
\cline{2-6}
 $\Pe\Pgt$  & \multicolumn{5}{c}{ $<$0.37  } \\[\cmsTabSkip]
\multicolumn{6}{c}{Observed limits~(\%)  } \\ \hline
                       &  \multicolumn{1}{c}{0-jet}   & \multicolumn{1}{c}{1-jet}    &  \multicolumn{1}{c}{2-jets} & \multicolumn{1}{c}{VBF}  & \multicolumn{1}{c}{Combined}                 \\  \cline{2-6}
$\Pe\Pgt_{\Pgm}$   		 & $<$1.22   	 & $<$1.66   	 & $<$2.25   	 & $<$1.10   	 & $<$0.78    \\
$\Pe\tauh$   		 & $<$0.73   	 & $<$0.81   	 & $<$1.94   	 & $<$1.49   	 & $<$0.72    \\
\cline{2-6}
  $\Pe\Pgt$  & \multicolumn{5}{c}{ $<$0.61  } \\[\cmsTabSkip]
\multicolumn{6}{c}{ Best fit branching fractions~(\%)} \\ \hline
               &  \multicolumn{1}{c}{0-jet}   & \multicolumn{1}{c}{1-jet}    &  \multicolumn{1}{c}{2-jets} & \multicolumn{1}{c}{VBF} &\multicolumn{1}{c}{Combined}                 \\  \cline{2-6}
$\Pe\Pgt_{\Pgm}$    		 & 0.47 $\pm$ 0.42  	 & 0.17 $\pm$ 0.79  	 & $-0.42$ $\pm$ 1.01  	 & $-1.54$ $\pm$ 0.44  	 & 0.18 $\pm$ 0.32  \\
$\Pe\tauh$    		 & $-0.13$ $\pm$ 0.39  	 & $-0.63$ $\pm$ 0.40  	 & 0.54 $\pm$ 0.53  	 & 0.70 $\pm$ 0.38  	 & 0.33 $\pm$ 0.24  \\
\cline{2-6}
 $\Pe\Pgt$  & \multicolumn{5}{c}{ 0.30 $\pm$ 0.18 } \\ \hline
\end{tabular}
\end{table}

\begin{table}[!hbtp]
 \centering
  \topcaption{Expected and observed upper limits at 95\% CL and best fit branching fractions in percent for  each individual jet category, and combined,  in the $\PH \to \Pe\Pgt$  process obtained with the $\mcol$ fit analysis.}
 \label{tab:expected_limits_CutBased_ETau}
\begin{tabular}{*{6}{c}}
\multicolumn{6}{c}{Expected limits~(\%)  } \\ \hline
                       &  \multicolumn{1}{c}{0-jet}   & \multicolumn{1}{c}{1-jet}    &  \multicolumn{1}{c}{2-jets} & \multicolumn{1}{c}{VBF}  & \multicolumn{1}{c}{Combined}                 \\  \cline{2-6}
$\Pe\Pgt_{\Pgm}$     & $<$0.94    & $<$1.21    & $<$3.73    & $<$2.76    & $<$0.71   \\
$\Pe\tauh$     & $<$1.52    & $<$1.93    & $<$3.55    & $<$1.76    & $<$0.97   \\
\cline{2-6}
 $\Pe\Pgt$  & \multicolumn{5}{c}{ $<$0.56  } \\[\cmsTabSkip]
\multicolumn{6}{c}{Observed limits~(\%) } \\ \hline
                       &  \multicolumn{1}{c}{0-jet}   & \multicolumn{1}{c}{1-jet}    &  \multicolumn{1}{c}{2-jets} & \multicolumn{1}{c}{VBF}  & \multicolumn{1}{c}{Combined}                 \\  \cline{2-6}
$\Pe\Pgt_{\Pgm}$       & $<$1.27      & $<$1.26      & $<$3.90      & $<$1.78      & $<$0.85    \\
$\Pe\tauh$       & $<$1.53      & $<$2.07      & $<$3.65      & $<$3.39      & $<$1.31    \\
\cline{2-6}
  $\Pe\Pgt$  & \multicolumn{5}{c}{ $<$0.72  } \\[\cmsTabSkip]
\multicolumn{6}{c}{ Best fit branching fractions~(\%) } \\ \hline
                       &  \multicolumn{1}{c}{0-jet}   & \multicolumn{1}{c}{1-jet}    &  \multicolumn{1}{c}{2-jets} & \multicolumn{1}{c}{VBF} &\multicolumn{1}{c}{Combined}                 \\ \cline{2-6}
$\Pe\Pgt_{\Pgm}$    		 & 0.46 $\pm$ 0.43  	 & 0.07 $\pm$ 0.39  	 & 0.13 $\pm$ 1.13  	 & $-1.38$ $\pm$ 1.03  	 & 0.21 $\pm$ 0.36  \\
$\Pe\tauh$    		 & 0.18 $\pm$ 0.35  	 & 0.45 $\pm$ 0.60  	 & 0.29 $\pm$ 1.13  	 & 2.03 $\pm$ 0.47  	 & 0.51 $\pm$ 0.41  \\
\cline{2-6}
 $\Pe\Pgt$  & \multicolumn{5}{c}{ 0.23 $\pm$ 0.24 } \\ \hline
  \end{tabular}
\end{table}

No evidence is found for either the $\PH \to \Pgm \Pgt$ or  $\PH \to \Pe \Pgt$ processes in this search.
The observed exclusion limits are a significant improvement over the 8\TeV results.
The new results exclude the branching fraction that corresponded to the best fit for the 2.4 $\sigma$ excess observed in the 8\TeV $\PH \to \Pgm \Pgt$ channel results at 95\% CL, in both the \mcol fit
and BDT fit analysis.
Table~\ref{tab:SummaryOfResults} shows a summary of  the new 95$\%$ CL upper limits.
The BDT fit analysis is  more sensitive than the \mcol fit analysis, with expected limits reduced
by about a factor of two. In both cases the results are dominated by the systematic uncertainties.

\begin{table}[!hbtp]
 \centering
  \topcaption{Summary of the observed and expected upper limits at the 95\% CL and the best fit branching fractions in percent for the $\PH \to \Pgm\Pgt$  and $\PH \to \Pe\Pgt$ processes, for the main analysis  (BDT fit) and the cross check (\mcol fit) method.}
 \label{tab:SummaryOfResults}
\begin{tabular}{c  cc  cc }
   \hline
    & \multicolumn{2}{c}{ Observed (expected) limits~(\%)} & \multicolumn{2}{c}{Best fit branching fraction~(\%) } \\ \hline
                       &  BDT fit &  \mcol fit & BDT fit & \mcol fit \\ \hline
$\PH\to \Pgm \Pgt$              & $<$0.25 (0.25)\%   &   $<$0.51 (0.49) \% & $0.00\pm0.12$ \% &   $0.02\pm0.20$  \%     \\
$\PH\to \Pe \Pgt$        & $<$0.61 (0.37) \%  &   $<$0.72 (0.56) \% & $0.30\pm0.18$ \% &   $0.23\pm0.24$  \%     \\ \hline
  \end{tabular}
\end{table}

\begin{figure}[htb]
\centering
\includegraphics[width=0.45\textwidth]{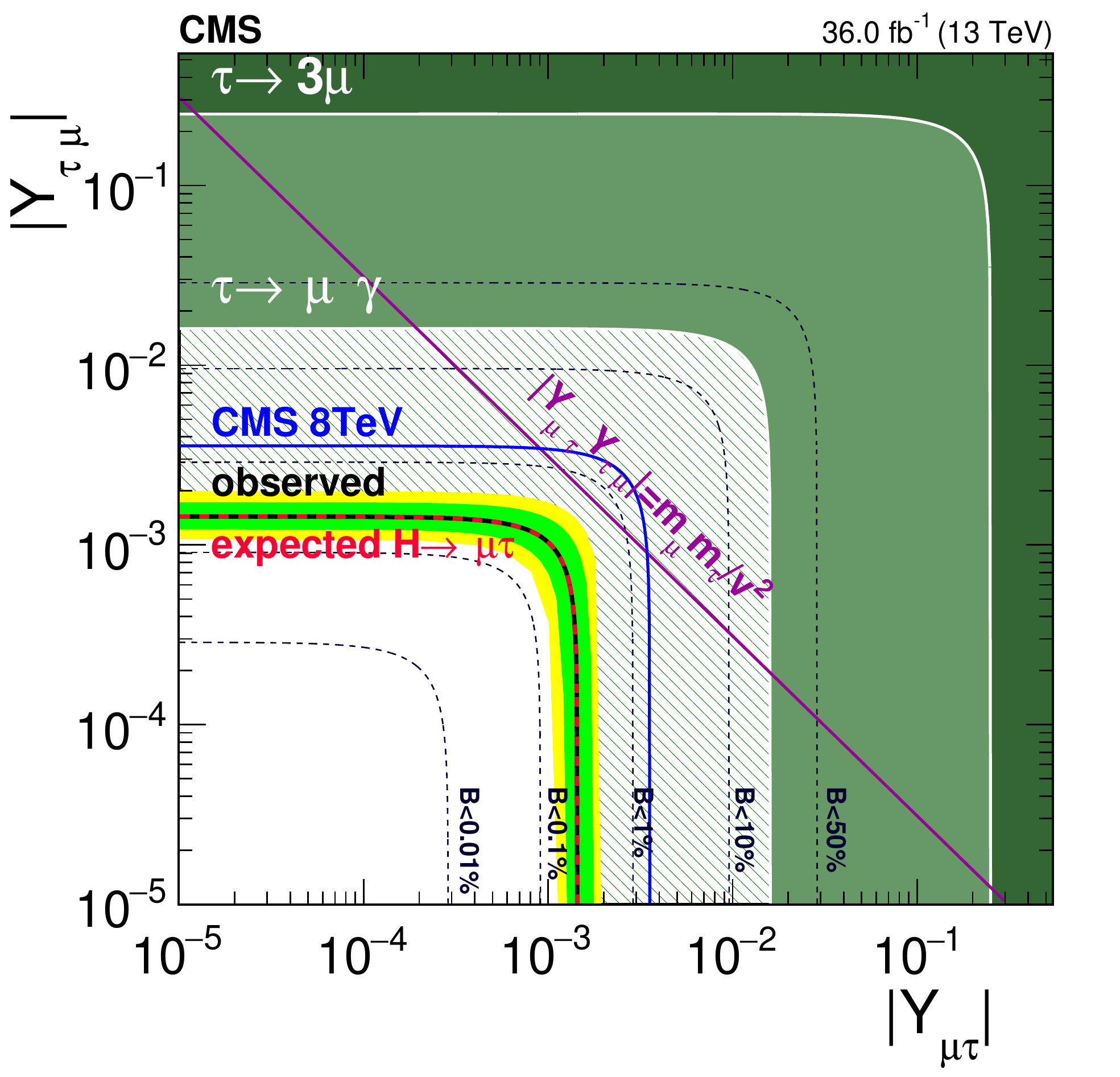}
\includegraphics[width=0.45\textwidth]{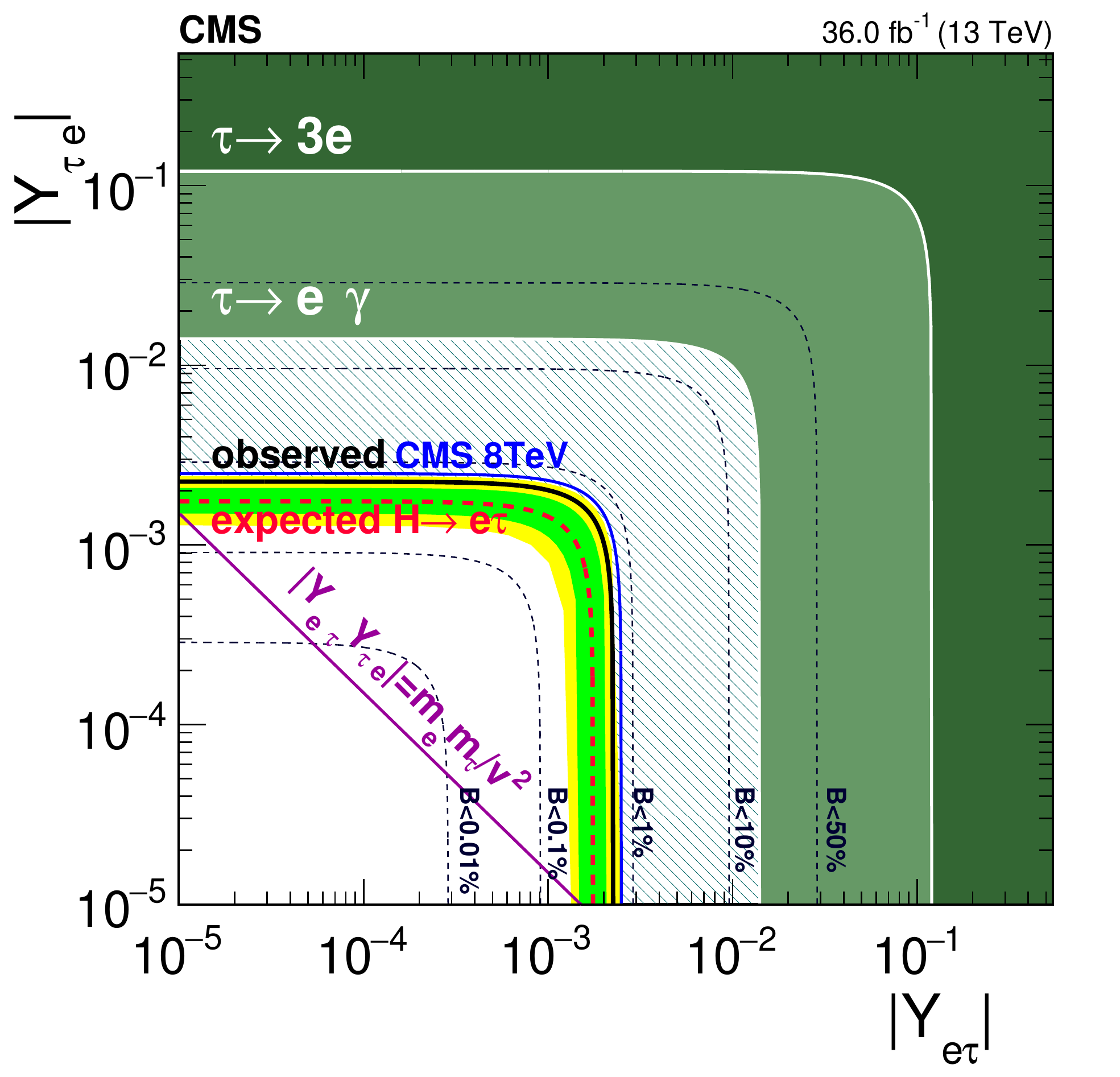}
\caption{Constraints on the flavour violating Yukawa couplings, $|Y_{\Pgm\Pgt}|, |Y_{\Pgt\Pgm}|$ (left) and $|Y_{\Pe\Pgt}|, |Y_{\Pgt\Pe}|$ (right), from the BDT result. The expected (red dashed line) and observed (black solid line) limits are derived from the limit on $\mathcal{B}(\PH \to \Pgm \Pgt)$ and  $\mathcal{B}(\PH \to \Pe \Pgt)$ from the present analysis. The flavour-diagonal Yukawa  couplings are approximated by their SM values. The green (yellow) band indicates the range that is expected to contain 68\% (95\%) of all observed limit excursions from the expected limit. The shaded regions are derived constraints from null searches for $\Pgt \to 3\Pgm$ or $\Pgt \to 3\Pe$ (dark green)~\cite{Hayasaka:2010np,Olive:2016xmw,Harnik:2012pb} and $\Pgt \to \Pgm \gamma$ or $\Pgt \to \Pe \gamma$ (lighter green)~\cite{Olive:2016xmw,Harnik:2012pb}. The green hashed region is derived by the CMS direct search presented in this paper. The blue solid lines are the CMS limits from~\cite{Khachatryan:2015kon} (left) and~\cite{HIG-14-040}(right).  The purple diagonal line is the theoretical naturalness  limit $|Y_{ij}Y_{ji}| \leq m_im_j/v^2$~\cite{Harnik:2012pb}.}
\label{fig:Yukawas}
\end{figure}

The constraints on $\mathcal{B}(\PH \to \Pgm \Pgt)$ and $\mathcal{B}(\PH \to \Pe \Pgt)$ can be interpreted in terms of LFV  Yukawa couplings~\cite{Harnik:2012pb}.
The LFV decays  $\Pe\Pgt$ and $\Pgm\Pgt$ arise at tree level from the assumed
flavour violating Yukawa interactions, $Y_{\ell^{\alpha}\ell^{\beta}}$ where $\ell^{\alpha},\ell^{\beta}$ denote the leptons, $\ell^{\alpha},\ell^{\beta}=\Pe, \Pgm, \Pgt$ and $\ell^{\alpha}\neq \ell^{\beta}$.
The decay width $\Gamma(\PH \to \ell^{\alpha}\ell^{\beta})$  in terms of the Yukawa couplings is given by:
\begin{equation*}
\Gamma(\PH \to \ell^{\alpha}\ell^{\beta})=\frac{m_{\PH}}{8\pi}\bigl(\abs{Y_{\ell^{\beta}\ell^{\alpha}}}^2 + \abs{Y_{\ell^{\alpha}\ell^{\beta}}}^2\bigr),
\end{equation*}
and the branching fraction by:
\begin{equation*}
\mathcal{B}(\PH \to \ell^{\alpha}\ell^{\beta})=\frac{\Gamma(\PH\to \ell^{\alpha}\ell^{\beta})}{\Gamma(\PH\to \ell^{\alpha}\ell^{\beta}) + \Gamma_{\mathrm{SM}}}.
\end{equation*}
The SM \PH decay width is assumed to be $\Gamma_{\mathrm{SM}}=4.1$\MeV~\cite{Denner:2011mq} for $m_{\PH}=125$\GeV.
The 95\% CL upper limit on the Yukawa couplings derived from the expression for the branching fraction above is shown in Table~\ref{tab:YukawaLimits}. The limits on the Yukawa couplings derived
from the BDT fit analysis results are shown in Fig.~\ref{fig:Yukawas}.
\begin{table}[!hbtp]
 \centering
  \topcaption{95\% CL observed upper limit on the Yukawa couplings,  for the main analysis  (BDT fit) and the cross check (\mcol fit) method.}
 \label{tab:YukawaLimits}
\begin{tabular}{c  cc }
   \hline
                        & BDT fit  &  \mcol fit \\ \hline
$\sqrt{\smash[b]{\abs{Y_{\Pgm\Pgt}}^{2}+\abs{Y_{\Pgt\Pgm}}^{2}}}$   & $<1.43\times 10^{-3}$ &  $<2.05\times 10^{-3}$  \\
$\sqrt{\abs{Y_{\Pe\Pgt}}^{2}+\abs{Y_{\Pgt\Pe}}^{2}}$     & $<2.26\times 10^{-3}$ &  $<2.45\times 10^{-3}$   \\ \hline
  \end{tabular}
\end{table}

\section{Summary}\label{sec:summary}

The search for lepton flavour violating decays of the Higgs boson in the $\Pgm\Pgt$ and $\Pe\Pgt$ channels,
with the 2016 data collected by the CMS detector, is presented in this paper. The data set analysed corresponds to an integrated
luminosity of 35.9\fbinv of proton-proton collision data recorded at $\sqrt{s}=13$\TeV.
The results are extracted by a fit to the output of a boosted decision trees discriminator trained to distinguish the signal from backgrounds. The
results are cross-checked with an alternate analysis that fits the collinear mass distribution after applying selection criteria on kinematic
variables. No evidence is found for lepton flavour violating Higgs boson decays.
The observed~(expected) limits on the branching fraction of the Higgs boson to $\Pgm\Pgt$ and to $\Pe\Pgt$ are less than 0.25\% (0.25\%) and 0.61\% (0.37\%), respectively, at 95\% confidence level. These limits constitute a significant improvement
over the previously obtained limits by CMS and ATLAS using 8\TeV proton-proton collision data corresponding to an integrated luminosity of about 20\fbinv.
Upper limits on the off-diagonal $\Pgm\Pgt$ and $\Pe\Pgt$ Yukawa couplings are derived from these constraints,
$\sqrt{\smash[b]{\abs{Y_{\Pgm\Pgt}}^{2}+\abs{Y_{\Pgt\Pgm}}^{2}}}<1.43\times 10^{-3}$ and $\sqrt{\abs{Y_{\Pe\Pgt}}^{2}+\abs{Y_{\Pgt\Pe}}^{2}}<2.26\times 10^{-3}$
at 95\% confidence level.

\begin{acknowledgments}
We congratulate our colleagues in the CERN accelerator departments for the excellent performance of the LHC and thank the technical and administrative staffs at CERN and at other CMS institutes for their contributions to the success of the CMS effort. In addition, we gratefully acknowledge the computing centres and personnel of the Worldwide LHC Computing Grid for delivering so effectively the computing infrastructure essential to our analyses. Finally, we acknowledge the enduring support for the construction and operation of the LHC and the CMS detector provided by the following funding agencies: BMWFW and FWF (Austria); FNRS and FWO (Belgium); CNPq, CAPES, FAPERJ, and FAPESP (Brazil); MES (Bulgaria); CERN; CAS, MoST, and NSFC (China); COLCIENCIAS (Colombia); MSES and CSF (Croatia); RPF (Cyprus); SENESCYT (Ecuador); MoER, ERC IUT, and ERDF (Estonia); Academy of Finland, MEC, and HIP (Finland); CEA and CNRS/IN2P3 (France); BMBF, DFG, and HGF (Germany); GSRT (Greece); OTKA and NIH (Hungary); DAE and DST (India); IPM (Iran); SFI (Ireland); INFN (Italy); MSIP and NRF (Republic of Korea); LAS (Lithuania); MOE and UM (Malaysia); BUAP, CINVESTAV, CONACYT, LNS, SEP, and UASLP-FAI (Mexico); MBIE (New Zealand); PAEC (Pakistan); MSHE and NSC (Poland); FCT (Portugal); JINR (Dubna); MON, RosAtom, RAS, RFBR and RAEP (Russia); MESTD (Serbia); SEIDI, CPAN, PCTI and FEDER (Spain); Swiss Funding Agencies (Switzerland); MST (Taipei); ThEPCenter, IPST, STAR, and NSTDA (Thailand); TUBITAK and TAEK (Turkey); NASU and SFFR (Ukraine); STFC (United Kingdom); DOE and NSF (USA).

\hyphenation{Rachada-pisek} Individuals have received support from the Marie-Curie programme and the European Research Council and Horizon 2020 Grant, contract No. 675440 (European Union); the Leventis Foundation; the A. P. Sloan Foundation; the Alexander von Humboldt Foundation; the Belgian Federal Science Policy Office; the Fonds pour la Formation \`a la Recherche dans l'Industrie et dans l'Agriculture (FRIA-Belgium); the Agentschap voor Innovatie door Wetenschap en Technologie (IWT-Belgium); the Ministry of Education, Youth and Sports (MEYS) of the Czech Republic; the Council of Science and Industrial Research, India; the HOMING PLUS programme of the Foundation for Polish Science, cofinanced from European Union, Regional Development Fund, the Mobility Plus programme of the Ministry of Science and Higher Education, the National Science Center (Poland), contracts Harmonia 2014/14/M/ST2/00428, Opus 2014/13/B/ST2/02543, 2014/15/B/ST2/03998, and 2015/19/B/ST2/02861, Sonata-bis 2012/07/E/ST2/01406; the National Priorities Research Program by Qatar National Research Fund; the Programa Severo Ochoa del Principado de Asturias; the Thalis and Aristeia programmes cofinanced by EU-ESF and the Greek NSRF; the Rachadapisek Sompot Fund for Postdoctoral Fellowship, Chulalongkorn University and the Chulalongkorn Academic into Its 2nd Century Project Advancement Project (Thailand); the Welch Foundation, contract C-1845; and the Weston Havens Foundation (USA).
\end{acknowledgments}

\bibliography{auto_generated}

\cleardoublepage \appendix\section{The CMS Collaboration \label{app:collab}}\begin{sloppypar}\hyphenpenalty=5000\widowpenalty=500\clubpenalty=5000\textbf{Yerevan Physics Institute,  Yerevan,  Armenia}\\*[0pt]
A.M.~Sirunyan, A.~Tumasyan
\vskip\cmsinstskip
\textbf{Institut f\"{u}r Hochenergiephysik,  Wien,  Austria}\\*[0pt]
W.~Adam, F.~Ambrogi, E.~Asilar, T.~Bergauer, J.~Brandstetter, E.~Brondolin, M.~Dragicevic, J.~Er\"{o}, M.~Flechl, M.~Friedl, R.~Fr\"{u}hwirth\cmsAuthorMark{1}, V.M.~Ghete, J.~Grossmann, J.~Hrubec, M.~Jeitler\cmsAuthorMark{1}, A.~K\"{o}nig, N.~Krammer, I.~Kr\"{a}tschmer, D.~Liko, T.~Madlener, I.~Mikulec, E.~Pree, N.~Rad, H.~Rohringer, J.~Schieck\cmsAuthorMark{1}, R.~Sch\"{o}fbeck, M.~Spanring, D.~Spitzbart, W.~Waltenberger, J.~Wittmann, C.-E.~Wulz\cmsAuthorMark{1}, M.~Zarucki
% ...
\vskip\cmsinstskip
% ...
1:~~Also at Vienna University of Technology, Vienna, Austria\\
% ...

\end{sloppypar}
\end{document}